\documentclass[aps,twocolumn,epsf,floats,pre,nofootinbib]{revtex4-1}
\usepackage{graphics,graphicx,epsfig}
\usepackage{amssymb,color}
\usepackage{epsf,epstopdf,wrapfig}
\usepackage{amsmath}
\usepackage{amsmath}
\usepackage{amssymb}
\usepackage{graphicx}
\usepackage{bm}
\usepackage{xr}
\usepackage{scrextend}
\usepackage{tikz-feynman}
\tikzfeynmanset{compat=1.0.0}

\usetikzlibrary{external}
\immediate\write18{mkdir -p pgf-img}
\newcommand{\de}{\partial}
\newcommand{\Ham}{\mathcal{H}}
\newcommand{\p}{\prime}

\newcommand{\mean}[1]{\bigl < #1 \bigl >}

\newcommand{\mc}[1]{\mathcal{ #1} }
\newcommand{\ddk}[1]{\frac{ d^d #1}{(2\pi)^d }}
	\newcommand{\dw}[1]{\frac{d #1}{2\pi}}
	\newcommand{\<}{\langle}
	\renewcommand{\>}{\rangle}
	\newcommand{\beq}{\begin{equation}}
	\newcommand{\eeq}{\end{equation}}
	\newcommand{\bea}{\begin{eqnarray}}
	\newcommand{\eea}{\end{eqnarray}}

	\newcommand{\bx}{ \mathbf{x}}

\newcommand{\bpsi}{\boldsymbol{\psi}}
\newcommand{\bsp}{\boldsymbol{s}}
\newcommand{\vm}{\boldsymbol{s}}
\newcommand{\bk}{ \mathbf{k}}
\newcommand{\bp}{ \mathbf{p}}
\newcommand{\bh}{ \mathbf{h}}
\newcommand{\bH}{ \mathbf{H}}

\usepackage{hyperref}
\hypersetup{
	pdfcreator = {},
	pdfproducer = {}
}

\begin{document}

\title{Renormalization group crossover in the critical dynamics of field theories with \\ mode coupling terms}

\author{
Andrea Cavagna$^{1,2}$, 
Luca Di Carlo$^{2,1}$,
Irene  Giardina$^{2,1,3}$, 
Luca Grandinetti$^{4}$,
Tomas S. Grigera$^{5,6,7}$,
Giulia Pisegna$^{2,1}$
}

\affiliation{$^1$ Istituto Sistemi Complessi, Consiglio Nazionale delle Ricerche, UOS Sapienza, 00185 Rome, Italy}
\affiliation{$^2$ Dipartimento di Fisica, Universit\`a\ Sapienza, 00185 Rome, Italy}
\affiliation{$^3$ INFN, Unit\`a di Roma 1, 00185 Rome, Italy}
\affiliation{$^4$ Dipartimento di Scienza Applicata e Tecnologia, Politecnico di Torino, Torino, Italy}
\affiliation{$^5$ Instituto de F\'\i{}sica de L\'\i{}quidos y Sistemas Biol\'ogicos CONICET -  Universidad Nacional de La Plata,  La Plata, Argentina}
\affiliation{$^6$ CCT CONICET La Plata, Consejo Nacional de Investigaciones Cient\'\i{}ficas y T\'ecnicas, Argentina}
\affiliation{$^7$ Departamento de F\'\i{}sica, Facultad de Ciencias Exactas, Universidad Nacional de La Plata, Argentina}


\begin{abstract}
Motivated by the collective behaviour of biological swarms, we study the critical dynamics of field theories with coupling between order parameter and conjugate momentum in the presence of dissipation. By performing a dynamical renormalization group calculation at one loop, we show that the violation of momentum conservation generates a crossover between a conservative yet IR-unstable fixed point, characterized by a dynamic critical exponent $z=d/2$, and a dissipative IR-stable fixed point with $z=2$. Interestingly, the two fixed points have different upper critical dimensions.  The interplay between these two fixed points gives rise to a crossover in the critical dynamics of the system, characterized by a crossover exponent $\kappa=4/d$. Such crossover is regulated by a conservation length scale, $\mc R_0$, which is larger the smaller the dissipation: beyond $\mc R_0$ the dissipative fixed point dominates, while at shorter distances dynamics is ruled by the conservative fixed point and critical exponent, a behaviour which is all the more relevant in finite-size systems with weak dissipation. We run numerical simulations in three dimensions and find a crossover between the exponents $z=3/2$ and $z=2$ in the critical slowing down of the system, confirming the renormalization group results. From the biophysical point of view, our calculation indicates that in finite-size biological groups mode-coupling terms in the equation of motion can significantly change the dynamical critical exponents even in the presence of dissipation, a step towards reconciling theory with experiments in natural swarms. Moreover, our result provides the scale within which fully conservative Bose-Einstein condensation is a good approximation in systems with weak symmetry-breaking terms violating number conservation, as quantum magnets or photon gases. 
\end{abstract}

\maketitle


\section{Introduction}

The success of the theory of critical phenomena is based upon a simple observation: systems with very different microscopic details behave in strikingly similar ways when correlations are sufficiently strong. This experimental fact eventually crossed over into theory with the formulation of the phenomenological scaling laws \cite{widom1965equation, kadanoff1966introduction, HH1967scaling, Ferrell1967}, whose key idea is that the only relevant scale ruling the spatio-temporal behaviour of a system near its critical point is the correlation length. Eventually, the great conceptual edifice of the Renormalization Group (RG) tied everything together, explaining why microscopically different systems shared so much at the macroscopic level, giving a demonstration of universality through the concept of attractive fixed points, and providing a method to calculate experimentally accessible quantities, most conspicuously the critical exponents \cite{wilson1971renormalization1, wilson1971renormalization2, wilson1972critical, wilson1974renormalization}.

Employing the same set of conceptual tools in collective biological systems could prove very helpful, given the recent massive flow of hugely diverse empirical data theory has to make sense of. In support of this strategy there is first an empirical observation regarding collective biological systems, 
namely systems in which a large numbers of units (cells, bacteria, insects, birds, mammals) interact locally in space and time giving rise to macroscopic patterns \cite{vicsek_review,marchetti_review}: these systems often exhibit unusually strong correlations, whose spatial range is significantly larger than the microscopic scales 
\cite{cavagna+al_10, zhang2010collective, attanasi2014collective, tang2017critical, mora+al_11}. Besides, recent experiments on natural swarms found evidence of dynamical scaling, a core mechanisms of statistical physics linking spatial correlation to temporal relaxation \cite{HH1967scaling, HH1969scaling}, whose validity in a biological context can hardly be considered a coincidence. Hence, despite the temptation, in front of the arresting complexity of biology, to confine ourselves to describing the specifics, we believe that exploring the path {\it correlation-scaling-RG} is a reasonable course of action. The hydrodynamic theory of flocking of Toner and Tu has led the way: it applied field-theoretical methods and the RG to bird flocks, namely collective biological systems in their strongly ordered phase \cite{toner_review,toner+al_95,toner1998flocks}. Here, we use the RG approach to study the other side of collective behaviour, namely the near-critical disordered phase of natural swarms.

In the biophysics of collective behaviour, a prominent role is played by a class of ferromagnetic theories with continuous symmetries, both in their symmetry-broken phase (flocks), and in the near-critical disordered phase (swarms) \cite{toner_review,cavagna2017dynamic}. When dynamics is taken into consideration, though, this universality class breaks down into smaller sub-classes, as there are different ways to implement the dynamics given the same static probability distribution of the system \cite{hohenberg1977theory,cardy1996scaling}. Dynamical diversity is regulated essentially by two distinct - though related - factors, namely conservation laws and symmetries. On the one hand, we have dynamical theories lacking symmetries and conservation laws (as in the classic Heisenberg model, or Model A of \cite{hohenberg1977theory}), or in which conservation is imposed despite the absence of an explicit symmetry (as in phase separation, or Model B of \cite{hohenberg1977theory}). On the other hand, we have theories ruled by symmetry and conservation laws, whose dynamics is characterized by the coupling between two fields, namely the order parameter and the conserved generator of the symmetry, i.e. the conjugate momentum. This second type of theories therefore have non-dissipative mode-coupling  terms in the equations of motions, and were originally introduced to describe systems displaying Bose-Einstein condensation (BEC), as superfluid helium, superconductivity, and quantum magnets (Models E, F, and G of \cite{hohenberg1977theory}).  Bizarre as it may seem, recent experiments suggest that some collective biological systems, as bird flocks \cite{cavagna+al_15} and insect swarms \cite{cavagna2017dynamic}, also have non-dissipative mode-coupling terms in their dynamical equations, and are thus akin to this second class of theories. The connection between BEC systems and flying animals reflects the great generality of the mathematical structure of collective dynamics governed by symmetry and conservation laws, whether the order parameter is the quantum phase of a condensate, or the direction of motion of a flock.

Here we will focus on this second class of theories, with the aim to study the critical dynamics of swarms. To make this introductory discussion more concrete, let us anticipate the actual dynamical field equations we are going to derive and analyze in detail in this work:
\begin{align}
\frac {\de \boldsymbol\psi}{\de t} &= -\Gamma_0 \frac {\delta \Ham} {\delta \bpsi}  + g_0\bpsi \times \frac {\delta \Ham} {\delta \bsp}  + \boldsymbol\theta 
\label{juemos}
\\
\frac {\de \vm} {\de t}  &= (\lambda_0 \nabla^2 - \eta_0) \frac {\delta \Ham} {\delta \vm } + g_0  \bpsi  \times \frac{\delta \Ham}{\delta \bpsi} + \boldsymbol \zeta \ ,
\label{burrito}
\end{align}
with effective Hamiltonian,
\begin{equation}
	\Ham= \int d^d x \, \biggl\{ \frac 12  (\nabla \bpsi)^2 +\frac 12 r_0 \psi^2 +  u_0 \psi^4  + \frac { s^2}{2 \chi_0} \biggr\}  \ .
\label{barrito}
\end{equation}
In the biological context the vector order parameter $\boldsymbol\psi({\bf x}, t)$ represents the velocity field, but it has different interpretations in BEC systems (for example, in liquid helium $\boldsymbol\psi$ is the expectation value of the Bose field). In all cases, though, the order parameter is coupled to its conjugate momentum, we call it spin, $\vm({\bf x}, t)$, which is the generator for rotations of $\boldsymbol\psi$, given the rotational symmetry of $\Ham$.\footnote{More precisely, the field {\it canonically} conjugate to $\bm s$ is the phase $\varphi$ of the order parameter, $\bm\psi$, not $\bm\psi$ itself; for example, in the planar case the order parameter is a complex field and $\psi= e^{i\varphi}$. This is the reason why cross products enter the dynamical equations; the relation between order parameter, phase and spin is similar to that between position, angle and angular momentum in standard rotational motion -- see \cite{cavagna2018physics} for a discussion of this point.}
The distinctive trait of this class of models are the mode-coupling cross terms, $\partial_t \boldsymbol\psi \sim  \delta_{\bsp} \Ham$ and $\partial_t \bsp \sim \delta_{\boldsymbol\psi} \Ham$, which generate a non-dissipative dynamics with the classic coordinate-momentum Hamiltonian structure; were it only for these terms, dynamics would be completely deterministic. On the other hand, the diagonal terms, $\partial_t \boldsymbol\psi \sim  \delta_{\boldsymbol\psi} \Ham$ and $\partial_t \bsp \sim \delta_{\bsp} \Ham$, give rise to the diffusion and transport phenomenology typical of stochastic statistical systems, and are thus complemented by the noises, $\boldsymbol\theta$ and $\boldsymbol\zeta$, whose variance is proportional to the kinetic coefficients, $2\Gamma_0$ and $2(-\lambda_0\nabla^2 +\eta_0)$, respectively.

The crucial feature of this theory is that, {\it in absence of dissipation}, namely when the effective friction $\eta_0$ is zero, the total integral of the spin is conserved: the cross term in \eqref{burrito} gives rise to a continuity equation for the symmetry generator, $\vm({\bf x}, t)$, prescribed by Noether's theorem, while the stochastic transport term in \eqref{burrito}, $\lambda_0 \nabla^2 \bsp$, is still the divergence of a current, leaving the continuity equation intact. This structure - symmetry and conservation - is a very profound feature of this class of models, as it leads to the existence of propagating hydrodynamic modes in the ordered phase, called spin waves; this mechanism give rise to `second sound' in liquid helium \cite{hohenberg1977theory}, it is responsible for linear information propagation in bird flocks \cite{attanasi+al_14}, and finally it explains spin-wave remnants in the near-critical phase of insect swarms \cite{cavagna2017dynamic}.

Why, then, introducing in equation \eqref{burrito} a dissipative term, $\eta_0$, which destroys spin conservation?
In the context of biological systems the answer is quite simple: the symmetry generator, or spin, is conjugated to the velocity field; hence, by rotating the velocity, the spin is what actually makes an animal to turn. Indeed, kinematically one can prove that the spin is related to the radius of curvature of the individual trajectories \cite{attanasi2015emergence}. Hence, at the individual level it is clear that there must be some dissipation relaxing the spin, thus making a trajectory straight in absence of external perturbations or interaction with the neighbours. On the other hand, in systems like superfluids or superconductors, the conservation law generated by the continuous symmetry of the quantum phase corresponds to number conservation and it cannot be violated. In other BEC systems, though, like quantum magnets \cite{bec_magnets}, exciton condensates \cite{eisenstein2004bose}, and photon gases \cite{klaers2010bose}, the Hamiltonian can contain terms that weakly violate the symmetry, hence dissipating the density in the continuity equation of the momentum. The effect of weak dissipation in the ordered phase is simply to generate a damping length scale on propagating spin-waves. However, in the near-critical phase the situation is more complicated: dissipative and non-dissipative models are known to have completely different critical exponents, hence what is the effect of dissipation in this case is unclear. This question is particularly relevant for biological swarms, as experiments found a dynamical critical exponent that cannot be reconciled with the prediction of purely dissipative theories \cite{cavagna2017dynamic}.

Here, by using a dynamical renormalization group approach, we study the effect of dissipation on the critical dynamics of a systems with mode-coupling terms. Our calculation shows that the dissipative term $\eta_0$ gives rise to an interesting crossover characterized by nontrivial critical exponents. The competition between conservative transport, $\lambda_0 \nabla^2 \bsp$, and dissipative friction, $-\eta_0 \bsp$, generates a novel conservation length scale, $\mc R_0$; beyond $\mc R_0$ the dynamics is ruled by a purely dissipative RG fixed point, so that the whole conservative (and propagating) nature of the theory is lost, whereas for distances smaller than $\mc R_0$, the conservative RG fixed point governs the dynamics, giving rise to the classic spin-wave phenomenology. We calculate the value of the dynamical critical exponents in these two regimes and of the crossover exponent, and we confirm our results through numerical simulations.

As we shall see, the conservation scale $\mc R_0$ is larger the smaller the dissipation. The implications of this fact are very important in the biophysical context. The presence of dissipation in the dynamical equations of biological groups may suggest that these systems are in the same universality classes as fully dissipative models, as dissipation always wins over conservative terms in the infinite-time and infinite-distance hydrodynamic limit. However, real biological groups are of course finite-size systems (and quite moderately sized, in the case of flocks and swarms), in which dissipation has been demonstrated by experiments to be quite low \cite{cavagna2017dynamic}. Therefore, the size of these systems may actually be {\it smaller} than the crossover scale $\mc R_0$, so that, even if dissipative terms are present in the equations of motion, critical dynamics is still ruled by the symmetric and conservative structure of the equations, and therefore have critical exponents drastically different from the dissipative ones. As we shall see, for natural swarms this theoretical mechanism produces a critical dynamics whose phenomenology is remarkable similar to that found in experiments.

Although our motivation is biological, it is worthwhile to remark that our results apply to any BEC system with weak dissipation, a relevant example of which are quantum magnets \cite{bec_magnets}, exciton condensates \cite{eisenstein2004bose} and photon gases \cite{klaers2010bose}. In quantum magnets Bose-Einstein condensation of magnons occurs at low temperature, due to the spontaneous breaking of the $U(1)$ symmetry; real quantum magnets, though, contain weakly symmetry-breaking terms in their Hamiltonian, thus violating the conservation of the conjugate momentum. Another BEC system our results could be applied to is that of excitons, bosonic hole-particle excitations created by laser pumps, whose number, though, is conserved only within their lifetime (which is finite, and dependent on many factors) \cite{eisenstein2004bose}. A similar situation arises within the context of photon gases, when a polariton condensate emerges \cite{klaers2010bose}; in this case, too, depending on the polariton lifetime, one can have a violation of the number conservation symmetry, which is equivalent to an effective dissipation.  In all these cases, our calculation could provide the crossover scale within which an exact BEC assumption is justified and it may describe the critical behaviour of the crossover.

Here is the plan of the paper. In Section II we will give a derivation of the microscopic dynamical equations in their biological context, whereas in Section III we will coarse-grain the microscopic equations and work out the dynamical field theory described by equations \eqref{juemos} and \eqref{burrito}. In Section IV we will perform a renormalization group calculation of critical dynamics in the momentum shell context; this Section will culminate with the formulation of the RG recursive equations, while the analysis of the crossover between the two different fixed points on the critical manifold, and the corresponding crossover of the critical dynamics, will be studied in Section V. In Section VI we will give an alternative derivation of our results using the more field-theoretical Callan-Symanzik approach. In Section VII we will perform numerical simulations to validate the RG results, and finally we will present our conclusions and discuss the outlook in Section VIII. Parts of the most technical material are contained in the Appendixes. A shorter account of our results can be found in \cite{companion_short}.


\section{Biophysical origin of the microscopic model}

In this Section we derive the microscopic model of collective behaviour that we will use to describe the dynamics of natural swarms. Because this model was first introduced in the context of flocks, rather than swarms, we will have to take a short detour in that direction. At the end of the Section we will discuss under what approximations we will be able to perform a field-theoretical RG study of the model.

\subsection{Collective behaviour and the Vicsek model} 
Collective behaviour in biological systems, and more specifically collective motion, is essentially a game of mutual imitation, in which each individual tries to make its own state of motion as close as possible to that of its neighbours \cite{vicsek_review}. From a physical point of view, such mechanism is clearly suggestive of a ferromagnetic-like interaction: if we focus our attention on the direction of motion of each individual, that is on the orientation of the velocity vector, such imitation game amounts to a local interaction due to which each (normalized) velocity vector tends to align to those of its neighbours, much as classical Heisenberg spins tend to align to each other. At variance with standard ferromagnets, though, in collective motion the positions of the particles change in time, as they are carried around by their own velocities, thus creating a non-equilibrium feedback between the alignment degrees of freedom and the interaction network \cite{toner+al_95, ginelli2016physics}.  The simplest yet most illuminating model describing this core mechanism of collective motion was introduced by Vicsek and co-workers \cite{vicsek+al_95}; it describes a set of self-propelled particles that interact with each other in a ferromagnetic way,
\begin{align}
\hat \eta\frac{d  \boldsymbol v_i }{dt} &= 
  \boldsymbol v_i \times  \hat J \sum_{j} n_{ij}(t) \boldsymbol {v}_j + \boldsymbol v_i \times \boldsymbol{\zeta}_i  
  \label{microVicsek} \\
\frac{d \boldsymbol r_i}{dt} &= \boldsymbol v_i  \ ,
\label{motion}
\end{align}
where $\boldsymbol r_i$ is the position of particle $i$, $\bm{v}_i$ its velocity, and $n_{ij}(t)$ is the (short-ranged) adjacency matrix (who is neighbour of whom) at time $t$. The interaction between individuals is given by the ferromagnetic term in \eqref{microVicsek}, where $\hat J$ gives the strength of the tendency to align to each other.\footnote{We use hatted parameters in the microscopic equations to distinguish them from their coarse-grained counterpart in the field equations later on.} Such alignment interaction is often called {\it social force} in the collective behaviour literature \cite{vicsek_review}. In the Vicsek model the speed is kept fixed, $|\boldsymbol v_i| = 1$, which is the purpose of the cross-products at the r.h.s. of \eqref{microVicsek}.
The term $\boldsymbol{\zeta}_i$ is a Gaussian white noise with variance,
\begin{equation}
\< \boldsymbol{\zeta}_i (t) \cdot \boldsymbol{\zeta}_j(t^\p) \> = 2d\, \hat\eta \,T\, \delta_{ij} \delta(t-t')  \ , 
\end{equation}
where $\hat \eta$ is a dissipation coefficient and $T$ a generalized temperature measuring the strength of the noise.

The power of the Vicsek model is that it describes collective motion in its two different phases. When noise is low (or density is high, in the metric case \cite{vicsek_review}), the alignment interaction produces long-range order across the system, forming a polarized moving {\it flock}. The interesting thing is that such ordering also occurs in two dimensions, which would be forbidden by the Mermin-Wagner theorem \cite{mermin+al_66} in an equilibrium ferromagnet with continuous symmetry; however, the Vicsek model has an off-equilibrium feedback between alignment and self-propulsion promoting long-range order \cite{toner+al_95}. On the other hand, when noise is large enough (or density is low, in the metric case \cite{vicsek_review}), the system is in a disordered (paramagnetic) phase, which reproduces quite well the statistical properties of real {\it swarms}. More precisely, it has been observed that natural swarms are disordered, but highly correlated systems \cite{attanasi2014finite}; the velocity static correlations are reproduced (at least qualitatively) by the Vicsek model close to its ordering transition. Hence, the Vicsek model captures rather well the static correlation functions of collective motion for both flocks and swarms. Dynamics is more problematic, though, at both the qualitative and the quantitative level.

\subsection{The Inertial Spin Model}
The first hint that the Vicsek equation of collective motion required some new ingredients came from experiments on flocks, in which it was observed that disturbances in the direction of motion of the birds (that is, turns) propagate linearly, with very low dissipation \cite{attanasi+al_14}. Although the hydrodynamic field-theoretical description of the Vicsek model introduced by Toner and Tu \cite{toner_review} contains linearly propagating `sound' modes, caused by the feedback between local density and phase fluctuations \cite{toner+al_98}, experiments indicate that flocks follow a different mechanism: during the propagating event the density displays very weak fluctuations, if any; moreover, the speed of propagation of the wave has been found to be higher the higher the polarization of the group, a feature absent in the hydrodynamic theory of the Vicsek model \cite{toner_review} (see also the discussion in \cite{cavagna2018physics}). It was therefore suggested in \cite{attanasi+al_14} and \cite{cavagna+al_15} that Vicsek dynamics had to be complemented with some non-dissipative inertial couplings between order parameter (the velocity) and a conjugate momentum, in order to reproduce the structure of a conservative Hamiltonian dynamics. The resulting microscopic dynamical equations give rise to the Inertial Spin Model (ISM) of collective motion \cite{cavagna+al_15},
\begin{equation}
\label{microISM}
\begin{split}
\frac{d  \boldsymbol v_i }{dt} &= \frac 1 {\hat \chi} \boldsymbol s_i \times \boldsymbol v_i \\
\frac{d \boldsymbol s_i}{dt} &=  \boldsymbol {v}_i \times \hat J \sum_{j} n_{ij}(t) \boldsymbol {v}_j  - \frac{\hat \eta}{\hat\chi} \boldsymbol {s_i} + \boldsymbol  {v}_i \times \boldsymbol{\zeta}_i   \\
\frac{d \boldsymbol r_i}{dt} &= \boldsymbol v_i  \ ,
\end{split}
\end{equation}
where the new variable $\bsp_i$ represents a generalized momentum conjugated to the velocities $\boldsymbol v_i$ and it is the generator of the rotational symmetry of the interaction; it is therefore called {\it spin}, in an analogy with quantum mechanics. Associated to the momentum $\bsp_i$ we have a generalized inertia, $\hat \chi$, which embodies the resistance of a particle to change its instantaneous radius of curvature \cite{cavagna_review}. One can show that, in the low noise, strongly polarized phase, the non-dissipative coupling between spin and velocity of the ISM generates linear propagating modes of the velocity fluctuations, which match quite accurately the experimental results, including the key relation between speed of propagation and polarization \cite{cavagna2015silent}.

In absence of a dissipative term, the Hamiltonian structure of the ISM would conserve the total spin, as it happens for any generator of a symmetry. However, one can show that the spin is essentially the instantaneous curvature of the particle's trajectory \cite{attanasi+al_14}, hence a single particle (or bird, in a flock) would maintain its radius of curvature forever, were the spin strictly conserved. This is quite unrealistic. Rather, it seems reasonable to expect curvature (and therefore spin) to be dissipated in the long run in absence of interaction or external perturbations. For this reason the ISM has also the dissipative term, $ - \hat \eta \bsp_i$, and stochastic noise, $ \boldsymbol{\zeta}_i$, granting relaxation of the spin for large times. If dissipation is small, though, and the biological group has finite size, linear waves will still propagate across the system, before dissipation kicks in \cite{cavagna+al_15}. In other words, although in the hydrodynamic limit (infinitely large times and distances) the conservative Hamiltonian structure always becomes irrelevant, on the finite-time and finite-size scales typical of biological phenomena the interplay between velocity and spin has crucial consequences on signal propagation. Note, finally, that once dissipation is included in the equations, one can recover the Vicsek model as the over-damped limit of the ISM \cite{cavagna+al_15}, which is quite reassuring.

The second hint that a model with non-dissipative dynamics was required came from swarms.
Swarms of insects are systems apparently completely different from flocks: they show no group-scale coordination, so that their net motion is zero: swarms `dance' above some landmark in seemingly random fashion \cite{attanasi2014collective}. In fact, experiments on natural swarms \cite{attanasi2014collective} showed that these systems have strong velocity correlations, indicating that, despite the lack of long-range order, the individuals within these groups are interacting with each other rather intensely, hence driving the system close to an ordering transition; indeed, such static correlations were qualitatively similar to those developed by the Vicsek model at its critical point \cite{attanasi2014finite}. More recent experiments \cite{cavagna2017dynamic} showed that swarms in their natural environment exhibit another important property of classical statistical physics, namely dynamic scaling \cite{HH1967scaling}: according to this law,  the dynamic correlation function of a system close to the critical point obeys the following relations,
\begin{equation}
\label{dynscal}
\begin{split}
C(k,t) &= C_0(k) \; F\left( t/\tau_k, k\xi\right) \\ 
   \tau_k &= k^{-z} f(k \xi) \ ,
\end{split}
\end{equation}
where $t$ is time, $k$ momentum, $C_0$ is the static correlation function, $\tau_k$ is the relaxation time of mode $k$, $F$ and $f$ are well-behaved scaling functions, and $z$ is the {\it dynamic critical exponent}, ruling how space and time scale with each other. The key idea of dynamic scaling is that the only relevant scale in ruling both spatial {\it and} temporal behaviour of a system close to the critical point, is the correlation length, $\xi$. For $k=0$ we obtain, $\tau \sim \xi^{z}$, a property known as {\it critical slowing down}: a system strongly correlated in space must also be strongly correlated in time \cite{hohenberg1977theory}. Experiments showed that swarms satisfy relations \eqref{dynscal} with a dynamic critical exponent $z \approx 1$, whereas numerical simulations of the Vicsek model in $d=3$ give $z\approx 2$ \cite{cavagna2017dynamic}. It must be noted that $z=2$ is the exact value of the dynamical critical exponent for a purely dissipative free theory (Gaussian model), and that even in the interacting case the exponent receives only very small (two loops) corrections to the value 2, if the dynamics has only dissipative terms (or even values larger than $2$, as in the case of Model B \cite{hohenberg1977theory}). On the other hand, dynamical models with non-dissipative inertial terms tend to have values of the exponent $z$ significantly {\it smaller} than $2$, as a result of the interplay between order parameter and conjugate momentum \cite{hohenberg1977theory}. Hence, the low value of $z$ in natural swarms was a further indication of the need for non-dissipative dynamics also in these non-polarized systems.
Finally, dynamic relaxation in the Vicsek model has a classic exponential form \cite{cavagna2017dynamic}, while natural swarms display a completely different shape, showing clear evidence of non-dissipative inertial behaviour for short times. More precisely, if we define the relaxation form factor \cite{cavagna2017dynamic}, $h=\dot C(t/\tau)/C(t/\tau)$, in the limit $t/\tau\to0$ we have that $h\to 1$ for the Vicsek model (as for any exponential correlation function), while experiments showed $h\to 0$ for natural swarms, as it would happen in a weakly damped harmonic system, where inertia dominates over dissipation \cite{cavagna2017dynamic}. Hence, the whole dynamical behaviour of swarms seems to require the existence of non-dissipative inertial terms in the equations of motion, which is exactly the extra ingredient the ISM has compared to the Vicsek model.

Our plan is therefore to study the critical dynamics of the ISM in its disordered yet near-critical phase, to describe the collective behaviour of natural swarms of insects, in order to try and reproduce the experimental results of \cite{cavagna2017dynamic}. Because the ISM was originally introduced to describe the dynamics of flocks, it has been studied extensively in its deeply ordered (i.e. polarized) phase, both numerically \cite{cavagna+al_15}, and theoretically \cite{cavagna2015silent}, while no previous studies of the ISM in the near-critical regime have been performed.

\subsection{Fixed network approximation}
Before we proceed with the coarse-graining of the model, though, we need to decide whether to attack directly the full-fledged off-equilibrium problem, which includes the self-propelled nature of collective motion, or whether we first take on the simpler (and yet non-trivial) equilibrium problem, in which particles sit on a fixed network and thus have a time-independent interaction matrix. For a number of reasons, we will follow the second strategy. The model we want to study differs from previous known cases in two main respects: {\it i)} it contains non-dissipative terms {\it and} effective friction, the interplay of which has never been studied before, not even at equilibrium; {\it ii)} the model is a self-propelled one, hence intrinsically off-equilibrium, which may seem particularly important in the swarm phase, in which each particle changes the local neighbours quite rapidly. Our central experimental concern is to reproduce the correct dynamical critical exponent $z$ and the correct relaxation form factor in natural swarms. The fact that the self-propelled, off-equilibrium Vicsek model in its swarm phase gives exactly the same exponent and form factor as equilibrium fully dissipative models (as the classical Heisenberg model), suggests that self-propulsion is not the primary source of the anomalous critical dynamics of swarms. Moreover, we believe that having under control the equilibrium problem puts us in a more solid position to tackle the off-equilibrium one in the future, much as knowing the physics of the equilibrium XY and Heinseberg models has been fundamental to fully understand and appreciate the Vicsek model.
Hence, we will study a fixed-network version of the ISM, in which the particles belong to a lattice and the connectivity matrix does not depend on time. In this context, the order parameter no longer has the role of a physical velocity, hence we will call it $\bm \psi$, the generic symbol for the order parameter, and write the microscopic model in the following way, 
\begin{align}
\frac{d \bm\psi_i}{d t} &=\frac{1}{\hat \chi} \bm s_i \times \bm \psi_i  
\label{ISM1_micro} \\
\frac{d \bm s_i}{d t} &=   \bm\psi_i \times  \hat J  \sum_{j} n_{ij} \bm\psi_j - \frac{\hat\eta}{\hat \chi}  \bm s_i + \bm\psi_i\bm\times \bm \zeta_i \ .
\label{ISM2_micro}
\end{align}
The modulus of the order parameter is still fixed to $|\bm\psi_i |^2=1$ and the adjacency matrix $n_{ij}$ now corresponds to a fixed interaction network. Thanks to this approximation, dynamics can now be rewritten in Hamiltonian terms,
\begin{align}
\frac{d \bm\psi_i}{d t} &=\ -\bm\psi_i \times \frac{\de H}{\de \bm s_i}  \ 
\label{ISM1_micro_2} \\
\frac{d \bm s_i}{d t} &=  -  \bm\psi_i \times  \frac{\partial H }{\de \bm \psi_i}- \hat\eta \frac{\partial H}{\de \bm s_i} +  \bm\psi_i \times  \bm \zeta_i \ ,
\label{ISM2_micro_2}
\end{align}
with microscopic Hamiltonian,
\begin{equation}
    H = - \hat J \sum_{i,j} n_{ij} \bm \psi_i \cdot \bm \psi_j +\sum_i \frac{s_i^2}{2 \hat \chi}  \ .
    \label{ham}
\end{equation}
This is the dynamical model we now proceed to coarse-grain in order to obtain a dynamical field theory.


\section{Coarse-grained field theory}

\subsection{Equations of motion}
Since we are interested in describing the large scale behavior of the system, it is convenient to pass from a microscopic description in terms of site dependent variables to a field description, where we consider smoothly varying velocity and spin fields ${\boldsymbol\psi}({\boldsymbol x},t)$, and ${\boldsymbol s}({\boldsymbol x},t)$, obtained by coarse graining the original variables over a small spatial volume. Upon coarse-graining, the original Hamiltonian (\ref{ham}) gives rise to an effective field Hamiltonian $\Ham[\bpsi,\vm]$ that - as in standard ferromagnetic systems - reads \cite{goldenfeld_lectures_1992,binney_book}
\begin{equation}
	\Ham[\bpsi,\vm] = \int d^d x  \biggl\{\frac 12  (\nabla \bpsi)^2 +  \frac 12 r_0 \psi^2 +  u_0 \psi^4  + \frac {s^2}{2 \chi_0}   \biggr\} \ ,\\
\nonumber
\end{equation}
where $r_0$ is the bare mass (negative in the ordered phase), $u_0$ is the bare static coupling constant and $\chi_0$ is the effective inertia.  Here, we remind, the gradient term comes from the mutual alignment interaction, which favours smoother configurations; the quadratic and quartic contributions for ${\boldsymbol\psi}$ represent a confining potential and derive from the original constraint on the ${\boldsymbol\psi}_i$ and the coarse-graining entropy; while the field $\vm$  remains Gaussian as its microscopic counterpart. 

When writing down the dynamical equation of motion for the fields, we need to take into account the presence of both the reversible and dissipative contributions present in the microscopic dynamics (\ref{ISM1_micro_2})(\ref{ISM2_micro_2}), and add additional dissipative terms, which might arise upon coarse graining. Under very general assumptions \cite{hohenberg1977theory, tauber2014critical}, we can therefore write
\begin{align}
\frac {\de \boldsymbol\psi}{\de t} &= -\Gamma_0 \frac {\delta \Ham} {\delta \bpsi}  + g_0\bpsi \times \frac {\delta \Ham} {\delta \bsp}  + \boldsymbol\theta 
\label{EqOfMotion1}\\
\frac {\de \vm} {\de t}  &= (\lambda_0 \nabla^2 - \eta_0) \frac {\delta \Ham} {\delta \vm } + g_0  \bpsi  \times \frac{\delta \Ham}{\delta \bpsi} + \boldsymbol \zeta \ ,
\label{EqOfMotion2}
\end{align}
where the noise correlations are chosen to have a Boltzmann-like static probability distribution, i.e.
\begin{equation}
\begin{split}
	\mean{\theta_\alpha(\bx,t) \theta_\beta(\bx^\p,t^\p )} &= 2\Gamma_0 \delta_{\alpha\beta} \delta^{(d)}(\bx-\bx^\p) \delta(t-t^\p)\\
\mean{\zeta_\alpha(\bx,t) \zeta_\beta(\bx^\p,t^\p )} &=  2(\eta_0 -\lambda_0 \nabla^2)\delta_{\alpha \beta} \delta^{(d)}(\bx-\bx^\p) \delta(t-t^\p)
\label{noisazzo}
\end{split}
\end{equation}
Here $\Gamma_0$, $\eta_0$ and $\lambda_0$ are the bare kinetic coefficient of the field $\bpsi$, the bare friction coefficient and transport coefficient of the field $\vm$, respectively, while $g_0$ is a mode-coupling constant that regulates the reversible dynamical terms and describes the symmetry properties relating the two fields: the fact that $\vm$ is the infinitesimal generator of rotations of $\bpsi$ is indeed specified by the Poisson commutation rules,
\begin{equation}
\{\psi_\alpha, s_\beta\}=g_0\, \epsilon_{\alpha \beta \gamma}  \psi_\gamma \ ; \quad \quad\{s_\alpha, s_\beta\}=g_0 \, \epsilon_{\alpha \beta \gamma}  s_\gamma \ ,
\end{equation}
where $\epsilon_{\alpha \beta \gamma}$ is the Levi-Civita antisymmetric symbol.\footnote{Note that we reabsorbed the minus sign in front of the cross products in (\ref{ISM1_micro_2}-\ref{ISM2_micro_2}) into the definition of the coarse-grained dynamical coupling constant, $g_0$, so to obtain in (\ref{EqOfMotion1}-\ref{EqOfMotion2}) the same field-theory notation as the classic reference papers, \cite{halperin1976renormalization} and \cite{de1978field}.}

The static properties of the model only depend  on the Hamiltonian $\Ham$. For the field $\bpsi$ they are therefore the same as in the Heisenberg model \cite{hohenberg1977theory}, with an ordering phase transition occurring for $r_0= r_c$. On the other hand, at the static level $\vm$ is a trivial, purely massive, Gaussian field. Since there is no static coupling term between this field and the order parameter, the inertia $\chi_0$ will not acquire any perturbative contributions; hence, in order to simplify our notation, we choose the units of $\vm$ such that $\chi_0 = 1$.
The dynamic properties are ruled by the transport coefficient $\lambda_0$, by the effective friction $\eta_0$, by the kinetic coefficient $\Gamma_0$ and by the dynamic coupling constant $g_0$; these quantities will take perturbative contributions arising from the dynamic interaction between $\vm$ and $\bpsi$, which is ruled by $g_0$.

Equations \eqref{EqOfMotion1}-\eqref{EqOfMotion2} have two additional dissipative terms compared to the microscopic theory of Eqs. \eqref{ISM1_micro_2}-\eqref{ISM2_micro_2}, namely $-\Gamma_0\delta{\cal H}/\delta\bpsi$ and $\lambda_0 \nabla^2\delta{\cal H}/\delta\vm$. The first term actually contains two contributions: first, a derivative of the confining potential, $\psi^2 +\psi^4$, which is the coarse-grained analogue of the microscopic sharp constraint, $|\bpsi_i|^2=1$; second, a diffusive piece, $\Gamma_0\nabla^2\bpsi({\bf x},t)$, which derives from a loss of reversibility due to the coarse-graining, and which describes the role of the fluctuations of the order parameter in the relaxation process; even though such fluctuations are negligible in the low temperature phase (where we recover the microscopic theory with $\Gamma_0 = 0$) they are crucial when considering the system close to the critical point. For this reason, even though we neglected the Laplacian of the order parameter in our previous analysis in the deeply ordered phase \cite{cavagna2015silent}, we need to take it into account in the present study of the critical regime.

On the other hand, the origin of the spin transport term, $\lambda_0\nabla^2 \vm({\bf x},t)$,  is perhaps less intuitive. In the microscopic model the spin is dissipated by the friction through the term $-\hat \eta \vm_i(t)$, hence one might have expected just a term $-\eta_0 {\vm}({\bf x},t) $ in the coarse-grained theory. Why then are we introducing the term in $\lambda_0 \nabla^2 \vm({\bf x},t)$? We will show in the following Sections that, in the context of perturbation theory and the renormalization group, such term arises naturally from the non linear interaction between the two fields once a momentum shell integration is performed. It is then necessary to include the transport coefficient $\lambda_0$ directly from the starting field equations. 

We notice that for $\eta_0=0$, equations \eqref{EqOfMotion1} and \eqref{EqOfMotion2} coincide with those of  Model G (antiferromagnet), or, in the planar case, of Model E (liquid helium), which have a fully conserved spin dynamics and whose critical dynamical properties have been studied long ago in a series of seminal papers \cite{halperin1976renormalization, de1978field}. The renormalization group a\-na\-ly\-sis described in the following sections will show that in an appropriate regime (when $\eta_0$ is small), the ISM displays the same critical behaviour as these fully conservative models.

\subsection{Free theory in Fourier space}

The starting point to build the perturbative expansion of the equations of motion is the free (or non-interacting) theory, which is obtained by setting to zero the non-linear coupling constants, namely $g_0=0$ and $u_0=0$. In Fourier variables, hence using momentum $k$ and frequency $\omega$, the free equations of motion become, 
\begin{align}
-i\omega\; \boldsymbol\psi({\boldsymbol k}, \omega) &= -\Gamma_0 (k^2 + r_0)\; \bpsi({\boldsymbol k}, \omega)  + \boldsymbol\theta({\boldsymbol k}, \omega) 
\label{sallo}\\
-i\omega\;\vm({\boldsymbol k}, \omega)   &= -(\eta_0+ \lambda_0 k^2 ) \; \vm({\boldsymbol k}, \omega)  + \boldsymbol \zeta({\boldsymbol k}, \omega) \ .
\label{salla}
\end{align}  
The free theory is linear and it is therefore possible to solve it exactly by merely inverting equations \eqref{sallo} and \eqref{salla}, 
\begin{align}
\boldsymbol\psi({\boldsymbol k}, \omega) &= G_{0, \psi}({\boldsymbol k}, \omega) \boldsymbol\theta({\boldsymbol k}, \omega)
\label{sao}\\
\vm({\boldsymbol k}, \omega)  &= G_{0, s}({\boldsymbol k}, \omega) \boldsymbol \zeta({\boldsymbol k}, \omega) \ ,
\label{saa}
\end{align}  
where the free propagators (or Green functions) are the inverse of the dynamical operators in Fourier space,
   	\begin{align}
	\label{freepropPsi}
		G^{-1}_{0, \psi}({\boldsymbol k}, \omega) =& -i \omega + \Gamma_0 (k^2 + r_0)
	\\
	\label{freepropS}
	G^{-1}_{0, s}({\boldsymbol k}, \omega) =& -i \omega +( \eta_0 + \lambda_0 k^2) \ .
	\end{align}
	The propagators describe the response of the fields to noise and to external perturbations \cite{lanczos1961linear}. We can also define the free dynamic correlation functions,
	\begin{align}
	C_{0,\psi}(k,\omega) &=\langle \boldsymbol\psi({\boldsymbol k}, \omega) \boldsymbol\psi(-{\boldsymbol k}, -\omega) \rangle \\
	C_{0,s}(k,\omega) &=\langle \vm({\boldsymbol k}, \omega) \vm(-{\boldsymbol k}, -\omega) \rangle \ .
	\end{align}
	By using \eqref{sao} and \eqref{saa}, and the noise correlators \eqref{noisazzo}, we get the relations, 
	\begin{align}
	C_{0,\psi} &= 2 \Gamma_0 | G_{0,\psi}|^2 \label{freecorrePsi}\\
    C_{0,s} &= 2 (\eta_0 + \lambda_0 k^2) | G_{0,s}|^2  \ .
    \label{freecorreS}
	\end{align}
These four quantities, propagators and correlation functions, are the building blocks of the perturbative expansion.

Calculations in the RG context are carried out in Fourier space, hence all relevant integrals are performed over the momentum, $k$; in the infinite size limit the lowest extreme of integration is $k=0$  (otherwise, in finite-size systems, it is of order $1/L$), whereas the upper extreme of integration is a momentum scale - the so-called {\it cutoff} - indicated by $\Lambda$, corresponding to the inverse of the length scale over which the coarse-graining has been performed; practically speaking, if the continuous field has been obtained by averaging the microscopic variables on a volume of linear size ${\cal L}$, we have $\Lambda=1/{\cal L}$. In principle the coarse-graining is supposed to be performed over a scale much larger than the lattice spacing, $a$; in practice, though, $\cal L$ is still a microscopic length scale of the system, so that, broadly speaking, one often assumes that $\Lambda$ is of order $1/a$.

The cutoff is an arbitrary scale, which therefore appears as an extra unknown parameter of the theory. In fact, {\it all} bare parameters in the theory, $r_0, u_0, \Gamma_0, \lambda_0, \eta_0, g_0$, depend on $\Lambda$, and therefore they are all equally unknown. As we shall see, the central idea of the renormalization group is to exploit constructively the arbitrariness of $\Lambda$, by studying how the bare parameters change when $\Lambda$ is changed; from this flow the critical properties of the theory will emerge.


\section{Renormalization Group in momentum shell}

Broadly speaking, the renormalization group is a set of symmetry transformations that are useful to determine the scale invariance properties of a system at its critical point \cite{wilson1971renormalization1}. An RG transformation unfolds through two stages: {\it i)} integration of the short wavelength details and, {\it ii)} rescaling of length and time.  The first operation amounts to integrating the fields over large values of the momentum, $\Lambda/b<k<\Lambda$, where $b$ is a rescaling factor larger than - but close to -  $1$; this integration interval is the so-called {\it momentum shell}. The effect of integration is to shift the cutoff from $\Lambda$ to $\Lambda/b$; the RG idea is that the long-distance physics of a system close to the critical point, where the correlation length is large, cannot change due to an arbitrary change of the cutoff. Hence, the second stage consists in rescaling space (and consequently time) in such a way to formally restore the original cutoff $\Lambda$ and to compare the newly obtained equations to the original ones. The compound effect of these two stages is to effectively change the parameters that appears in the equations of motion, hence determining a flow in the space of parameters.  At the critical point, where the correlation length is infinite, the RG transformation must have left the system exactly the same, and therefore the fixed points of the RG flow provide all the important information on the large scale physical properties of the system. 

The RG technique is nowadays standard and discussed in the literature both for static and dynamical problems \cite{cardy1996scaling,parisi_book, amit2005field, tauber2014critical}. In this section we adopt a momentum shell renormalization scheme \cite{wilson1971renormalization1}, as this is the approach that was used in the original papers on critical dynamics \cite{halperin1972calculation, halperin1976renormalization, hohenberg1977theory}. This will allow us to immediately spot differences with respect to the fully conservative case \cite{halperin1976renormalization}. In Section VI we will illustrate how the same results can be obtained using a Callan-Symanzik approach, more common in recent applications of the dynamical renormalization group \cite{tauber2014critical}.


\subsection{Integration of the short-wavelength details}

In the first stage of the RG we integrate out short wavelength fluctuations, namely modes with $\Lambda/b<k<\Lambda$. This operation (described in Appendix D) leads to a new effective theory that only depends on fields fluctuating over larger wavelengths, $k<\Lambda/b$. In the free theory, modes at different wave vectors are independent, so this operation has no practical effects. On the other hand, when non-linear interactions are present, the coupling between long and short wavelength modes makes it impossible to carry our exactly this operation, which therefore requires a perturbative expansion that we describe in detail in Appendix A. The bottom line result of shell integration is to produce additional terms in the equations of motion that effectively modify the coefficients of both the linear and the non-linear terms. We start with the linear dynamical coefficients, namely $\Gamma_0, \lambda_0$ and $\eta_0$. These parameters are contained in the free propagators, eqs. \eqref{freepropPsi} and \eqref{freepropS}. Due to the shell integration, the propagators acquire some new contributions, the so-called on-shell self-energies, $\Sigma_b$ and $\Pi_b$, so we can write,
\begin{align}
G_{\psi}^{-1}({\bf k},\omega) &\ =  -i\omega + \Gamma_0(k^2 + r_0) - \Sigma_b({\bf k},\omega) \ 
 \label{prop1} \\
G_{s}^{-1}({\bf k},\omega) &\ = -i\omega + (\eta_0 + \lambda_0 k^2) -\Pi_b({\bf k},\omega) \ .
\label{prop2}
\end{align}
As we said, to compute the self-energies one uses perturbation theory. To carry out the expansion we follow the generating functional approach of Martin-Siggia-Rose \cite{MartinSiggiaRose1973,de1978field,cardy1996scaling}, where averages of physical observables over the stochastic dynamics are rewritten as thermal averages over a functional measure. The complication is that new auxiliary fields  must be introduced and the effective field theory therefore involves four fields, rather than two. The advantage is that the standard Feynman technique can be used to perform a diagrammatic expansion, and perturbation theory can be carried out in the same way as in equilibrium statistical field theory. The details can be found in Appendix A. 
To one loop order, the self-energies read, 
\begin{widetext}
\begin{align}
\Sigma_{b}(\bk,\omega)= &
-2 g_0^2 \int_{\Lambda/b}^\Lambda\ddk p   \frac {(k^2+r_0)}{(p^2+r_0)(-i\omega + \Gamma_0(p^2+r_0) +\lambda_0(\bk-\bp)^2+\eta_0 ) } \label{sigma} \\
\Pi_{b}(\bk,\omega) = &
 -g_0^2 \int_{\Lambda/b}^\Lambda \ddk p 
 \frac{[p^2-(\bk-\bp)^2]^2}{(p^2+r_0)[(\bk-\bp)^2+r_0] [-i\omega + \Gamma_0(p^2+(\bk-\bp)^2 +2r_0)]} \label{Pi} \ .
\end{align}
\end{widetext}
The self-energies modify the poles of the propagators in the frequency plane, therefore affecting the way response and correlation functions decay in time. In particular, the $k\to 0$ and $\omega\to 0$ expansion of the self-energies and of their derivatives modify the kinetic and transport coefficients, so that we can define their renormalized values,
\begin{align}
\Gamma_R \equiv \ & \  \frac{\partial G_{\psi}^{-1}}{\partial k^2}\bigg|_ {k=0 \atop  \omega=0} = \Gamma_0 \left( 1 -  \frac{1}{\Gamma_0}\frac{\partial \Sigma_b}{\partial k^2}\bigg|_{k=0 \atop \omega=0} \right) \\
\lambda_R \equiv \ &\  \frac{\partial G_{s}^{-1}}{\partial k^2}\bigg|_{k=0 \atop \omega=0} =
\lambda_0 \left( 1 -  \frac{1}{\lambda_0}\frac{\partial \Pi_b}{\partial k^2}\bigg|_{k=0 \atop \omega=0} \right) \   \end{align}
\begin{equation}
\eta_R \equiv \ \ G_{s}^{-1}\bigg|_{k=0 \atop \omega=0}=\eta_0  \left( 1 -  \frac{1}{\eta_0} \Pi_b\bigg|_{k=0 \atop \omega=0} \right) \ .
\label{RenormalizedParameters}
\end{equation}
First of all, we notice an important point: from \eqref{Pi} we immediately see that $\Pi_b(k=0)=0$, and therefore we conclude that the effective friction $\eta_0$ has no perturbative corrections. As discussed in Appendix A, from the diagrammatic point of view this is a consequence of the structure of the vertex, which makes all perturbative contributions to $\Pi_b(k=0)$ equal to zero; therefore this result is valid to all orders in perturbation theory. Physically, this fact is a consequence of the rotational symmetry of the theory: even though $\eta_0$ breaks the conservation law of the spin, the symmetry is still at work, implying that it is impossible for the conservative theory to produce a non-conservative friction through coarse-graining.

The accuracy of the perturbation expansion is increased by substituting the bare mass $r_0$ with its renormalized value  $r$ \cite{amit2005field}, which represents the inverse static susceptibility and goes to zero when the systems approaches the critical temperature. 
Since we are interested in the critical behavior, from now on we will evaluate all integrals at $r=0$, namely at the critical point; given that integrals are on the shell there are no infrared singularities and the self-energies are finite. We thus have,
\begin{align}
\Gamma_R&=\Gamma_0\left[ 1+ 2\, \frac{g_0^2}{\Gamma_0} 
\int_{\Lambda/b}^\Lambda \ddk p\ \frac{1}{p^2[(\Gamma_0+\lambda_0)p^2 +\eta_0]}
\right] \label{buffo}
\\
\lambda_R &= \lambda_0\left[ 
1+\frac{1}{2}\frac{g_0^2}{\Gamma_0\lambda_0}\int_{\Lambda/b}^\Lambda \ddk p \ \frac{1}{p^4}
\right] 
\label{baffo}\\
\eta_R &= \eta_0  \ .
\label{bogo}
\end{align}
These equations show that there is a great difference in the role of the two parameters $\lambda_0$ and $\eta_0$. If we start from a frictionless model that has $\eta_0=0$, the coarse-graining of the RG will not generate a friction, $\eta_R \neq 0$, through shell integration. On the contrary, even if we start from a model without spin transport coefficient, $\lambda_0=0$, integration over short wavelengths {\it inevitably} generates a transport term $\lambda_R \neq 0$. In other words, the interaction between the spin $\vm$ and the primary field $\bpsi$ generates a non-zero transport coefficient $\lambda_R$ even if $\lambda_0 = 0$ in the original microscopic theory. For this reason, we included from the outset the parameter  $\lambda_0$ in the coarse-grained field equations\footnote{More to the point: $\eta_0=0$ is an RG fixed point (although unstable, as we shall see), whereas $\lambda_0=0$ is not  a fixed point.}.

To make further progress we must address a rather crucial algebraic detail. While the integral in \eqref{baffo} is straightforward, the one in \eqref{buffo} requires some care. The cutoff $\Lambda$ is large, while the rescaling factor $b$ is close to $1$; hence, the shell integration is performed over large values of the internal momentum $p$. If the effective friction $\eta_0$ is finite (or zero), then the term of order $p^2$ dominates over $\eta_0$ at the denominator, so that the overall integrand will have a $1/p^4$ behaviour for large momentum. As we shall see later on, the hypothesis that $\eta_0 < \infty$ is by no means harmless, and we shall need to return over this point. Yet, for now we will work under this hypothesis, and recognise that it is therefore convenient to rewrite the integral as,
\begin{widetext}
\begin{equation}
\Gamma_R =\Gamma_0\left[ 1+ 2\, \frac{g_0^2}{\Gamma_0 (\Gamma_0+\lambda_0)} 
\int_{\Lambda/b}^\Lambda \ddk p\ \frac{1}{p^2\left(p^2 +\frac{\eta_0}{\Gamma_0+\lambda_0}
\right)} \right] \ .
\end{equation}
We can now change variables, defining $p=\Lambda x$, and obtain, 
\begin{align}
\Gamma_R &=\Gamma_0 \left[ 
1+ 2\, \frac{g_0^2 \Lambda^{d-4}}{\Gamma_0\lambda_0 (1+\frac{\Gamma_0}{\lambda_0})} 
\int_{1/b}^1 \ddk x\ 
\frac{1}{x^2\left(x^2 +\frac{\eta_0}{\lambda_0}\frac{\Lambda^{-2}}{1+\Gamma_0/\lambda_0}
\right)}
\right] 
\label{buffo2}
\\
\lambda_R &= \lambda_0\left[ 
1+\frac{1}{2}\frac{g_0^2\Lambda^{d-4}}{\Gamma_0\lambda_0}\int_{1/b}^1 \ddk x \frac{1}{x^4}
\right] \ ,
\label{baffo2}
\end{align}
It is now convenient to introduce a set of effective parameters, through which we can express these integrals in a simpler way (so to speak),
\begin{equation}
f_0 = \frac{g_0^2}{\lambda_0 \Gamma_0}K_d \Lambda^{d-4}, \quad w_0 = \frac{\Gamma_0}{\lambda_0}, \quad
\mc R_0 = \sqrt{\frac {\lambda_0}{\eta_0}} \ ,
\label{def_parameters}
\end{equation}
where $K_d$ is the unit sphere volume in dimension $d$. We can thus finally write the perturbative expression of all three kinetic parameters after shell integration,
\begin{align}
\Gamma_R&=\Gamma_0\left[ 1+ \frac{2 f_0}{1+w_0}\int_{1/b}^1 \frac{d^d x}{x^2} \ \frac{1}{x^2+ (\mc R_0\Lambda)^{-2}(1+w_0)^{-1}}
\right] 
= \Gamma_0\left[1+  \frac{2f_0}{1+w_0} X_0\log b \right]
\label{ren_gamma}\\
\lambda_R &= \lambda_0\left[
1+\frac{f_0}{2}\int_{1/b}^1 \frac{d^d x}{x^4}
\right] 
=\lambda_0\left[1+ \frac 12 f_0 \log  b \right]
\label{ren_lambda}\\
\eta_R &= \eta_0  \ ,
\label{ren_eta}
\end{align}
\end{widetext}
where we have introduced the dimensionless crossover parameter $X_0$,
\begin{equation}
X_0 = \frac{(\mc R_0\Lambda)^2 (1+ w_0)}{1+ (\mc R_0\Lambda)^2(1+w_0) }  \ ,
\label{Xdefinition}
\end{equation}
and where to compute the integrals we have exploited the fact that in the limit $b\to1$, that is for an infinitesimal RG transformation, the shell becomes 
infinitesimal, so we have written the integrals as the shell thickness, $1-1/b\sim \log b$, times the integrand evaluated at $x=1$. 

The dimensionless parameter $w_0$ is rather harmless, and it will play only a moderate role in what follows; quite conveniently, it will remain finite in all fixed points we will find. On the other hand, $f_0$ is crucial, as it acquires the role of the {\it effective dynamical coupling constant}: the perturbative expansion, which naively one would think as a series in powers of $g_0$, is in fact a series in powers of $f_0$. From the dimensional form of $f_0$, and in particular from the fact that it contains a term $\Lambda^{d-4}$, RG connoisseurs can already deduce that the dynamical upper critical dimension of the theory will be $d_c=4$, the same as the static one. This will be made explicit once we will have solved the recursive RG equation for $f_0$ further on. We recall that this is a consequence of the fact that both
integrands go like $1/p^4$ for large momenta, hence giving a logarithmic behaviour at $d=4$, and that, in turns, this is a consequence of having assumed that $\eta_0$ is finite in the integral of equation \eqref{buffo}. We will return on this hypothesis later on.

The second important effective parameter emerging from the equations is the length scale, $\mc R_0$, given by the ratio between the transport coefficient and the effective friction of the spin.  Because of its definition, we can intuitively expect that if $\mc R_0$ is large ($\eta_0$ small) the dynamics of the spin is ruled by a conservative diffusion mechanism. On the contrary, if $\mc R_0$ is small ($\eta_0$ large), we expect the dynamics of the spin to be ruled by a dissipation mechanism. It is worth noticing that for $\mc R_0=\infty$, namely $\eta_0=0$, the crossover parameter is equal to $1$, and one correctly gets the same equations as the fully conservative Model G. On the other hand, when the conservation length scale is very small, $\mc R_0\sim0$, which happens for $\eta_0 \gg \lambda_0$, one gets $X_0\sim0$, so that $\Gamma_0$ receives very weak perturbative corrections at one loop. We will return on this crucial point later on and we will see that this interplay between non-conservative dissipation and conservative transport coefficient of the spin plays a key role, giving rise to a non-trivial crossover between two different RG fixed points with different dynamic critical exponents.

Up to now we focused on the coefficients of the linear terms in the equations of motion, $\Gamma_0, \lambda_0, \eta_0$. However, in general shell integration produces corrections to all terms, including the non-linear ones. Therefore, the dynamical coupling constant, $g_0$, could in principle get a perturbative correction from shell integration, in particular from the renormalized vertex. However, it can be shown that - due to the structure of the interaction vertices - there are no perturbative corrections at all orders (see Appendix B about vertex corrections),
\begin{equation}
g_R = g_0 \ .
\end{equation}
As in the case of the lack of corrections to $\eta_0$, this result is a consequence of the symmetry properties of the system. Indeed, we remind that the field $\vm$ is the generator of the rotational symmetry of the $\bpsi$ field. Even though the global spin is not conserved in our case, the symmetry still generates some Ward identities that protect $g_0$ at all orders (see Appendix C).  

Finally, let us note that the static coupling constant, $u_0$, is renormalized as usual in the standard equilibrium theory \cite{amit2005field}. However, the lowest order corrections to the dynamical coefficients due to the vertex $u_0$ are at {\it two} loops  \cite{hohenberg1977theory}, whereas we perform here a {\it one} loop calculation. Therefore we do not need to address static renormalization any further in what follows. This is actually a nice feature of all theories with mode-coupling terms: the dynamical vertex is triple (see Appendix A), hence one obtains very sizeable corrections to the critical exponents already at one loop order, without using the two-loop corrections of the static vertex.


\subsection{Rescaling of space and time}

After integration over the shell, we are left with a theory which has new renormalized parameters, and also a new, smaller cutoff, $\Lambda/b$. In order
to compare the new theory with the old one, and therefore to be able to write a set of recursive equations for the parameters, we rescale space, and therefore
momentum $k$, by a factor $b$, in such a way to formally restore the original cutoff, $\Lambda/b \to \Lambda$. It must be noted that frequency does not have a similar cutoff, hence in principle we would have no formal need to rescale $\omega$; however, this is deceiving: in order to reabsorb all powers of $b$ in the novel equations of motions one can see that it is necessary also to rescale frequency \cite{hohenberg1977theory}. Physically, this means that to the rescaling of space and time {\it cannot} proceed independently: space and time are tied together by the - yet unknown - \textit{dynamic critical exponent $z$}, through the following rescaling relations,
 \begin{equation}
          k \to bk \quad \qquad \omega \to b^z \omega  \ .
 \end{equation}
Due to the non-linear form of the equations of motion, we do not a priori know how the spatial integration affects the dynamics and we therefore allow for a generic dynamic critical exponent, $z$. As we shall see, its value determines how the order parameter relaxes close to criticality.

Rescaling momentum and frequency actually means changing physical units, hence each parameter will also rescale according to its naive physical dimensions,
which can be expressed in powers of $b$ in the following way, 
\begin{equation}
\begin{split}
\Gamma_{0}  \to  b^{z-2}\Gamma_0 
\quad &, \quad
\lambda_{0}  \to b^{z-2} \lambda_0 \ 
\\ 
\eta_{0}  \to b^z \eta_0 
\quad &, \quad \ 
g_{0}  \to b^{z-d/2} g_0  \ .
\end{split}
\end{equation}
Because the perturbative contribution of the shell integrals are given in terms of the effective parameters, $f_0, w_0, {\cal R}_0$, it is also necessary to write their corresponding rescaling laws, 
\begin{equation}
f_0 \to  b^{4-d}  f_0\quad ,\quad
w_0  \to w_0 \quad , \quad
{\cal R}_0 \to b^{-1} {\cal R}_0 \ .
\end{equation}
For $d>4$ the naive dimension of the effective coupling constant $f_0$ becomes negative; as we shall see this implies that at its non-trivial fixed point the effective coupling constant will be of order $\epsilon$, with,
\begin{equation}
\epsilon = 4-d \ ,
\end{equation} 
as it happens to the static coupling constant, $u_0$ \cite{wilson1974renormalization}, confirming the fact that the dynamical upper critical dimension is $d_c=4$. We notice that, in general, the rescaling of $k$ and $\omega$ implies also a rescaling of the fields, which also carry some physical dimensions. However, once again we remark that the current calculation is at one-loop level, whereas the fields renormalize at two-loop level; hence, in the following, for all practical purposes all will proceed as if the fields do not renormalize.\footnote{For this same reason the anomalous dimension of the field $\bm\psi$, normally called $\eta$ (not to be confused with the friction), will be set to zero in the present calculation.}

\subsection{Renormalization group recursive equations}

The two stages described above, shell integration and rescaling, must now be put together to define one step of the RG transformation; in this step, a generic parameter $\cal P$ is brought from its initial bare value, ${\cal P}_0$, to a new value, ${\cal P}_1$ through an RG equation with the structure, 
\beq
{\cal P}_1 = b^{D_{\cal P}} \, {\cal P}_0 \left(1+ \delta_{\cal P} \log b \right)  \ ,
\eeq
where the power of $b$ comes from the rescaling step, so that ${D_{\cal P}}$ is the physical dimension of $\cal P$, whereas the term in the bracket comes from the shell integration. For an infinitesimal RG transformation $b\sim 1$, hence we can write $1+ \delta_{\cal P} \log b = b^{\delta_{\cal P}}$, so that $\delta_{\cal P}$   is an effective correction to the naive scaling dimension ${D_{\cal P}}$ of the parameter. Of course, $\delta_{\cal P}$, which is the result of the shell integration, will depend on all the other parameters of the theory. One can then iterate this step $l$ times, giving rise to a recursive RG equation for $\cal P$, 
\beq
{\cal P}_{l+1} = b^{D_{\cal P}} \, {\cal P}_l \left(1+ \delta_{{\cal P}_l} \log b \right)  \ ,
\eeq
where we emphasise that all integrals that appear at the r.h.s. through the factor $\delta_{{\cal P}_l}$, must be evaluated at the {\it running} value of the parameters, namely at their value at the RG step $l$, whereas the naive physical dimension ${D_{\cal P}}$ is fixed once and for all. By using this procedure for the dynamical parameters of our theory, we obtain the following RG recursive relations,
\begin{align}
\label{Gammaricorsione}
\Gamma_{l+1} &= b^{z-2} \, \Gamma_l \left(1+ \frac{2 f_l}{1+w_l} X_l \log b \right) \\
\lambda_{l+1} &= b^{z-2} \, \lambda_l \left(1+\frac 12 f_l \log b \right) \\
\eta_{l+1} &= b^z \, \eta_l  \\
g_{l+1} &= b^{z-d/2} \, g_l   \  ,
\end{align}
and we recall that we are working at $T=T_c$, namely on the critical manifold. From these equations we can finally write a {\it closed} set of recursive relations for the effective coupling constant $f_l$, the dimensionless parameter $w_l$, and the conservation length scale $\mc R_l$,
\begin{equation}
\begin{split}
f_{l+1} &= f_l \; b^{\epsilon}  \biggl[ 1-f_l\biggl( \frac 12 + \frac {2X_l}{1+ w_l} \ \biggl)\log b \biggl]\\
w_{l+1} &= w_l\ \ \biggl[1 - f_l\biggl(\frac 12  -\frac{2X_l}{1+w_l}  \biggl)\log b   \biggl]\\
\mc R_{l+1} &= \mc R_l \; b^{-1} \biggl[1+\frac 14 f_l \log b \biggl]  \ , \\
\end{split}
\label{parameters_flow}
\end{equation}
where $X_l$ depends on $\mc R_l$ and $w_l$ through equation \eqref{Xdefinition}. We note that the full scaling dimension of the conservation length scale $\mc R$ is determined by its naive dimension, $b^{-1}$, plus a perturbative contribution, $1+\frac 14 f_l \log b = b^{\frac 14 f_l}$, hence developing an anomalous scaling dimension that will be crucial in ruling the crossover.

The derivatives of $f$, $w$ and $\mc R$ with respect to $(-\log b)$ are called \textit{beta-functions}, and measure how the parameters change when performing an infinitesimal RG transformation,\footnote{
Normally, in momentum shell, the $\beta$-functions are defined as derivatives wrt $\log b$; however, in that way one ends up with the opposite sign of the Callan-Symanzik approach, within the context of which they are defined as derivatives wrt the arbitrary momentum scale, $\mu$, which is morally $1/b$. No big deal; we use this convention so to have, in the end, the same set of $\beta$-functions.}
\begin{equation}
    \begin{split}
        \beta_f &= - f \ \left[ \epsilon- f\left ( \frac 12 + \frac{2X}{1+w}\right ) \right]\\
        \beta_w &= w f \left[\frac 12 - \frac{2X}{1+w} \right]\\
        \beta_{\mc R} &=   \mc R \ \left[1- \frac 14 f \right]  \ .
        \label{betafunc}
    \end{split}
\end{equation}
The zeros of these functions define the fixed points of the RG flow and thus have a crucial role in the theory. The beta-functions also will provide a link between the momentum shell RG approach followed so far and the Callan-Symanzik approach described in Section VI.

\subsection{Fixed points and dynamic critical exponent}

The fixed point values of the RG equations (that we are going to indicate with an asterisk) rule the critical behaviour of the system.
The exponent $z$ can be found by requiring that the fixed point value of the kinetic coefficient of $\boldsymbol{\psi}$, namely $\Gamma^*$, is finite \cite{hohenberg1977theory,cardy1996scaling}. This condition, as we shall see, is what we need to  investigate the relaxation behavior of the field $\bpsi$  close to criticality.
From equation \eqref{Gammaricorsione} we obtain $z$ as: 
\begin{equation}
\label{eqzeta}
\Gamma^* = \mc O (1)\quad \Rightarrow  \quad z = 2 - \frac{2 f^*}{1+w^*} X^* \ .
\end{equation}
The dynamic critical exponent is therefore given by the fixed point values of the parameters $f$, $w$ and $X$.

From the corresponding recursion equation \eqref{parameters_flow} 
it is evident that  $\mc R$ can have  two fixed points, namely:
\begin{equation}
\mc R^* = 0 \qquad \mc R^* = \infty \ .
\end{equation}
Since the fixed point of $f$ is expected to be of order $\epsilon$ (see Eqs. (\ref{betafunc})), the scaling dimension of $\mc R$ is negative. Therefore, the $\mc R ^* = 0$ fixed point is {IR-stable} while the $\mc R ^* = \infty$ fixed point is {IR-unstable}: any large, but finite, initial value of $\mc R_0$,  decreases under the RG equation \eqref{parameters_flow}, driving the systems to the  $\mc R^* = 0$ fixed point.
Inserting back the possible values of $\mc R ^*$ in the other equations, we therefore find two fixed points for the global set of parameters.

\subsubsection{The IR-unstable conservative fixed point}
The first fixed point with $\mc R^* = \infty$ and $X^* = 1$, which we call IR-unstable (or conservative), is:
\begin{equation}
f^* = \epsilon  \quad
 w^* = 3 \quad 
 \mc R^* = \infty \quad
 X^*=1
 \quad
\Rightarrow
   z=d/2 \ .
   \nonumber
\end{equation}
This fixed point describes a dynamics with $z=d/2$,  typical of conservative models such as Model G and Model E \cite{halperin1976renormalization}. Indeed,
dissipation becomes irrelevant ($\eta^* = 0$), and the conservation law expressed by the symmetries of the Hamiltonian, rules the dynamics at all scales.
If the system has $\mc R_0=\infty$ (i.e. $\eta_0 = 0$) the RG flow will converge to this fixed point, the only stable one for zero dissipation. However, as mentioned above, any other value of  $\mc R_0$ will cause the flow to eventually converge to the other fixed point.

\subsubsection{The IR-stable dissipative fixed point}
The second fixed point is characterized by $\mc R^* = 0$, or equivalently $X^*=0$, and we call it IR-stable (or dissipative):
\begin{equation}
f^* = 2 \epsilon \quad 
w^* = 0 \quad
\mc R^* = \infty \quad
X^*=0
\quad
\Rightarrow
   z=2  \ .
   \nonumber
\end{equation}
In this case, dissipation takes over ($\eta^* = \infty$) and the dynamic critical exponent that we obtain is $z=2$, which is common for models with a completely dissipative dynamics \cite{hohenberg1977theory}. 
What we have depicted here is a scenario that includes the presence of two fixed points with different dynamical behaviors and different dynamic critical exponents, namely $z=d/2$ (conservative dynamics) and $z=2$ (dissipative dynamics). Even though one of such fixed points is unstable along one direction, it is stable along the others and - as it will be discussed in the next section - it can rule the RG flow at intermediate iterations. In other terms, there is a crossover in the RG flow in parameter space and, as a consequence, also in the behavior of physical observables.


\section{Renormalization group crossover}

\subsection{RG flow on the critical manifold}
To investigate the dynamic crossover we studied the RG flow from a numerical point of view. In the limit of infinitesimal RG transformation ($b\to 1$), the recursion relations \eqref{parameters_flow} become a system of coupled differential equations. We introduce the continuous variable $x = l \log b$; Eqs. \eqref{parameters_flow} can then be rewritten in the continuum limit (replacing, for instance, $f(x) = f_l$):
\begin{equation}
\begin{split}
f'(x) &= \beta_f(f,w,\mc R)\\
w'(x) &= \beta_w(f,w,\mc R)\\
\mc R'(x) &= \beta_{\mc R}(f,w,\mc R)\ ,\\ 
\end{split}
\label{2:FlowDifferential} 
\end{equation}
where the prime stands for a derivative with respect to $x$.

\begin{figure}
	\includegraphics[scale = 0.38 ]{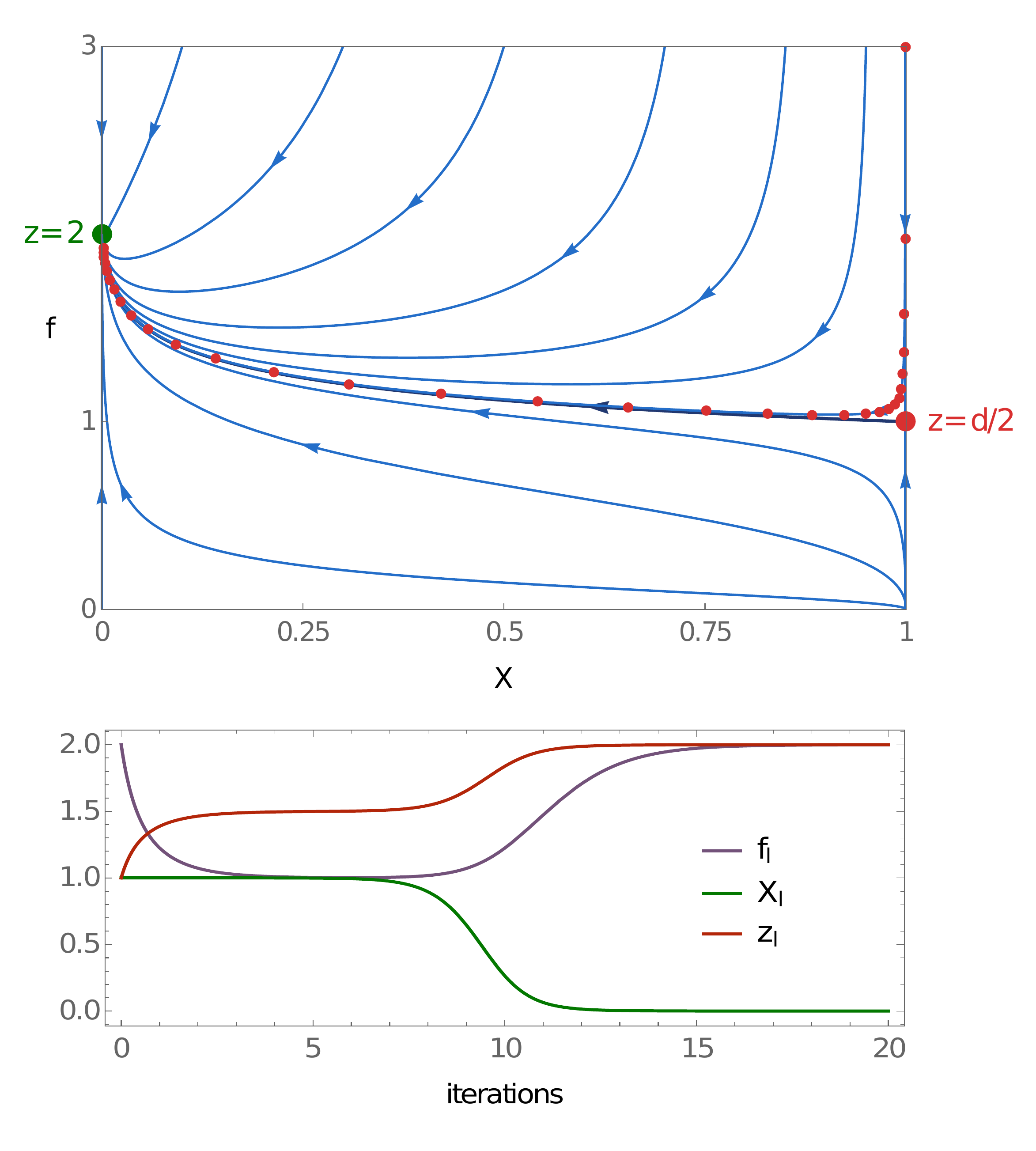}
				\caption{{\bf Renormalization group flow and crossover.} Top: Flow
		diagram on the  $(X_l,f_l )$ plane for $d = 3$. When the initial
		friction $\eta_0$ is small, $X_0 \sim 1$, the flow converges towards the
		unstable fixed point, $z = d/2$, and remains in its proximity for
		many iterations, before crossing over to the stable $z = 2$ fixed
		point. Bottom: running parameters and critical exponent $z$ as a function of the iteration step along
		a flow line at small $\eta_0$.
}
	\label{2:fluxOnCm}
\end{figure}

The set of Equations (\ref{2:FlowDifferential}) can be studied numerically, for any given initial condition. In Fig.\ref{2:fluxOnCm} (upper panel), we show the resulting flow in the $(X,f)$ plane, each line corresponding to a different set of initial values of $\eta_0,\lambda_0,\Gamma_0$ (and therefore of $f_0$ and $X_0$). The flow always proceeds from the conservative to the dissipative fixed point, as expected. However, how fast it does so, depends on the initial condition $X_0$. When this value is close to $1$, which means that  the friction $\eta_0$ is small, the flow of parameters approaches the $z=d/2$ fixed point and remains close to it for many RG iterations. Then, it eventually moves towards the stable fixed point with $z=2$.  
In the lower panel of FIG.\ref{2:fluxOnCm} we show the same dynamic crossover in terms of $z$, of the coupling constant $f$ and of $X$. From this figure we clearly see that there is a well defined intermediate regime where the flow is regulated by the unstable fixed point, the parameter $X$ driving the dynamic critical exponent from one value ($z=d/2$) to the other ($z=2$).

\subsection{Crossover in the critical dynamics}
What we have discussed so far is the crossover taking place along the RG flow in parameters space. This crossover has important observable consequences in the  relaxational behavior of the system. In the previous section we showed that there is a parameter with the dimensions of a lenght-scale, ${\mathcal R}$, which plays a crucial role in the RG flow.  As we shall see, it is precisely the interplay between ${\mathcal R}$ and the relevant physical lenght-scales in the system, to determine the way it relaxes. 

\subsubsection{Crossover in $k$ at $\xi=\infty$}

Let us start by considering the system at the critical point ($r =0$). In this case the correlation length is infinite and the only physically relevant length scale is the one at which we are observing the system, namely $1/k$. To study relaxation at this scale, we can measure the characteristic frequency of the order parameter, which is directly connected to the dynamic critical exponent through the dynamic scaling hypothesis (\ref{dynscal}), which, for $\xi = \infty$ and for small $k$, reads,
\begin{equation}
\omega_c (k) = \tau_k^{-1}\sim k^z \ ,
\end{equation}
where we have introduce the characteristic frequency, $\omega_c$, as the inverse of the relaxation time.
Since there are two possible values of $z$ one can wonder at this point which is the one to consider in this relationship. It turns out that this depends on $k$ vs. $\mathcal R$.
To see this, we remind that the characteristic frequency is the pole of the propagator of the field $\bpsi$ and we can therefore study its infrared behavior by looking at $G^{-1}_\psi$ at small $k$. This is very convenient because - by construction - propagators along the RG flow are related to each other, and we can link what happens in parameters space to the physical behavior of the system.

\begin{figure}[t]
	\centering
	\includegraphics[scale = 0.4]{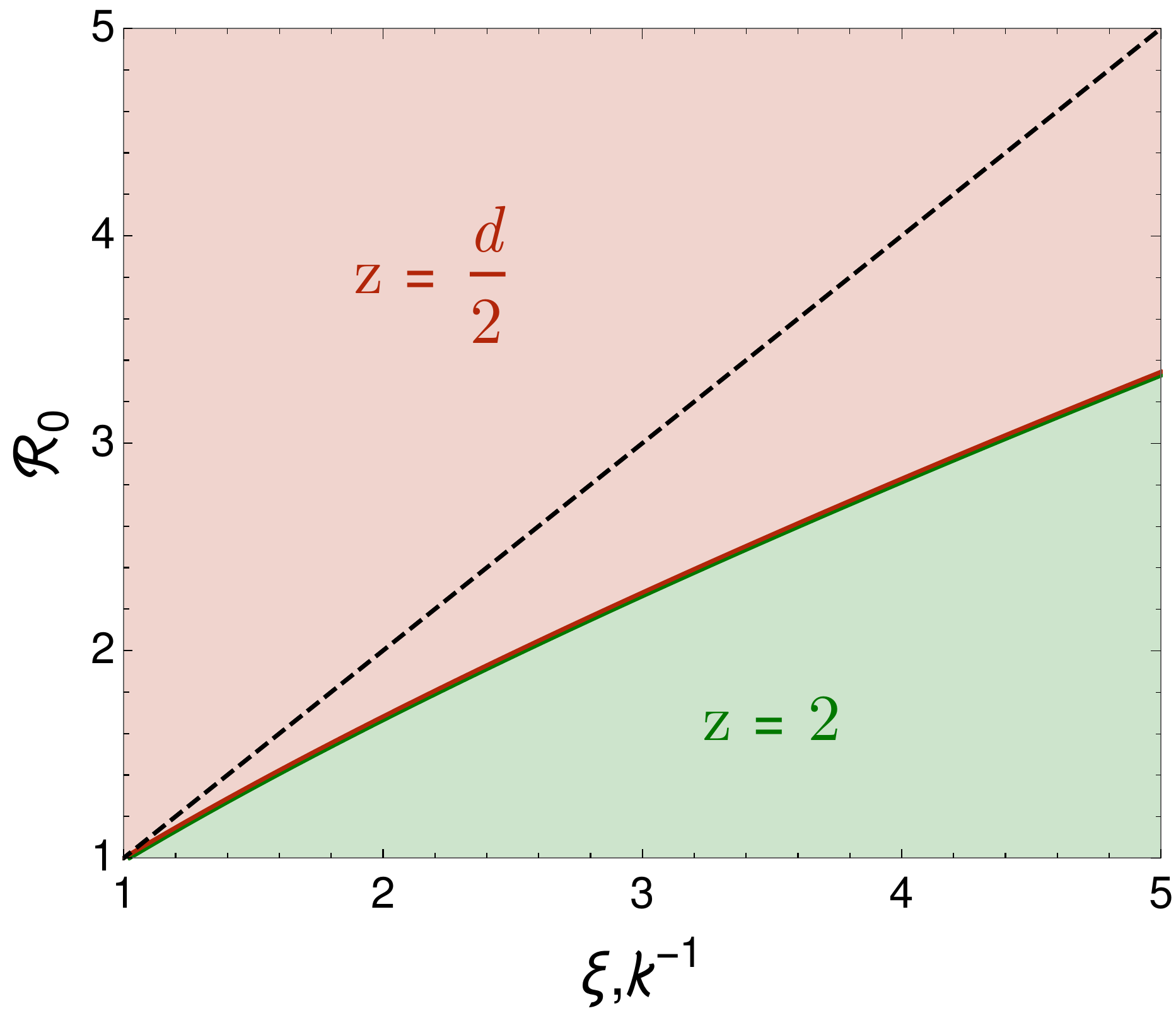}
	\caption{\textbf{Different critical regions}. Different values of $k$, $\xi$ and $\mc R_0$ correspond to different critical behaviors. Red region corresponds to conservative critical dynamics with $z = d/2$, while green region corresponds to dissipative critical dynamics with $z=2$. We set $\Lambda=1$ so that physical values for lengths are $k^{-1}>1$, $\xi>1$ and $\mc R_0>1$.  On the critical manifold relaxation is studied in the  $(k^{-1},{\cal R}_0)$ plane:  the two different regimes are separated by the curve $\mc R_0 = k^{-d/4}$.  Off the critical manifold  relaxation is studied in the $(\xi,{\cal R}_0)$ plane: the two different regimes are separated by the curve $\mc R_0 = \xi^{d/4}$. The black dashed line represents, respectively, $\mc R_0 =  k^{-1}$ or $\mc R_0 = \xi$. The figure refers to the $d=3$ case.
	}
	\label{fig:regions}
\end{figure}

At every step $l$ of the RG, the physical propagator verifies the relation,
\begin{equation}
\label{physicalG}
G_\psi(k,\omega, \mc P) = (b^{l})^{ z_l} G_\psi (b^l k, \omega b^{z_l}, \mc P_l) \ ,
\end{equation}
where with $\mc P_l$ we indicate the set of the  parameters after $l$ steps of RG (and $\mc P = \mc P_0$ in the l.h.s.). The scaling factor on the right side of \eqref{physicalG} is just the scaling dimension of the propagator. 
What we are doing is to consider an initial point in parameters space corresponding to our physical system (l.h.s.), and then follow the RG flow in the critical manifold starting at that point.
As $l$  increases, the propagator on the r.h.s. is evaluated at farther points along the RG line.
Since we know that there is a crossover along the RG flow we are writing this expression with a dynamic critical exponent, which explicitly depends on the recursion step $l$. 
If we choose $l$ such that $b^l = \Lambda/k$, i.e. the maximum possible value, the inverse of the propagator satisfies:
\begin{equation}
\label{physicalG*}
G^{-1}_\psi(k,\omega=0, \mc P) = \left( \frac{\Lambda}{k}\right)^{- z^*} G^{-1}_\psi (\Lambda, \omega =0, \mc P^*) \ .
\end{equation}
Here  we have evaluated the function on the right side at the fixed point values of the parameters $\mc P$. This is justified if $l$ is large enough to approach the vicinity of a fixed point (i.e. small $k$).  Which one of the two fixed points is reached - and therefore the value of $z^*$ above - depends on the starting point (i.e. the set $\mathcal P$) and on the precise number of iterations.
More precisely, the condition that discriminates between the two possible fixed points is: 
\begin{equation}
\label{sogliaR}
\mc R_l \simeq \Lambda^{-1} \ ,
\end{equation}
because it determines the value of the variable $X_l$ in expression \eqref{eqzeta}. Let us consider the situation, which interests us more,  where the starting point of the flow is close to the IR-unstable fixed point. The initial value of $\mathcal R$ is therefore large, corresponding to a system with low $\eta_0$. If  $\mc R_l \gg  \Lambda^{-1}$  holds for all the iterations, the flow will explore only the neighborhoods of the unstable fixed point and the values of $\mc P^*$ in (\ref{physicalG*}) are the ones of the conservative dynamics. Therefore, in this case:
\begin{equation}
G^{-1}_\psi(k,\omega=0, \mc P)  \sim k^{d/2} \implies \omega_c \sim k^{d/2} \ .
\end{equation}
However, it may happen that, even starting at the same point in parameters space, the number of iterations is so large that eventually the condition $\mc R_l \ll \Lambda^{-1}$ becomes satisfied, and the flow approaches the stable fixed point corresponding to $z=2$. In this case:
\begin{equation}
G^{-1}_\psi(k,\omega=0, \mc P)  \sim k^2 \implies \omega_c \sim k^2 \ .
\end{equation}
Since the number of iterations is fixed by the value of the wave-number $k$ ($b^l = \Lambda/k$), the condition  $\mc R_l \simeq \Lambda^{-1}$ can be translated into a condition on $k$. The recursion relation for the conservation length scale gives $\mc R_l = \mc R_0 b^{l(-1+f^*/4)}$; since we are considering a flow starting close to the conservative fixed point, we can set $f^* = \epsilon$, which gives the anomalous scaling dimension of $\mc R$ at the conservative fixed point,
\beq
\mc R \sim b^{-d/4}  \ ,
\label{blunt}
\eeq
from which we see that the length scale $\mc R$ has a scaling dimension equal to its naive dimension at the upper critical dimension, $d=4$, as expected.
From the relation  $\mc R_l = \mc R_0 b^{l(-d/4)}$ we can finally identify a threshold value $k_c$ 
marking the limit between the two different scenarios described above, namely:
\begin{equation}
k_c = \Lambda (\Lambda \mc R_0)^{-4/d} \ .
\end{equation}
To summarize, we therefore find that at criticality  the relaxation behavior of the order parameter - as captured by the critical exponent $z$  - depends on the relation between the scale at which we observe the system and the value of the length-scale ${\cal R}_0$, i.e.
\begin{equation}
\begin{split}
k &\ll \Lambda (\Lambda \mc R_0)^{-4/d} \quad \rightarrow \quad z=2 \\
k &\gg \Lambda (\Lambda \mc R_0)^{-4/d} \quad \rightarrow \quad z=d/2 \ .
\end{split}
\end{equation}
We therefore have found the third non-trivial critical exponent of the theory, namely the {\it crossover exponent} \cite{cardy1996scaling}, 
\beq
\kappa =4/d \ ,
\eeq
which, as we have seen, is intimately related to the anomalous dimension of the conservation length scale. In Fig.\ref{fig:regions} we show the regions corresponding to the two dynamical behaviors in the $(k^{-1},{\cal R}_0)$ plane. 

\subsubsection{Crossover in $\xi$ at $k=0$}

In many cases, and in particular when looking at experimental data, real systems are not {\it exactly} at the critical point. For all practical purposes we need to extract information on the critical behavior of the system also when its correlation length $\xi$ is finite, even if large. Predictions can be obtained following a reasoning much similar to the one above, but taking explicitly  into account the dependence of the propagator on temperature, i.e. on the correlation length. Besides, since there is a characteristic length-scale, it is convenient to set $k=0$, $\omega = 0$. Instead of Eq. (\ref{physicalG}), the relevant equation for the propagators then becomes
\begin{equation}
\begin{split}
G_\psi(\xi) &= G_\psi(k=0, \omega = 0, \xi, \mc P) \\
&= \left( b^l \right)^{z_l} G_\psi(k=0, \omega=0, \xi_l, \mc P_l) \ .
\end{split}
\end{equation}
What we are doing is, again, to consider a point in parameters space corresponding to our physical system, and to relate the physical propagator with propagators of models along an RG line starting at that point. The difference with the previous case is that now the RG flow takes place off the critical manifold, therefore not only the parameters change upon iteration, but also the correlation length, i.e.
\begin{equation}
\xi_{l+1} = \xi_l/b  \ ,
\end{equation}
with $\xi_0 = \xi$ (i.e. the correlation length of the physical system). We can choose the number $l$ of iterations such that $b^l = \xi \Lambda$. If $\xi$ is large enough that the system comes close to a fixed point, then the inverse propagator satisfies the relation:
\begin{equation}
\begin{split}
G^{-1}_\psi(\xi) = (\Lambda \xi)^{-z^\star} G^{-1}_\psi(k=0, \omega=0 ,  \mathcal{P}^*) \ .
\end{split}
\end{equation}
Since the pole of the propagator for $k=0$ is the global characteristic frequency of the system, we immediately get the relaxation behavior as $\omega_c(\xi) \sim \xi^{-z^\star}$. As before, the value of $z$ depends on which one of the two fixed points is approached at the end of the RG flow after $l$ iterations. The discriminating  condition is always $\mc R_l \simeq \Lambda^{-1}$. For $\mc R_l \gg \Lambda^{-1}$ the fixed point is characterized by $z=d/2$, then the characteristic frequency diverges as:
\begin{equation}
\begin{split}
G^{-1}_\psi(\xi)  \sim \xi^{-d/2} \implies \omega_c \sim \xi^{-d/2} \ .
\end{split}
\end{equation}
For $\mc R_l \ll \Lambda^{-1}$, the other fixed point is reached ($z=2$) and we have:
\begin{equation}
\begin{split}
G^{-1}_\psi(\xi)  \sim \xi^{-2} \implies \omega_c \sim \xi^{-2} \ .
\end{split}
\end{equation}
Since the number of iterations is fixed by $\xi$ (i.e. $b^l = \xi \Lambda$), the discriminating condition $\mc R_l \simeq \Lambda^{-1}$ now identifies a threshold value $\xi_c$  for the correlation length that can be obtained using the recursion relations of both $\mathcal R$ and $\xi$:
\begin{equation}
\xi_c \simeq \bigl( \mc R_0 \Lambda \bigl)^{4/d} \Lambda^{-1} \ .
\end{equation}
To conclude, we therefore find that critical slowing down is ruled by two different critical exponents depending on how large the correlation length is (i.e. how close the system is to the critical point) with respect to the conservation length-scale ${\cal R}_0$, i.e.
\begin{equation}
\begin{split}
\xi \gg \bigl( \mc R_0 \Lambda \bigl)^{4/d} \Lambda^{-1} \quad \implies \quad   z&=2 \\
\xi \ll \bigl( \mc R_0 \Lambda \bigl)^{4/d} \Lambda^{-1} \quad \implies    \quad z&=d/2  \ ,
\end{split}
\end{equation}
thus giving the same crossover exponent as in the $k$ description.
A graphical representation of the different critical regimes can be found in Fig.\ref{fig:regions}. To summarize, it is therefore the interplay between the correlation length $\xi$ and the conservation length-scale ${\mathcal R}_0$ that defines what kind of critical dynamical behavior is observed. We also note that - due to the non-trivial recursion relation for ${\mathcal R}$ (see Eqs.(\ref{parameters_flow})) - these two lenghtscales rescale differently upon RG transformations. As a consequence, the region corresponding to the conservative critical dynamics is {\it larger} than in the case of naive scaling.


\subsection{A new upper critical dimension}
So far we have been studying the RG flow in the vicinity of the conservative, $z=d/2$, fixed point. Our original motivation was indeed to describe experimental findings on swarms of insects, where a low-dissipation critical dynamics has been observed. As we have seen, in the neighbourhood of this fixed point we have an upper critical dimension $d_c = 4$ and the effective dynamic coupling constant is the parameter $f_0$. However, we also showed that the conservative fixed point is unstable, hence the RG flow inevitably brings the system to the dissipative fixed point, $z=2$. The problem is that, in the vicinity of this fixed point, the on-shell self-energy $\Sigma_b$ has to be treated quite differently form the previous case, and $f_0$ does not play the role of the effective dynamic coupling constant anymore. Let us see this in more detail.

In  proximity of the IR-stable fixed point, the running effective friction $\eta_l$ becomes very large, eventually diverging. In this regime, our previous assumption to have a mild, finite value of the friction $\eta_l$ in equation \eqref{buffo} must be revised, and the integral must be rearranged differently,
\begin{widetext}
\begin{equation}
\Gamma_{l+1}=b^{z-2} \,\Gamma_l\left[ 1+ 2\, \frac{g_l^2}{\Gamma_l} 
\int_{\Lambda/b}^\Lambda \ddk p\ \frac{1}{p^2[(\Gamma_l+\lambda_l)p^2 +\eta_l]}  \right] 
=
b^{z-2} \,\Gamma_l\left[ 1+ 2\, \frac{g_l^2}{\Gamma_l\eta_l} 
\int_{\Lambda/b}^\Lambda \ddk p\ \frac{1}{p^2\left(\frac{\Gamma_l+\lambda_l}{\eta_l}p^2 + 1\right)}  \right] \ .
\label{buffone}
\end{equation}
\end{widetext}
We see that, as the running friction $\eta_l$ goes to infinity, approaching the stable fixed point, the large $p$ behaviour of the integrand turns from $1/p^4$ to $1/p^2$, thus giving,
\begin{equation}
\Gamma_{l+1}=b^{z-2} \,\Gamma_l
\left[ 1+ 2\, \frac{g_l^2}{\Gamma_l\eta_l} 
\int_{\Lambda/b}^\Lambda \ddk p\ \frac{1}{p^2}  \right] \ .
\label{meravigliao}
\end{equation}
From this last equation it is clear that, in the proximity of the IR-stable fixed point, the actual effective coupling constant in the perturbative expansion of $\Gamma_l$, is no longer $f_l$, but $q_l = g_l^2\Lambda^{2-d}/\Gamma_l \eta_l$, whose naive scaling dimension is $d-2$, not $d-4$; accordingly, the integral now has a logarithmic UV divergence at $d=2$. We conclude that  the upper critical dimension for this fixed point is no longer $4$, but $\tilde d_c = 2$, and that the actual parameter of expansion is,
\begin{equation}
\tilde\epsilon=2-d \ .
\end{equation}
In $d=3$, which is the case of interest for us, the dimensions of $q$ is negative, which is equivalent to say that the only stable fixed point is $q^*=0$ (this can also be seen explicitly by writing the RG recursive equations for $q_l$). Therefore, the self-energy contribution in \eqref{meravigliao} vanishes and the kinetic coefficient has no perturbative contributions (at one loop), thus giving, 
\begin{equation}
\Gamma_{l+1} = b^{z-2} \Gamma_l \ ,
\end{equation}
so that the only way to keep finite the kinetic coefficient at its fixed point is to have, 
\beq z=2 \ ,
\eeq
in agreement with the previous result. In this regime $\boldsymbol \psi$ behaves dynamically as an independent field, i.e. its relaxation has no contributions from the mode-couplings term in the equations of motion. This very non-trivial crossover between two different upper critical dimensions will be made more explicit in the 
Callan-Symanzik approach, which we describe the following Section.


\section{Callan-Symanzik approach}

\def\cR{\mathcal{R}}

In this section we derive the RG results within a different
renormalization approach, in which the large scale properties of the
system are deduced from a differential equation (called
Callan-Symanzik equation or renormalization group equation).  This
equation in turn follows, as we explain below, from a
reparametrization invariance of the renormalized dynamic theory,
which is introduced to deal with the strong cutoff ($\Lambda$)
dependence of the original theory (which leads to divergences in
the $\Lambda\to\infty$ limit).  This approach is complementary to the
momentum-shell renormalization developed in Sec.~IV.  Its principles
are described in several texts, e.g. \cite{le1991quantum,
  binney_book, Itzykson1989, zinnjustin_QFTCF}.  Our treatment of the
ISM under the Callan-Symanzik approach follows the lines of the dynamic
renormalization study of model E by De Dominicis and Peliti
\cite{de1978field} (see also \cite{zinnjustin_QFTCF}).
The CS approach involves the following steps:
\begin{enumerate}
\item Write a renormalized theory, i.e.\ reparametrize the original
  dynamic functional in a way that all $\Lambda$-dependence
  (equivalently, divergencies that appear for $\Lambda\to\infty$) of physical observables is
  absorbed into a finite set of constants.  This is done at  
  an arbitrary momentum scale $\boldsymbol\mu$.
\item Using the fact that the renormalization can be done at arbitrary
  values of $\boldsymbol \mu$, write a differential equation describing how
  relevant renormalized observables (in our case the response and correlation
  functions) change as $\mu=|\boldsymbol \mu|$ is varied.  This is the RG equation,
  sometimes called CS equation.  Combining this with dimensional
  analysis, one finally obtains a differential equation that describes
  the change of the renormalized observable as the \emph{observation
    scale} (external momentum) is varied.  The coefficients of this
  equation are the $\beta$-functions, which are computed from the
  renormalization constants at a given order in perturbation theory.
  The equation is solved by the standard \emph{method of
    characteristics.}
\item The solution by the method of characteristics shows that 
  the behavior of the response and correlations functions at large scales can be
  obtained by studying the response/correlation at a reference observation scale
  but with scale-dependent coupling constants.  How the coupling
  constants change when increasing the observation scale is ruled by
  the $\beta$ functions, and the trajectories in parameter space
  induced by a change of scale are called \emph{RG flow.}  Thus one
  finally studies the RG flow, with particular attention to fixed
  points, which will lead to scaling behavior of the response.
\end{enumerate}

We describe these steps in the following subsections.  Many aspects of
the calculation are identical to Models E and G and for these we refer to the
article by De Dominicis and Peliti \cite{de1978field}.  We only
describe in detail the aspects that are novel in the ISM.

\subsection{Renormalized theory and renormalization factors}
\label{sec:diverg-regul-renorm}

The diagrammatic expansion of the dynamical action involves integrals
in momentum space that are divergent (in the space dimension of
interest) for large integration momenta (ultraviolet divergences)
unless some regularization procedure is adopted (like the cutoff $\Lambda$ for
large momentum we used in the momentum-shell calculation, see Appendix A).  To
construct a renormalized theory means to reparameterize the functional
in terms of a different set of coupling constants and fields in such a
way that the divergences (or equivalently the details of the
regularization procedure) are confined in a finite set of constants.

Instead of using a cutoff, here we renormalize according to the
dimensional regularization plus minimal subtraction prescription
\cite{zinnjustin_QFTCF}: diverging integrals are evaluated in a
dimension low enough that they are convergent, then analytically
extended to non-integer dimension.  The original divergences then show
up as poles in the dimension variable.  The minimal subtraction
procedure consists in introducing the renormalized parameters so that
they absorb only those poles.

Renormalization thus starts with the identification of all the ultraviolet
divergences of the theory and with the definition of the renormalized
constants to absorb them.  Looking at the perturbative expansion, we
see that the introduction of $\eta_0$ leaves the free propagator of
the $\bpsi$ field \eqref{freepropPsi} unchanged with respect to the
model G case, while in the free propagator of the $\vm$ field
$G_{0,s}(k,\omega)$ a $k$-independent term is added, so that the
$k\to\infty$ behavior of the free propagators is unchanged.  Then,
since the structure of the diagrams is identical to that of models G
and E (because the interacting part is the same), the divergences in
ISM arise in the same diagrams.  Then from ref.\
\cite{de1978field} we know that the theory is renormalizable in
$d=4$ (which is the upper critical dimension of the theory).  The
divergent diagrams relevant to the dynamic renormalization arise in
the expansion of $G_\psi(k,\omega)$ and $G_s(k,\omega)$, in particular
in the derivatives
\begin{align}
  &\frac{\partial G_\psi^{-1}}{\partial k^2}, &&
                                                 \frac{\partial
                                                 G_s^{-1}}{\partial
                                                 k^2}.
                                                 \label{div}
\end{align}
Both divergences are logarithmic in $d=4$.  There are two additional
divergences in $G_\psi(k,\omega)$ that we do not need to consider.
One is the quadratic divergence in $G^{-1}_\psi(k=0,\omega=0)$ that is
absorbed into a renormalized mass (susceptibility) in the static 
theory.  Since we work here at the critical point defined by $r=0$, in
practice this means setting $r_0=0$ in all the diagrams we consider.
There is also a logarithmic divergence in $\partial
G^{-1}_\psi(k,\omega)/\partial \omega$ that however does not arise at
the 1-loop level (and which is related to field renormalization).

The divergences are taken care in the following way: we consider the relevant divergent quantities (e.g. the derivatives in Eq. (\ref{div})) and evaluate them at $\omega=0$  and at a given value of the momentum $\bf k=\boldsymbol{\mu}$ (so as to eliminate infrared divergences). We then replace the original kinetic/transport coefficients and coupling constants by renormalized counterparts that absorb the divergences, in a such a way that - once expressed in terms of the new parameters - the quantities of interest are finite. The renormalized parameters are defined through multiplication by $Z$-factors; when considering the derivatives in (\ref{div}) this amounts to introduce renormalized kinetic coefficients
\begin{align}
  \Gamma &= Z_\Gamma \Gamma_0, & \lambda &=Z_\lambda\lambda_0.
  \label{CS-renorm-bare}
\end{align}
The two remaining dynamic couplings, $g_0$ and $\eta_0$ do not pick up
perturbative renormalization.  In the case of $g_0$ this is a
consequence of a Ward identity deriving from the fact that $
\vm$
generates the rotations of $\bpsi$ (Appendix B and C) \cite{de1978field}.
In the case of $\eta_0$ the reason is that it is not involved in
absorbing divergences due to the fact that $G^{-1}_s(k=0,\omega=0)$ is finite
(see next section).  We introduce however $\eta$ and $g$ as
adimensional counterparts of $\eta_0$, and $g_0$,
\begin{align}
  g^2 &= K_d \mu^{d-4} g_0^2, &   \eta &= \frac{\eta_0}{\mu^2},
\end{align}
where $K_d=2\pi^{d/2}(2\pi)^{-d}/\Gamma(d/2)$ is introduced for
convenience and $\mu$ is the arbitrary momentum scale used 
to evaluate the propagators during renormalization.
(Note that in this section we choose the frequency units so that
$\Gamma_0$ and $\lambda_0$ are adimensional, i.e.\ $[\omega]=[k^2]$).

The $Z$-factors now have to be determined at a given order in
perturbation theory so that all the renormalized propagators (and in
consequence correlation and response functions) are finite, i.e.\ the
$Z$-factors are divergent in a such way that all observable quantities
(expressed as averages with the renormalized theory) are finite.  
One can then in principle determine all the renormalized parameters of the
model in terms of a finite number of observations (at fixed wavenumber
and frequency).  Finally, since at one-loop, as mentioned above,
the fields are not renormalized, the relation between original and renormalized 
propagators and correlations is
\begin{align}
G_\psi(k,\omega,u_0,g_0,\Gamma_0,\lambda_0,\eta_0) =
  G_\psi^R(k,\omega,u,g,\Gamma,\lambda,\eta;\mu) \label{Gr}\\
C_\psi(k,\omega,u_0,g_0,\Gamma_0,\lambda_0,\eta_0) =
  C_\psi^R(k,\omega,u,g,\Gamma,\lambda,\eta;\mu) \label{Cr}
  \end{align}

We discuss the determination of the $Z$-factors in
below.  These play a leading role in the CS procedure since they give
the nontrivial contributions to the RG equation coefficients
(sec.~\ref{sec:rg-equation-r_psir}).

\subsection{RG equation and dynamic critical exponent}
\label{sec:rg-equation-r_psir}

In this approach the dynamic critical exponent is identified after
finding the dynamic scaling form of the correlation functions.  The
procedure is the standard one, which we briefly recall.  From the
perturbation expansion at $\omega\neq0$, one can see that the
correlation function $C_\psi$ can be written in terms of the static
coupling $u_0$, the effective dynamic couplings ($f_0$, $w_0$,
$\cR_0$) and $\Gamma_0$, where $\Gamma_0$ and $\omega$ always appear
in the combination $\omega/\Gamma_0$. So we can rewrite Eq.(\ref{Cr}) as
\begin{equation}
  C_\psi(k,\omega,u_0,g_0,\Gamma_0,\lambda_0,\eta_0) =
  C_\psi^R(k,\omega/\Gamma,u,f,w,\cR;\mu).
\end{equation}
Since the lhs is independent of the arbitrary scale $\mu$, deriving
with respect to $\log\mu$ one obtains the RG equation:
\begin{equation}
  \label{eq:CS-RG-Rpsi}
  \left\{ \mu\frac{\partial}{\partial\mu} + \sum_l \beta_l
    \frac{\partial}{\partial l} + \nu_\Gamma
    \Gamma\frac{\partial}{\partial\Gamma} \right\} C_\psi^R
      = 0,
\end{equation}
where $l=u,f,w,\cR$ and the $\beta$ and $\nu$ functions are
\begin{equation}
  \beta_l(u,f,w,\cR) = \mu\frac{\partial l}{\partial\mu}, \qquad
  \nu_X = \mu\frac{\partial \log Z_X}{\partial\mu}.
  \label{def-beta}
\end{equation}
The only dimensional arguments are $k$, $\omega$ and $\mu$, and the dimension
of $C_\psi^R$ is 2.  Then dimensional analysis leads to an Euler
equation which can be used to eliminate the $\mu$ derivative:
\begin{equation}
  \left[\mu\frac{\partial}{\partial\mu} + {\bf k}\cdot\nabla_k + 2
    \omega \frac{\partial}{\partial\omega}\right] C_\psi^R = -2 C_\psi^R.  
  \label{eq:CS-Euler-dimensional}
\end{equation}
Restricting ourselves to changes in the scale of ${k}$, i.e.\
${\bf k}=\boldsymbol{\mu}/b$, we have that
${\bf k}\cdot\nabla_k = -b\partial/\partial b$.  Then combining
\eqref{eq:CS-RG-Rpsi} and \eqref{eq:CS-Euler-dimensional} we get
\begin{equation}
  \left\{\sum_l \beta_l
    \frac{\partial}{\partial l} + \nu_\Gamma
    \Gamma\frac{\partial}{\partial\Gamma} 
    - 2 -2\omega\frac{\partial}{\partial\omega}
    +b\frac{\partial}{\partial b} \right\} C_\psi^R
      = 0
  \label{eq:CS-final-eq-RG}
\end{equation}
(we have omitted terms that only appear beyond one loop). We solve
\eqref{eq:CS-final-eq-RG} using as initial condition $b=1$, i.e.\ the
value of the correlation at a reference ${\bf k}=\boldsymbol{\mu}$, at some
frequency $\omega$, and at the physical values of the couplings
$\Gamma$, $u$, $f$, $w$, $\cR$.  The solution, found by the method of
characteristics, is
\begin{multline}
  C_\psi^R\left(\frac{\boldsymbol{\mu}}{b}, \omega, \Gamma, u, f, w,
     \cR\right) = \\
  b^2  C_\psi^R\left({\bf k}_0,\frac{\omega b^2}{\hat\Gamma(b)},\hat
    u(b),\hat f(b),\hat w(b),\hat \cR(b)\right),
  \label{eq:rg-solution}
\end{multline}
with
\begin{equation}
   \hat\Gamma(b) = \Gamma
              \exp\left[-\int_1^b\!\!\frac{\nu_\Gamma(b')}{b'}\,db'\right], \label{eq:CS-Gamma-solution}
\end{equation}
and where $\nu_\Gamma$ depends on $b$ through the couplings $f,w,\cR$.
The dependence of these on $b$ is given by the functions $\hat u(b)$
etc.\ (the \emph{running coupling constants}), which are the solution
of the system
\begin{subequations}
  \begin{align}
    b\frac{d\hat u}{db} &= -\beta_u(\hat u), & \hat u(1)=u,
    \label{CS-flow-u}\\
    b\frac{d\hat f}{db} &= -\beta_f(\hat f, \hat w, \hat\cR), & \hat
                                                                f(1)=f,
    \label{CS-flow-f} \\
    b\frac{d\hat w}{db} &= -\beta_w(\hat f,\hat w,\hat \cR), &\hat
                                                               w(1)=w,
    \label{CS-flow-w} \\
    b\frac{d\hat \cR}{db} &= -\beta_\cR(\hat\cR,\hat f), &
                                                           \hat\cR(1)=\cR, \label{CS-flow-R}
  \end{align}\label{CS-flow-all}
\end{subequations}
where the $\beta$ functions must be computed perturbatively from the
relation between the bare and renormalized couplings
\eqref{CS-renorm-bare}.  At one loop, the flow of the static coupling
constant $u$ is completely uncoupled from the dynamic couplings, so we
do not take it into account in the following.

The meaning of \eqref{eq:rg-solution} is that the correlation function at
the physical values of the couplings $\vec{u}\equiv(u,f,w,\cR)$ and at
a rescaled wave vector $\boldsymbol{\mu}/b$ is equal to the response function
at the original scale $\boldsymbol{\mu}$ but evaluated for different
couplings $\hat{\vec{u}}(b)$.  Fixed points thus are  sets of
coupling values $\vec{u}^*$ such that all $\beta$ functions vanish
simultaneously: it is clear that if the flow starts at such a point,
or approaches it for some large value of $b$, it will stay there for
all larger $b$.  If in addition the function
$C_\psi^R(k,\omega/\Gamma,\vec{u})$ is continuous at
$\vec{u}=\vec{u}^*$, then all the $k$-dependence at large $b$ (small
$k$) is contained in $\omega b^2/\Gamma(b)$: equation
\eqref{eq:rg-solution} is then the scaling law we seek, and we can
read off the scaling behavior from its second argument even if we
don't know the form of $C_\psi^R$.  For example, if the flow is near a
fixed point for $b>b^*$, then
$\nu_\Gamma(b)=\nu_\Gamma(\vec{u}(b))\approx
\nu_\Gamma(\vec{u}^*)\equiv \nu_\Gamma^*$.  Then
\eqref{eq:CS-Gamma-solution} gives
\begin{equation}
  \hat\Gamma(b) \approx \Gamma \exp\left[ -\int_1^{b^*}
    \!\!\frac{\nu_\Gamma(b')}{b'}\,db' - \nu_\Gamma^* \log(b/b^*)
  \right] \sim b^{-\nu_\Gamma^*}. 
  \nonumber
\end{equation}
Since $k=\mu/b$, we have $b\sim k^{-1}$ and
\begin{equation}
  \frac{\omega b^2}{\hat\Gamma(b)} \sim \omega b^{2+\nu_\Gamma^*} \sim
  \omega k^{-z}, \qquad z=2+\nu_\Gamma^*,
\end{equation}
i.e.\ the value of $\eta_\Gamma$ at the fixed point gives the
correction to the naive dynamic critical exponent.

So we proceed next (secs.~\ref{sec:determ-z-fact}
and~\ref{sec:treatm-cross-determ}) to deterine the $Z$-factors that
furnish the $\beta$-functions, and then
(sec.~\ref{sec:fixed-points-dynamic}) to find the fixed points of the
flow \eqref{CS-flow-all} and their infrared (i.e.\ $b\to\infty$)
stability.  The infrared stable fixed points will rule the scaling
behavior at large lengthscales.  Unstable fixed points may, depending
on initial conditions, lead to transient scaling laws observable in
certain regimes.

\subsection{Determination of $Z$-factors}
\label{sec:determ-z-fact}

We must determine the two dynamic $Z$-factors $Z_\lambda$ and
$Z_\Gamma$ (in a two or higher loops calculation a third factor,
related to field renormalization, would arise, but we do not need it
here). First $Z_\lambda$ is fixed by requiring that
$\partial G^{-1}_s/\partial k^2|_{\omega=0,k=\mu}$ be finite.  We
have
\begin{equation}
  G^{-1}_s (k,\omega) = 
  \left( -i\omega+ \eta_0  + \lambda_0 k^2 - \Pi(k,\omega) \right),
\end{equation}
where $\Pi$ is the same self-energy\footnote{
Notice, however, that from now on all the integrals in $k$ in the self-energies will no longer be performed on-shell, but rather between $0$ and $\infty$, and for this reason we drop the subscript $b$ from the self-energy symbols.}
as in equation \eqref{Pi}. From Eq.(\ref{Gr}) we then have 
\begin{equation}
  \frac{\partial (G_s^R)^{-1}}{\partial
      k^2}
      \bigg|_{k=\mu \atop \omega=0} 
      =
    \frac{\lambda}{Z_\lambda} - \frac{\partial \Pi}{\partial
        k^2} 
      \bigg|_{k=\mu \atop \omega=0}
  \label{CS-tombola}
\end{equation}
From this equation two conclusions follow: the first is that the
$\lambda_0 k^2$ term  \emph{cannot be left out} from a
renormalizable theory.  This is a consequence of the fact that, in an
expansion of $\Pi$ in the external wavevector, it is the $k^2$
coefficient that is divergent, not that of $k^0$ (in fact
$\Pi(k=0)=0$).  Thus if $\lambda_0$ is absent, it is useless to define
$\eta=Z_\eta \eta_0$ and try to absorb the pole of $\Pi$ into
$Z_\eta$, because $\eta$ drops out from \eqref{CS-tombola}.  This is
equivalent to the finding, in the momentum-shell scheme, that the
renormalization transformation generates a $\lambda$ coefficient
even if it is absent in the original theory.

The second conclusion is that $Z_\lambda$ is determined solely by the
behavior of $\Pi$, and, since at one loop this self energy is
unchanged with respect to model G, we can without further discussion
write it from the model G result \cite{de1978field}:
\begin{equation}
Z_\lambda = 1 + \frac{f}{2\epsilon}.
\end{equation}

At one loop, the differences between model G and ISM are only found in
$Z_\Gamma$, which we proceed to compute now.  The propagator of the
$\psi$ field is
\begin{equation}
  G^{-1}_\psi(k,\omega) = \delta_{\alpha\beta} \left[
  - i\omega + \Gamma_0( k^2+r_0) -\Sigma(k,\omega) \right],
\end{equation}
where $\Sigma$ is the self-energy \eqref{sigma}, which we recall here for
convenience:
\begin{align}
  \Sigma &=-2 g_0^2  \int\!\!\frac{d^dp}{(2\pi)^d}
                        \frac{k^2+r_0}{p^2+r_0} \times \notag\\
  & \qquad \times \frac{1}{ -i\omega
                        +\Gamma_0(p^2+r_0) +\lambda_0({\bf k}-{\bf p})^2+\eta_0}.
 \label{CS-Sigma-psi-bare}
\end{align}
We must now consider the renormalized derivative
\begin{equation}
  \frac{\partial (G^R_\psi)^{-1}}{\partial k^2}
  \bigg|_{k=\mu \atop \omega=0}
   =
 \frac{\Gamma}{Z_\Gamma} -
   \frac{\partial \Sigma}{\partial k^2} \bigg|_{k=\mu \atop \omega=0}
  \label{CS-ren-psi}
\end{equation}
and choose $Z_\Gamma$ so that it is finite.  In the dimensional
regularization procedure this means that the $Z$-factor cancel the
poles that appear for $d=4$ (i.e.\ terms proportional to $1/\epsilon$,
$\epsilon=4-d$), so that \eqref{CS-ren-psi} is  free of poles.
Computing the derivative from \eqref{CS-Sigma-psi-bare} for general
dimension $d=4-\epsilon$ at the critical point and ignoring the
contribution from a convergent integral one has
\begin{widetext}
  \begin{align}
    \frac{\partial (G^R_\psi)^{-1}}{\partial k^2}\bigg|_{k=\mu \atop \omega=0}
    &=\Gamma_0 \left\{ 1 + K_d \frac{\Gamma(d/2)}{\pi^{d/2}\Gamma_0}
      \frac{g_0^2\mu^{d-4}}{\Gamma_0+\lambda_0} \int
      \!\!\frac{d^dx}{x^2}  \frac{1}{x^2+[-2\lambda_0\hat{\boldsymbol \mu}\cdot
      {\bf x} + \lambda_0 + \eta_0/\mu^2]/[\Gamma_0+\lambda_0]} \right\} \notag\\
    &=\Gamma_0
      \left\{ 1  + K_d \Gamma(d/2) \frac{g_0^2}{\Gamma_0\lambda_0}
      \frac{\mu^{-\epsilon}}{1+\Gamma_0/\lambda_0} \Gamma(2-d/2)
     \  I_\psi \left( \cR,w \right)
      \right\}, \label{CS-deriv-bare-1} 
      \end{align}
  \end{widetext}    
where $\cR^2= \lambda/\eta = \mu^2 Z_\lambda \lambda_0 / \eta_0$, $\hat{\boldsymbol \mu}=\boldsymbol\mu/\mu$ and,      
      \beq
  I_\psi(\cR,w)   \equiv  \int_0^1\!\!d\beta\, \left[ \frac{1+\cR^{-2}}{1+w} \beta
      - \frac{1}{(1+w)^2}\beta^2 \right]^{d/2-2}  \ .
      \label{CS-Gamma-deriv-bare-2}
  \eeq
The renormalized counterpart can therefore be written as,
\begin{multline}
  \frac{\partial (G^R_\psi)^{-1}}{\partial k^2}\bigg|_{k=\mu \atop \omega=0}
  =
      \Gamma
     \left\{ \frac{1}{Z_\Gamma} + \right. \\ \frac{f}{1+w} \Gamma(d/2)
     \left. \Gamma(2-d/2) I_\psi(\cR,w) \right\}, \label{CS-GammaR}
\end{multline}
The one-loop term has a pole in $d=4$ (from the second $\Gamma$
function): this is how the original divergence of the integral in
$d=4$ manifests itself in dimensional regularization.  The minimal
renormalization prescription stipulates that this pole be identified
so that an equivalent pole but with opposite residue can be added to
$1/Z_\Gamma$, thus making the renormalized vertex finite.  So we
expand the second term: setting
${\cal I}(\cR,w)=(1+\cR^{-2})\beta/(1+w) - \beta^2/(1+w)^2$,
\begin{align}
  \frac{\partial (G^R_\psi)^{-1}}{\partial k^2} 
   &=  \Gamma
     \left\{ \frac{1}{Z_\Gamma} +  \frac{f}{1+w} \left[
     \frac{2}{\epsilon} + O(\epsilon^0) \right] \right. \times \notag \\
  &\quad \times \left. \left[ 1 - \frac{\epsilon}{2} \int_0^1\!\!d\beta\, \log {\cal I}(\cR,w)
    + O(\epsilon^2) \right]  \right\} \notag \\
   &=   \Gamma
     \left\{ \frac{1}{Z_\Gamma} + \frac{2}{\epsilon} \frac{f}{1+w}
     + \ldots \right\},
   \label{eq:CS-1}
\end{align}
As long as $\cR\neq0$, all singular behavior is contained in the pole
at $d=4$, i.e.\ the term proportional to $1/\epsilon$ in the last line
of \eqref{eq:CS-1}.  Then defining $Z_\Gamma=1 + (2/\epsilon)f/(1+w)$
renders $G^R_\psi$ finite.  Thus naively one finds that $Z_\Gamma$ is
independent of $\cR$, and, since $Z_\lambda$ is also independent of
$\cR$ at one loop, this leads to $\beta$-functions for the parameters
$f$ and $w$ that are independent of $\cR$, and thus to flow equations
identical to model G for $f$ and $w$, uncoupled to the flow of $\cR$.
However, this is wrong: we have already seen, in the momentum-shell
scheme, that the presence of $\eta_0$ profoundly affects the flow of
$f$ and $w$, with the notable macroscopic consequence of a change in
the dynamic critical exponent.

Even without the insight we have from the momentum-shell calculation,
one could guess that the naive expectation cannot be right: since
$\cR_0$ has the dimensions of a length, one expects that its stable
fixed point is 0, and indeed below we shall find from the
$\beta$-function \eqref{CS-betaR} that $\cR \sim b^{-1+f^*/4}$ for
$b\to\infty$, with $f^*$ of order $\epsilon$.  In the above equations,
one sees that the limit $\cR\to0$ requires special treatment: in
\eqref{eq:CS-1} a logarithmic divergence appears in the $\epsilon$
expansion of $I(\cR,w)$, and even before expanding one sees that
\eqref{CS-Gamma-deriv-bare-2} is problematic because $I_\psi(\cR\to0,w)$
vanishes for $d<4$, while $\lim_{\cR\to0} \lim_{d\to4} I_\psi(\cR,w)=1$.

The difficulty here is that the length scale $\cR$ occurs in the
Gaussian part of the dynamical functional, and that it appears in the
loop integrals in such a way that their convergence properties change
at one of the fixed points of $\cR$.  To deal with this we use a
generalized minimal substraction as discussed by Frey and others
\cite{frey1995crossover, frey1990renormalized}.  This method involves
incorporating the singular $\cR$ dependence into the renormalization
$Z$-factors (and consequently into the $\beta$-functions): we thus
``enrich'' the pole with a crossover factor $X(\cR,w)$ extracted from $I(\cR,w)$
which absorbs the singular $\cR\to0$ behavior.  Let us rewrite
\eqref{CS-GammaR} as
\begin{align}
  \frac{\partial (G^R_\psi)^{-1}}{k^2} &=
   \Gamma \biggl\{ \frac{1}{Z_\Gamma} +
                                         \frac{f}{1+w}
                                         \Gamma(d/2)
                                         \Gamma(2-d/2) \times \notag\\
  & \qquad \times  X(\cR,w) X^{-1}(\cR,w) I_\psi(\cR,w) \biggr\},
    \label{CS-GammaR-enriched}
\end{align}
where $X(\cR,w)$ is such that $X^{-1}(\cR,w)I_\psi(\cR,w)$ (and in
consequence all coefficients of its $\epsilon$-expansion) is
well-behaved for all values of $\cR$ (including $\cR=0$,
$\cR=\infty$).  We discuss in the next subsection how to fix this
factor, but before let us write the renormalization factors including
the as-yet unknown $X(\cR,w)$:
\begin{align}
  Z_\lambda &= 1 + \frac 1\epsilon f/2 ,  \\
  Z_\Gamma &= 1 + \frac 1\epsilon 2 X(\cR,w) \frac{f}{1+w}.
\end{align}
We conclude this subsection writing $\nu_\Gamma$, $\nu_\lambda$, and
the $\beta$ functions for the couplings $f$, $w$ and for $\cR$.  These
functions determine the RG flow and the asymptotic scaling properties
of the observables (sec.~\ref{sec:rg-equation-r_psir}). Recalling the definition (\ref{def-beta}) of $\beta$ functions and $\nu$ exponents, and developing the to first order in $f$ we get 
\begin{align}
  \nu_\Gamma & =  -2X(\cR,w) \frac{f}{1+w},\\
  \nu_\lambda & = -\frac{f}{2}, 
  \end{align}
  so that the $\beta$-functions (to be compared to \eqref{betafunc}) are the following,
  \begin{align}
  \beta_f &\equiv \mu\frac{\partial f}{\partial \mu} = 
  -f (\epsilon + \nu_\Gamma + \nu_\lambda) \notag\\
  & = -f \left[ \epsilon - f \left(\frac{1}{2} + \frac{2X}{1+w} \right)\right] \label{CS-betaf}\\
  \beta_w &\equiv \mu\frac{\partial w}{\partial \mu} = w
            (\nu_\Gamma-\nu_\lambda) = 
            w f \left[ \frac{1}{2} -
            \frac{2X}{1+w}\right],\label{CS-betaw}\\
  \beta_\cR & \equiv \mu\frac{\partial \cR}{\partial\mu} = \cR(1+\nu_\lambda/2) =
            \cR \left[ 1 - \frac{1}{4} f \right]. \label{CS-betaR}
\end{align}

\subsection{Determination of the crossover factor $X$}
\label{sec:treatm-cross-determ}

To determine $X(\cR,w)$, the idea is to to fix it in such a way that
the renormalization factor $Z_\Gamma$ contains all the singularities
near both fixed points $\cR=0$ and $\cR=\infty$.  When $\cR$ is
nonzero, the only singularity in \eqref{CS-GammaR} is the pole at
$d=4$ originating in the Gamma function $\Gamma(2-d/2)$.  Thus the
first condition we impose is that
\begin{equation}
  \lim_{\cR\to\infty}X(\cR,w) = 1.
\end{equation}
To find another condition, we must study $G^{-1}_{\psi}$ for
vansishing $\cR$. To do this let's define a new parameter
\begin{equation}
  q \equiv \frac{g^2}{\Gamma \eta} = \frac{K_d g_0^2 \mu^{d-4}}{Z_\Gamma
    \Gamma_0 \mu^{-2}\eta_0} = f\cR^2,
\end{equation}
and rewrite \eqref{CS-deriv-bare-1} as
\begin{align}
  \frac{\partial (G^R_\psi)^{-1}}{\partial k^2} &= 
                                                 \Gamma \left\{ \frac{1}{Z_\Gamma} + 
  q \frac{\Gamma(d/2)}{\pi^{d/2}} \int \!\!\frac{d^dx}{x^2+m^2}
                                                 \times\right.\notag\\
  &\qquad\times \left.\frac{1}{\cR^2wx^2 + \cR^2({\bf x}-\hat{\boldsymbol \mu})^2 +1
                                                 }\right\},
  \label{CS-GammaR-with-q}
\end{align}
where we have introduced a constant $m^2$ to avoid an infrared
divergence in $d=2$.  We can now set $\cR=0$ in the integral to find
\begin{equation}
    \frac{\partial (G^R_\psi)^{-1}}{\partial k^2}
   = \Gamma\left\{ \frac{1}{Z_\Gamma} + q
     \Gamma(1-d/2)\Gamma(d/2) m^{d-2} \right\}.
\end{equation}
We now find a pole at $d=2$, corresponding to the fact that for
$\cR=0$ the integral in \eqref{CS-GammaR-with-q} diverges for
$d\geq2$: \emph{the critical dimension for $q$ is $d_c=2$,} not 4.  We
then expand around $d_c=2$, and find
\begin{align}
  Z_\Gamma &= 1 + q\frac{2}{2-d} = 1 + \frac{2}{\tilde \epsilon}q, \\
  \beta_q &= -(2-d) q + 2 q^2 = -\tilde \epsilon q + 2 q^2,  \label{CS-beta-q-limit-r0}
\end{align}
where,
\beq 
\tilde \epsilon=2-d \ .
\eeq
  We note that $\beta_q$ above is correct
up to second order \emph{in $\tilde\epsilon$, i.e.\ around the new
  critical dimension $d_c=2$.} Equation \eqref{CS-beta-q-limit-r0} can give us the condition to impose on
$X(\cR,w)$ near the other fixed point: from \eqref{CS-betaf} and
\eqref{CS-betaR} we can write
\begin{equation}
  \beta_q = \cR^2\beta_f + 2\cR f \beta_\cR =
  -q \left[ 2-d - 2X(\cR,w)\frac{f}{1+w}\right],
  \label{CS-beta-q-with-X}
\end{equation}
so that to recover \eqref{CS-beta-q-limit-r0} we impose
\begin{equation}
  \lim_{\cR\to0}\lim_{d\to2}X(\cR,w) = (1+w) \cR^2.
\end{equation}
A simple choice for $X(\cR,w)$ is then
\begin{equation}
  X(\cR,w) = \left[ \frac{(1+w)\cR^2}{1 + (1+w)\cR^2} \right]^{a(d)}.
  \label{CS-enrichment-factor}
\end{equation}
with $a(d=2)=1$.
Referring to \eqref{CS-GammaR-enriched}, we find
\begin{widetext}
  \begin{align}
    X^{-1}(\cR,w) I_\psi(\cR,w) &= \int_0^1\!\!d\beta\, \left[ \frac{ (1+w)
        \cR^2}{1+(1+w)\cR^2} \right]^{-a} 
    \left[ \frac{1+1/\cR^2}{1+w} \beta
      - \frac{1}{(1+w)^2}\beta^2 \right]^{-\epsilon/2}
      \nonumber \\
    &= \int_0^1\!\!d\beta\, \left[ \frac{ (1+w)
        \cR^2}{1+(1+w)\cR^2} \right]^{-a+\epsilon/2}
     \left[ \frac{1+\cR^2}{1+(1+w)\cR^2} \beta -
      \frac{\cR^2}{1+(1+w)\cR^2} \frac{\beta^2}{1+w}
    \right]^{-\epsilon/2},
  \end{align}
\end{widetext}
so that $X(\cR,w)$ defined as in \eqref{CS-enrichment-factor} with
$a=\epsilon/2$ fulfills the three conditions i) $X(\cR\to\infty)=1$,
ii) $X(\cR\to0,d\to2)=(1+w)\cR^2$, and iii) makes $X^{-1}(y,w) I(y,w)$
finite and non-vanishing for all values of $\cR$, so that all terms
not included in $Z_\Gamma$ (i.e terms of order $\epsilon^0$ or higher)
are regular and don't cause further trouble.  In particular
$\lim_{\cR\to0} \lim_{d\to4} X^{-1} I_\psi = \lim_{d\to4} \lim_{\cR\to0}
X^{-1} I_\psi=1$.

A subtle but important point however remains to be made.  At face
value, our choice of $X(\cR,w)$ recovers the flow
\eqref{CS-beta-q-limit-r0} only at $d$ exactly equal to 2, while
\eqref{CS-beta-q-limit-r0} has actually been obtained in general
dimension.  Thus it seems that one would want $a=1$, which however
accounts for the $\cR\to0$ behavior of $I(\cR,w)$ only at $d=2$.  The
way out of this seeming inconsistence is to remember that
\eqref{CS-beta-q-limit-r0} is valid in general dimension \emph{but
  only up to second order in $\tilde\epsilon=2-d = \epsilon -2$}.
This means that one should actually write this exponent as
$a=1+\tilde\epsilon/2$, and \eqref{CS-enrichment-factor} as
\begin{multline}
  X(\cR,w) =  \frac{(1+w)\cR^2}{1 + (1+w)\cR^2} \biggl\{ 1 + \\
    \frac{\tilde\epsilon}{2} \log \left[ \frac{(1+w)\cR^2}{1 +
       (1+w)\cR^2} \right]
   + \ldots \biggr\}.
\end{multline}
It is then clear that (since the nonzero fixed point of $q$ will be of
order $\tilde\epsilon$) \eqref{CS-beta-q-with-X} indeed recovers
\eqref{CS-beta-q-limit-r0} in the limit $\cR\to0$ for general
dimension \emph{up to order $\tilde\epsilon^2$}.  But then it also
becomes clear that \eqref{CS-beta-q-limit-r0} \emph{cannot fix
  $X(\cR,w)$} at order $\tilde\epsilon$ and beyond.  So
$a=1+\tilde\epsilon/2$ is fine near $d=2$ and also near $d=4$ (where
\eqref{CS-GammaR} must be expanded).

From these considerations, in what follows, we will simply set $a=1$
when writing the $\beta$ functions also in $d=3$, which is the
dimension of interest here.  We do so because this is the simplest
choice that gives a correct description of the flow at the one-loop
level: when $\cR\to0$ the $O(\tilde\epsilon)$ contributions to
$X(\cR,w)$ cannot be fixed without going to two loops so we may as
well omit them, and on the other hand near the $\cR\to\infty$ fixed
point $X(\cR,w)\to 1$ independently of the exponent.  So the final
form of $\beta$ functions we find within the CS scheme is identical to
equations \eqref{betafunc} we previously obtained with the momentum-shell
technique.

\subsection{Fixed points}
\label{sec:fixed-points-dynamic}

Equations~\eqref{CS-betaR} and~\eqref{CS-flow-R} show that $\cR^*=0$ and
$\cR^*=\infty$ are fixed points of $\cR$.  The flow can be solved
formally as
\begin{equation}
  \hat\cR(b) = \cR \exp\left[ \int_1^b\!\!\frac{1-f(b')/4}{b'}\,db'
  \right].
\end{equation}
Since we expect (and confirm below) that the fixed point of $f$ will
be of order $\epsilon$, we see that the integrand within the
exponential is negative (for large $b$) at least, and that $\cR^*=0$
is IR-stable while $\cR^*=\infty$ is IR-unstable.

\subsubsection{$\cR^*=\infty$ - conservative fixed point}
\label{sec:cr=infty-fixed-point}

The $\cR^*=\infty$ fixed point corresponds to $X(\cR,w)=1$.  Since it
is unstable, it is only relevant for $b\to\infty$ when the system
starts at $1/\cR=0$.  This corresponds to a very important special
case, namely $\eta_0=0$, i.e.\ model G (equivalent to model E for what
concerns the scaling properties).  It is also relevant at moderately
large scales for $1/\cR$ very small, when the flow stays near $X=1$
long enough that the other couplings approach the model E fixed point
before $\cR$ becomes so small that $X$ is significantly different from
1 (see secs.\ V.D and~\ref{sec:crossover}).

Equations \eqref{CS-flow-f} and \eqref{CS-flow-w} for $X=1$ were
studied by DeDominics and Peliti \cite{de1978field}, who considered
model E at two loops, and we refer to them for the analysis of the
fixed points and their stability.  In summary the relevant (IR-stable
in the $(f,w)$ subspace) fixed point is
\beq
  f^*= \epsilon \quad , \quad  w^*=3 \quad , \quad   \cR^*=\infty   
\eeq
implying,
\beq  
  \nu_\Gamma^*=-\frac{\epsilon}{2} \quad , \quad  z=2-\frac{\epsilon}{2}=\frac{d}{2}.
\eeq
It is interesting to remark that the result for $z$, although obtained
here at one loop, has to be valid at all orders in perturbation theory
as long as $w^*$ and $f^*$ are different from 0: if $w$ is non-null,
\eqref{CS-betaw} implies $\nu_\Gamma=\nu_\lambda$, and
\eqref{CS-betaf} then gives $2\nu_\Gamma=-\epsilon$.

\subsubsection{$\cR^*=0$ - dissipative fixed point}
\label{sec:cr=0-fixed-point}

When $\cR=0$, $X(\cR,w)=0$ which gives immediately
\begin{align}
  \nu_\Gamma^*&=0, & z&=2,
\end{align}
regardless of $w$ and $f$ (as long as they are finite).  Setting
$X(\cR,w)=0$ in \eqref{CS-flow-f} and \eqref{CS-flow-w} one finds two
solutions:
\begin{align}
  f^*&=0,\\
  \intertext{and}
  f^*&=2\epsilon, & w^*&=0.
\end{align}
\vskip 1 cm

The stability of the two fixed points can be studied linearizing the flow
\eqref{CS-flow-all} around the fixed point $\vec{u}^*$:
\begin{equation}
  b\frac{d\vec{u}}{db} = -W \bigl(\vec{u}-\vec{u}^*\bigr),
\end{equation}
where $W$ is the Jacobian matrix
\begin{equation}
  W =
  \begin{pmatrix}
    \frac{\partial \beta_f}{\partial f} & \frac{\partial \beta_f}{\partial w} &
    \frac{\partial\beta_f}{\partial\cR}\\
    \frac{\partial \beta_w}{\partial f} & \frac{\partial \beta_w}{\partial w} &
    \frac{\partial\beta_w}{\partial\cR}\\
    \frac{\partial \beta_\cR}{\partial f} & \frac{\partial \beta_\cR}{\partial w} &
    \frac{\partial\beta_\cR}{\partial\cR}
  \end{pmatrix}
\end{equation}
evaluated at the fixed point.  The fixed point is stable if $W$ is
positive definite, i.e.\ its eigenvalues are all positive.  We find
for the two cases above
\begin{equation}
  W(f=0,\cR=0) =
  \begin{pmatrix}
    -\epsilon & 0 & 0\\
    1/2 & 0 & 0\\
    0 & 0 & 1
  \end{pmatrix},
\end{equation}
which has eigenvalues 0, 1, and $-\epsilon$, while
\begin{equation}
  W(f=2\epsilon,w=0,\cR=0) =
  \begin{pmatrix}
    \epsilon & 0 & 0\\
    1/2 & \epsilon/2 & 0\\
    0 & 0 & 1-\epsilon
  \end{pmatrix},
\end{equation}
with eigenvalues $\epsilon$, $\epsilon/2$ and $1-2\epsilon$.  So the
only IR stable fixed point (at 1 loop, near $d=4$) is $f^*=2\epsilon$,
$w^*=0$, $\cR^*=0$, which implies as we have seen that the critical
exponent is $z=2$.

\subsection{Crossover}
\label{sec:crossover}

We have just concluded that the only IR-stable fixed point gives
$z=2$, so that for large observation scales ($k\to0$) the critical
dynamics is like that of a purely dissipative model (like model A)
when the starting (physical) value of $\eta$ is nonzero.  However, for
non zero but small $\eta$, such that the starting $\cR$ is very large,
the initial value of $X(\cR,w)$ is very close to 1 and will stay so
until $\cR(b)$ is of order one (e.g.\ for $w=3$, $X$ is larger than
0.99 for $\cR>5$).  So one can expect that $f$ and $w$ will at first
move as if $X=1$, i.e.\ towards the conservative (model G) fixed
point, staying in its neighborhood until $\cR$ decreases
significantly, and in effect the numerical study of the flux
(Fig.~\ref{2:fluxOnCm}) confirms this expectation.  Then
experimentally one will observe model G critical behavior ($z=d/2$)
for moderate (i.e.\ not too small) values of $k$, possibly lasting a
rather wide interval, until at some point for $k\to0$ the asymptotic
$z=2$ exponent will be seen.  We show here how to obtain the scaling
of the wavevector $k_c$ (marking the end of the model G behavior) with
the physical value of $\cR$ (cf.\ sec.\ V E).

Assume then that the physical value of $\cR$ is
$\hat\cR(b=1)\equiv\cR_1\gg1$ so that $X(\cR_1,w(1))\approx 1$.
Assume also that one is observing at a scale $k=\mu/b$ such that the
flow has already reach the neighborhood of the $z=d/2$ fixed point
(which in particular implies $\nu_\Gamma \approx -\epsilon/2$).  We
ask how small we must make $k$ so that the system moves away from this
fixed point and the scaling law changes.  For this to happen, $X$ must
be significantly less than 1, so let's impose that $X<c$, with
$c=0.99$ say.  This requires $\cR(k)<\cR_c=\sqrt{c/[4(1-c)]}$.  Now
since we are near the conservative fixed point we can use \eqref{CS-flow-R}
with $\nu_\Gamma=\nu_\Gamma^*$ to obtain
\begin{equation}
  \cR(k) \approx \cR_1 (k/\mu)^{1+\nu_\lambda^*/2}.
\end{equation}
Now the crossover wavevector will be such that $\cR(k_c)=\cR_c$, so
that
\begin{equation}
  k_c \sim \cR_1^{-1/(1+\nu^*_\lambda/2)} = \cR_1^{-1/(1-\epsilon/4)}=\cR_1^{-4/d} \ .
\end{equation}
Hence, we find the same crossover exponent as in momentum shell, namely, 
\beq
\kappa= 4/d \ .
\eeq
Let us notice that the crossover exponent $\kappa$ is nontrivial: from naive dimensional analysis one
would have guessed $k_c \sim \cR_1^{-1}$.  However, the renormalized
$\cR$ is dimensionless, and the RG result is actually taking into
account the nontrivial effects of the hidden microscopic lengthscale.

Finally, let us note that the crossover exponent derives its value from
$\nu_\lambda$ at the model G fixed point which, as we have mentioned
before, takes the value $-\epsilon/2$ at all orders in perturbation
theory \cite{de1978field}, and thus so must the crossover exponent.


\section{Numerical Simulations}
To test our results, we performed numerical simulations of the microscopic ISM model on a fixed lattice in $d=3$. We implemented the dynamical equations   \eqref{ISM1_micro} \eqref{ISM2_micro} using a generalized Verlet integrator \cite{allen1987,allen_brownian_1980,swope_computer_1982-1} for second order equations. Details of the algorithm can be found in \cite{cavagna2016spatio}. The lattice spacing is $\Lambda^{-1}=1$. We fixed the parameters $\hat J=1$, $\hat \chi = 1$, and performed simulations at several values of the temperature $T$ and of the friction coefficient $\hat\eta$. Since the temperature sets the correlation length and the friction regulates the conservation length scale $\mc R_0$ we can in this way explore the $(\xi,{\cal R}_0)$ plane of Fig.\ref{fig:regions}. For all values of $T$ and $\hat\eta$ considered, we computed  the correlation length $\xi$ and the relaxation time $\tau$, and inferred the exponent $z$ from the scaling relation Eq. \eqref{dynscal} between them. In this way, we could investigate the dynamical critical behavior and compare results with the predictions of the RG computation. Before illustrating the results, let us briefly explain the procedure followed to compute the main quantities required for our analysis, namely $\xi$ and $\tau$.
\begin{figure*}[t]
\centering
{\includegraphics[width=.30\textwidth]{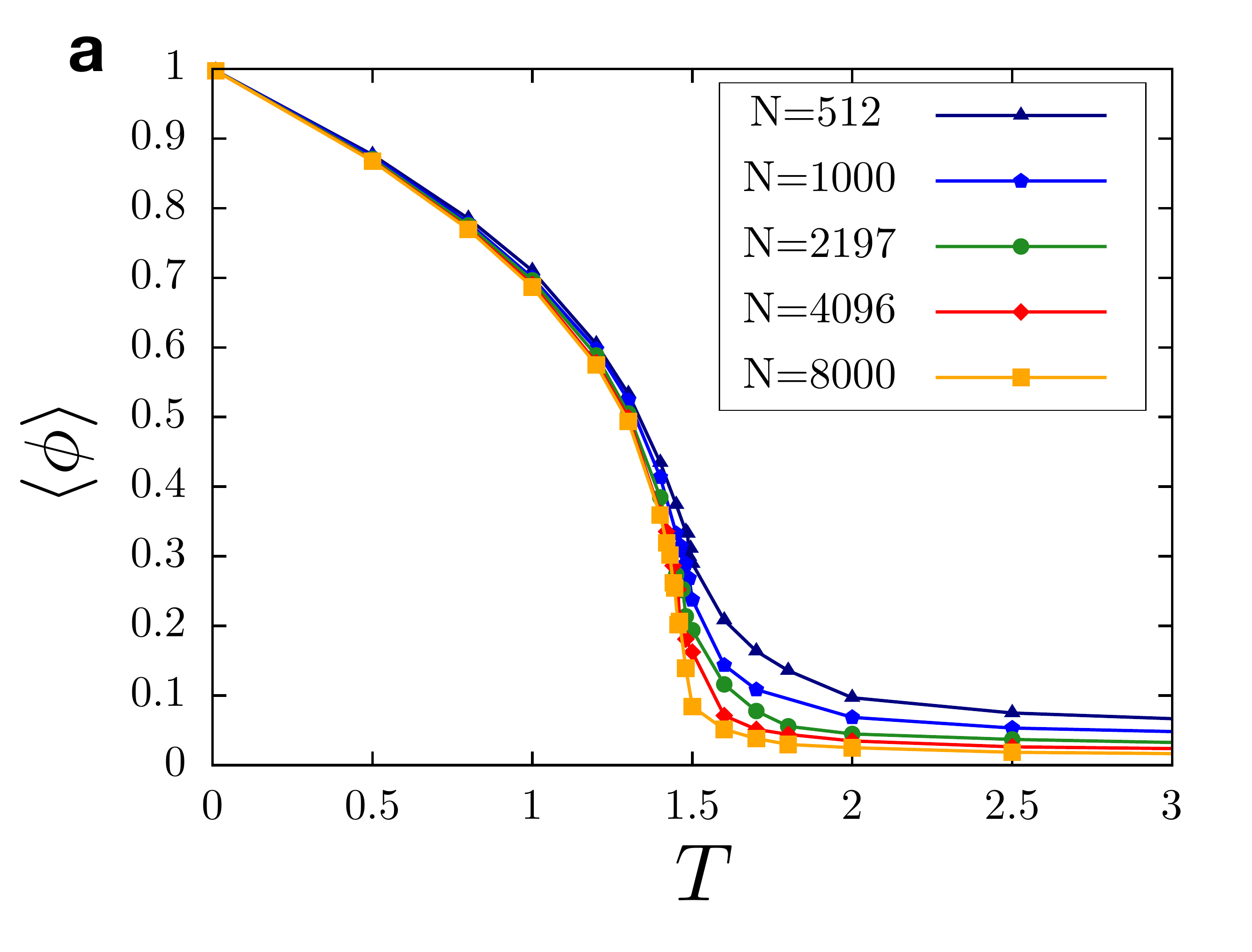}} 
 \quad
{\includegraphics[width=.325\textwidth]{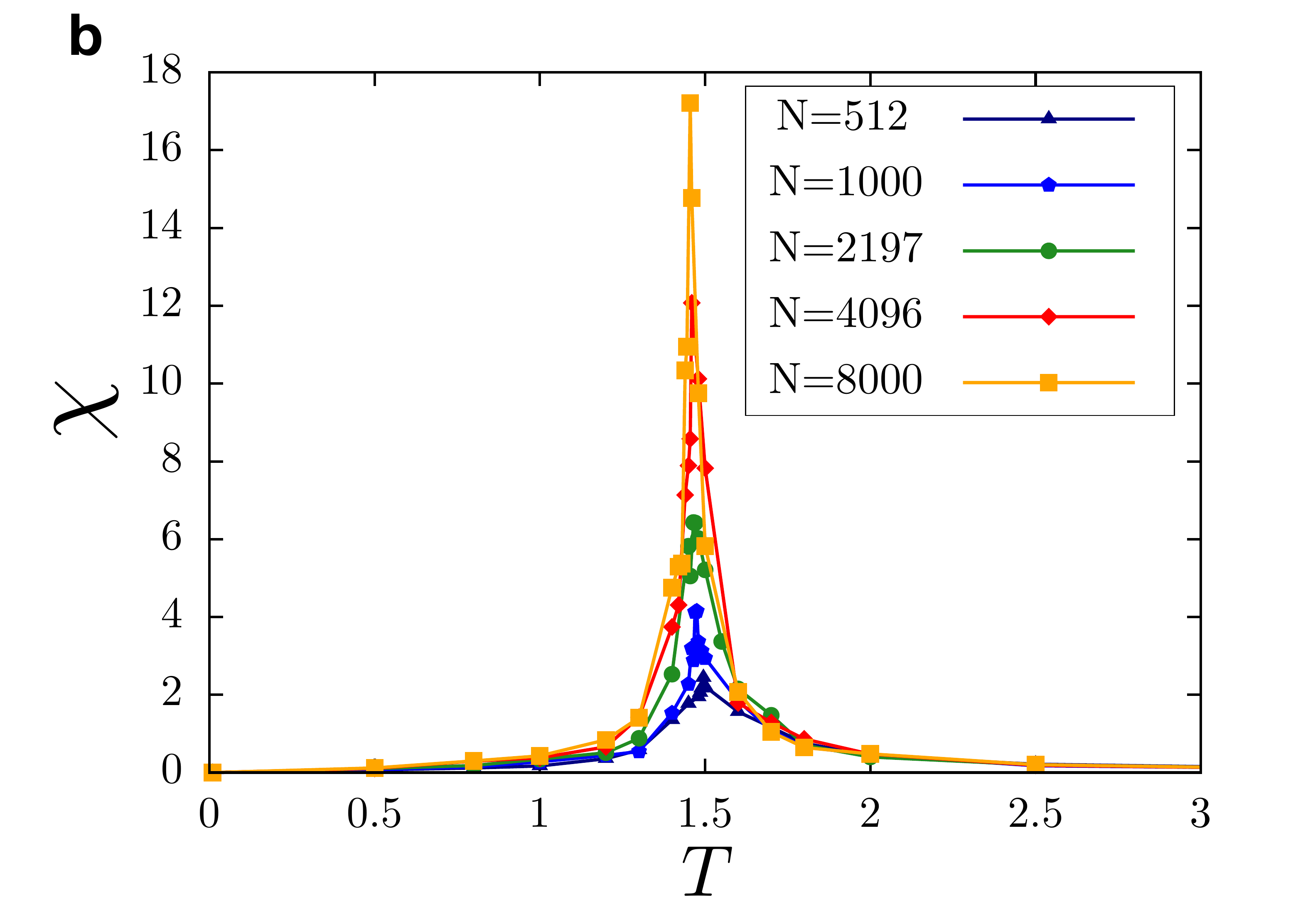}}
\quad
  \includegraphics[width = .32 \textwidth]{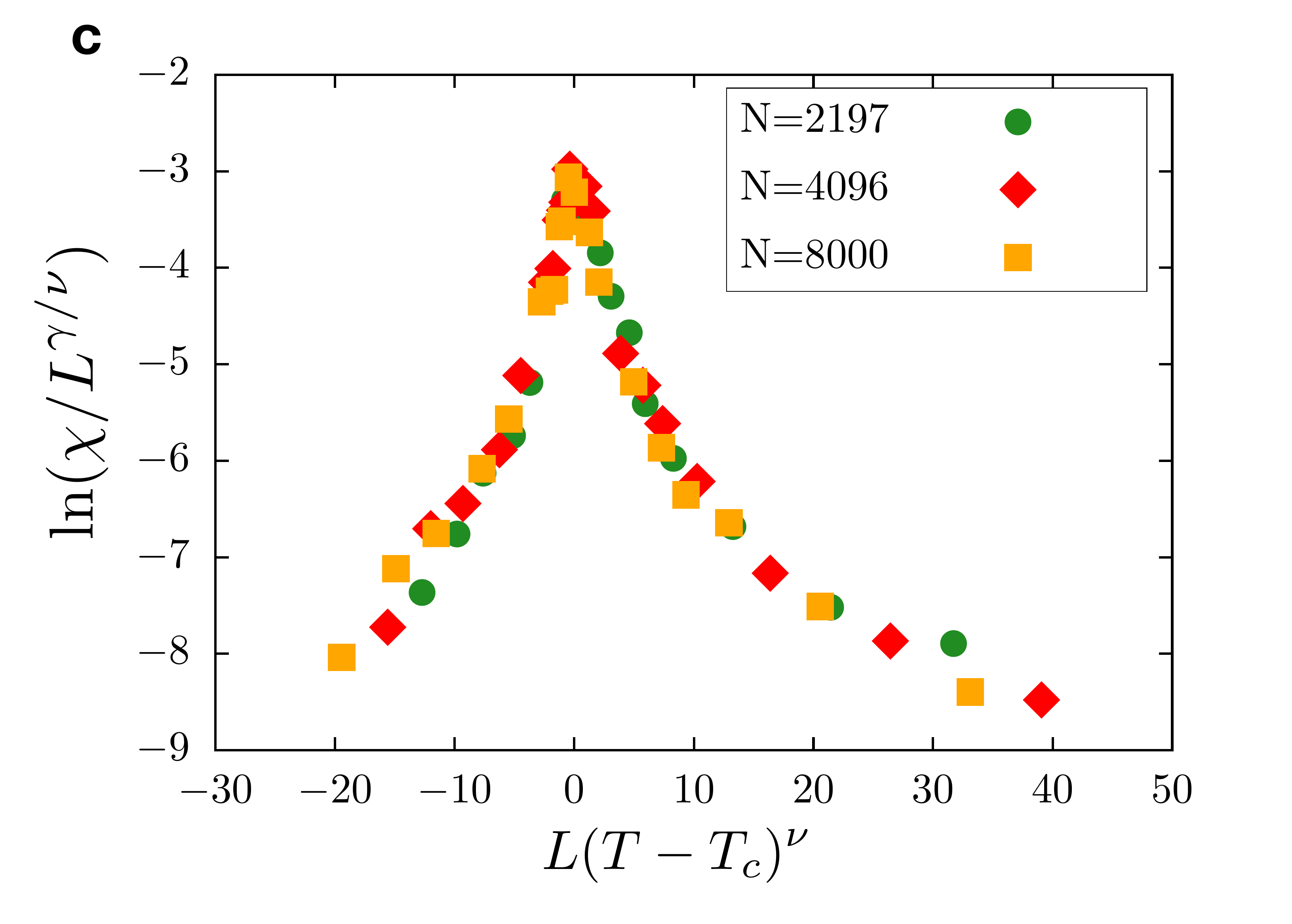}
\caption{ \textbf{Static critical behaviour.}
a): Average scalar polarization for temperatures $0.1 \leq T \leq 3.0$ and for different sizes ($N=512,1000,2197,4096,8000$). An ordering transition occurs at approximately  $T_c \simeq 1.5$. b) Susceptibility as a function of temperature, same sizes as in panel a); the maximum of each curve is located at a temperature that decreases with increasing the size of the system, and approaches the critical temperature $T_c$ in the thermodynamic limit. c) Finite size scaling of the susceptibility. Curves at $N=2197,4096,8000$ satisfy finite size scaling with  exponents $\nu = 0.707$ and $\gamma/\nu= 1.973$, as predicted by the theory of the Heisenberg model \cite{binney_book}.}
\label{fig:chi_polarization}
\end{figure*}

\subsection{Static behaviour and determination of $\xi$}
Since we are interested in the critical behavior of the system, we need first of all to locate the transition temperature and characterize the static critical properties of the system. To do so, we perform numerical simulations of  Eqs.\eqref{ISM1_micro} \eqref{ISM2_micro}  in the stationary regime, and use decorrelated dynamical configurations to compute equal time equilibrium averages (from now on indicated with $\langle \cdots \rangle$). From the static point view the ISM on a lattice is completely equivalent to a standard ferromagnetic model, we therefore expect static properties to reproduce the well known results of the Heisenberg model.
For a system of $N$ velocities/vectors, it is possible to define the polarization as:
\begin{equation}
    \boldsymbol{\Phi} = \frac{1}{N} \sum_i \boldsymbol{\psi}_i  \ ,
\end{equation}
measuring the degree of global alignment, and its modulus, the scalar polarization $\phi$.   The average value of this quantity is plotted in 
Fig.\ref{fig:chi_polarization}a as a function of temperature, and clearly shows the occurrence of an ordering transition. The critical temperature can be conveniently located by looking at the fluctuations of $\langle \phi\rangle$, namely the susceptibility
\begin{equation}
\chi = \beta N [ \langle  \phi ^2 \rangle - \langle \phi \rangle^2 ]  \ ,
\end{equation}
where $\beta$ is the inverse of the temperature. We analyzed these quantities for a wide range of temperatures ($0.1 \leq T \leq 5.0$) and sizes ($N=512, 2197, 4096,  8000$).
From Fig.\ref{fig:chi_polarization}a,b we can see that the critical temperature is located approximately at $T_c \simeq 1.5$. The critical point moves towards lower temperatures as the linear size of the system increases, in according to finite size scaling. 

To measure the correlation length $\xi$ we first computed the static connected correlation function $C(r)$:
\begin{equation}
    C(r) = \frac{\sum_{i,j} \langle \delta \boldsymbol{\psi}_i  \cdot \delta \boldsymbol{\psi}_j \rangle \delta(r-r_{ij}) }{\sum_{i,j} \delta(r-r_{ij})}  \ ,
\end{equation}
where $r_{ij}$ is the distance between two sites $i$ and $j$, and  $\delta \boldsymbol{\psi}_i = \boldsymbol{\psi}_i - \langle \boldsymbol{\psi}_i \rangle $. Since we are mostly interested in the paramagnetic phase of the model, the relevant one to describe experimental data of insect swarms, we focused on temperatures $T>T_c$, approaching the critical point from above. The behaviour of the correlation function (not displayed) is as expected for a Heisenberg model, we therefore computed the correlation length from the expression $rC(r) = \exp(-r/\xi)$, exploiting the fact that the anomalous dimension is small \cite{binney_book}. We combined this information with data on the susceptibility to obtain an estimate of the ratio between critical exponents $\gamma/\nu$ for sizes $N=2197,4096,8000$. Simulations at $N=8000$ give $\gamma/\nu = 1.905$,  in agreement with the literature \cite{binney_book}. We therefore used this size of the system for all following analysis. Finally, to further test the equivalence of the static properties of ISM with the Heisenberg model we performed a finite size scaling analysis on the susceptibility, as displayed in Fig.\ref{fig:chi_polarization}c.

\subsection{Dynamic behaviour and determination of $\tau$}
To investigate the dynamical behavior of the system one has to look at time dependent quantities. In particular, the characteristic time scale $\tau$ is by definition the scale over which fluctuations of the order parameter become decorrelated. To compute it, we introduce the spatio-temporal correlation function, that is:
\begin{equation}
    \label{ckt}
    \begin{split}
    C(k,t) &= \frac{1}{N} \sum_{i,j} \frac{\sin ( k r_{ij})}{k r_{ij}} \langle \delta \boldsymbol{\psi}_i (t_0) \cdot \delta \boldsymbol{\psi}_j (t_0+t) \rangle _{t_0} \\
    \langle ( \cdot ) \rangle_{t_0} &= \frac{1}{T_{max} -t} \sum_{t_0 =1}^{T_{max}-t} ( \cdot )  \,
    \end{split}
\end{equation}
with $T_{max}$ the length of the simulation. The number of operations needed to calculate this quantity is in general $\sim T_{max}N^2$; however what we actually need for the scaling analysis is the correlation function at $k=0$ (see previous section), which is numerically less demanding:
\begin{equation}
\label{C0t}
    \begin{split}
        C(k=0,t) &= \frac{1}{T_{max}-t}\sum_{t_0 =1}^{T_{max}-t} \overline{\delta \boldsymbol{\psi}(t_0)} \cdot \overline{\delta\boldsymbol{\psi} (t_0+t)} \\
        \overline{\delta \boldsymbol{\psi}(t_0)} &= \frac{1}{N} \sum_i \delta \boldsymbol{\psi}_i (t_0)  \ .
    \end{split}
\end{equation}
From this quantity, we computed the characteristic time scale $\tau$ from the condition,
\begin{equation}
    \frac{1}{2 \pi} = \int_0^\infty dt \frac{1}{\tau} \sin \left(\frac{t}{\tau} \right) \frac{C(k=0,t)}{C(k=0,t=0)}  \ .
\end{equation}
This condition corresponds to requiring that half of the total integrated area of the dynamic correlation function in the frequency domain comes from the interval $-\omega_c<\omega<\omega_c$, with $\omega_c=1/\tau$. This definition of $\tau$  has the advantage of capturing the relevant time-scale both when relaxation is dissipative, and when propagating modes are present, and it is the standard definition adopted in the literature on dynamic critical phenomena \cite{HH1969scaling}.

\subsection{Dynamic crossover}
\begin{figure}
    \centering
    \includegraphics[width = 0.45 \textwidth]{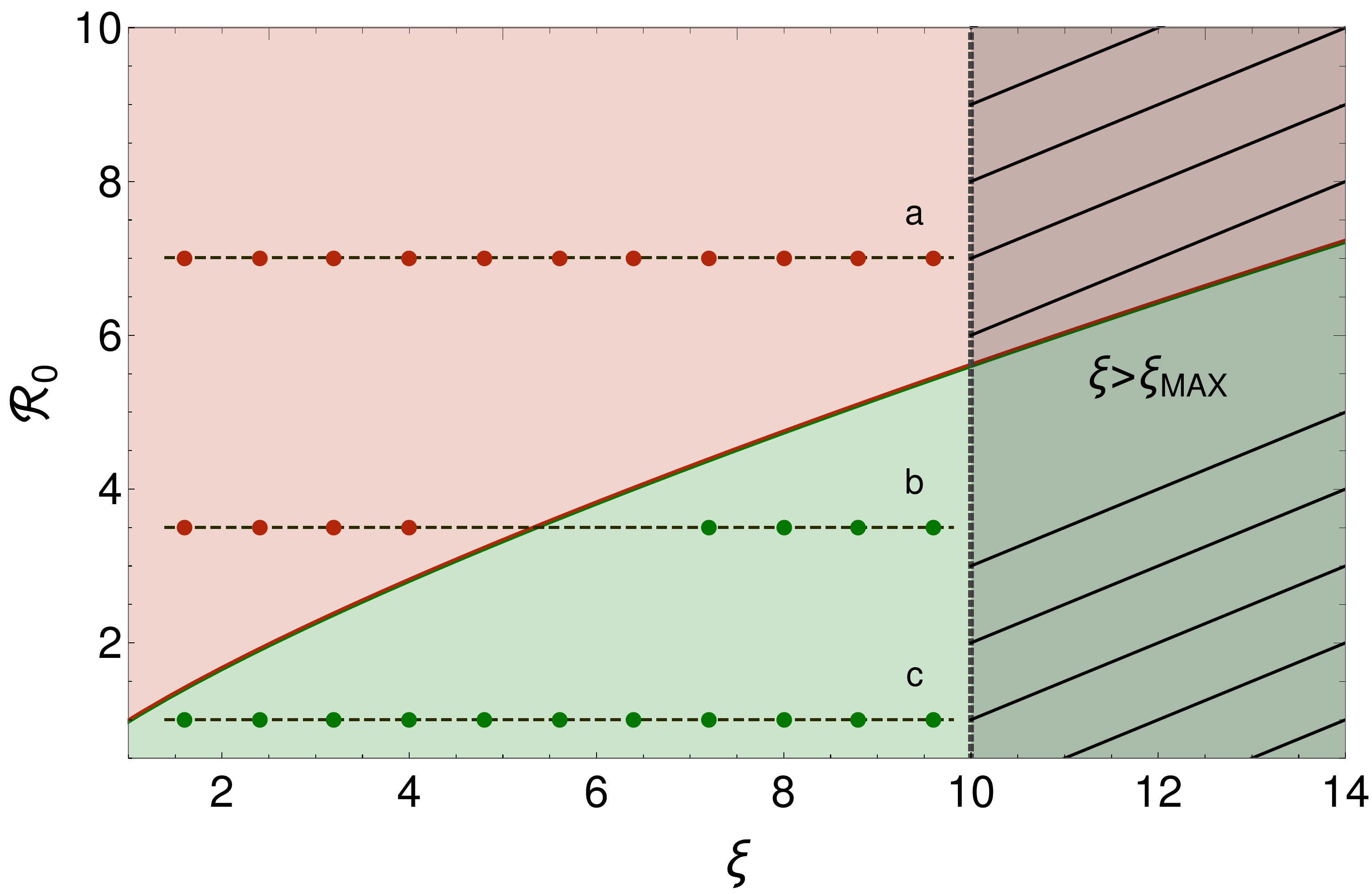}
    \caption{ {\bf Numerical protocol in the $(\xi,{\mathcal R}_0)$ plane}.
Simulations performed at fixed $\hat\eta$ and different $T$ correspond to exploring the $(\xi,{\mathcal R}_0)$ plane along horizontal segments. Since the size of the system is finite ($L\leq 20$), only a limited window of $\xi$ can be accessed and the length of such segments is finite ($1=\Lambda^{-1}<\xi<\xi_{max}=10$). According to the RG prediction, for all values of $\hat\eta$ corresponding to ${\cal R}_0> 10^{3/4}$ the segments belong entirely to the conservative region (segment a). For larger values of $\hat\eta$, such that $1<{\cal R}_0<10^{3/4}$, the segments cross from the conservative region to the dissipative one (segment b): in this case there is no sufficient span  in each region to extract the exponent $z$ from the $\tau$ vs $\xi$ plot. Since the minimum physical value of ${\cal R}_0$ is $\Lambda^{-1}=1$, larger values of $\hat\eta$ are all equivalent to the ${\cal R}_0=1$ case (segment c).}
\label{fig:protocols}
\end{figure}

 Our primary objective is to observe the crossover in the dynamic critical behavior predicted by the RG computation. The simplest protocol to do that would seem to fix  the value of the dissipation coefficient (and therefore ${\cal R}_0$), and extensively vary the correlation length by tuning $T$. In the $(\xi, {\cal R}_0)$ plane of Fig.\ref{fig:regions} it corresponds to a straight horizontal line crossing from the red conservative region on the left to the dissipative green one on the right. 
 In numerical simulations, when plotting $\tau$ vs. $\xi$, we should then observe two different power laws, one with exponent $z=d/2$ for small $\xi$, and another one with $z=2$ at large $\xi$.  The problem with this protocol is that  to see a power-law crossover one should span several orders of magnitudes in $\xi$; three decades is the very minimum, but $L=10^3$, gives $N=10^9$ in $d=3$, which is quite awful, considering that the largest relaxation time would be of order, $\tau \sim \xi^2 \sim L^2 \sim 10^6$. This is not possible, and the maximum size we used is well below ($L\leq 20$). 
 In other terms, as illustrated in Fig.\ref{fig:protocols}, we can in practice explore only a finite horizontal interval in $\xi$ included between the lattice spacing and the maximum achievable $\xi$, which is not long enough to appropriately measure the exponents in the two dynamical regimes (segment b in Fig.\ref{fig:protocols}). Interestingly, the same figure shows that for many values of ${\cal R}_0$ the accessible interval does not even intercept the crossover line, but entirely lies within the same dynamic regime (segments a and c in Fig.\ref{fig:protocols}). In this case,  in numerical simulations at fixed $\hat\eta$ we should expect that only one power-law is observed in the $\tau$ vs. $\xi$ plot. For this reason, the best way to capture the presence of the crossover is to run simulations at different values of the effective friction. If the picture above is correct, when  $\hat\eta$ is small, corresponding to large ${\cal R}_0$, we should measure $z=3/2$ (segment a), while for large enough $\hat\eta$ we should measure $z=2$ (segment c).

\begin{figure}[t]
    \centering
    \includegraphics[scale = 0.28]{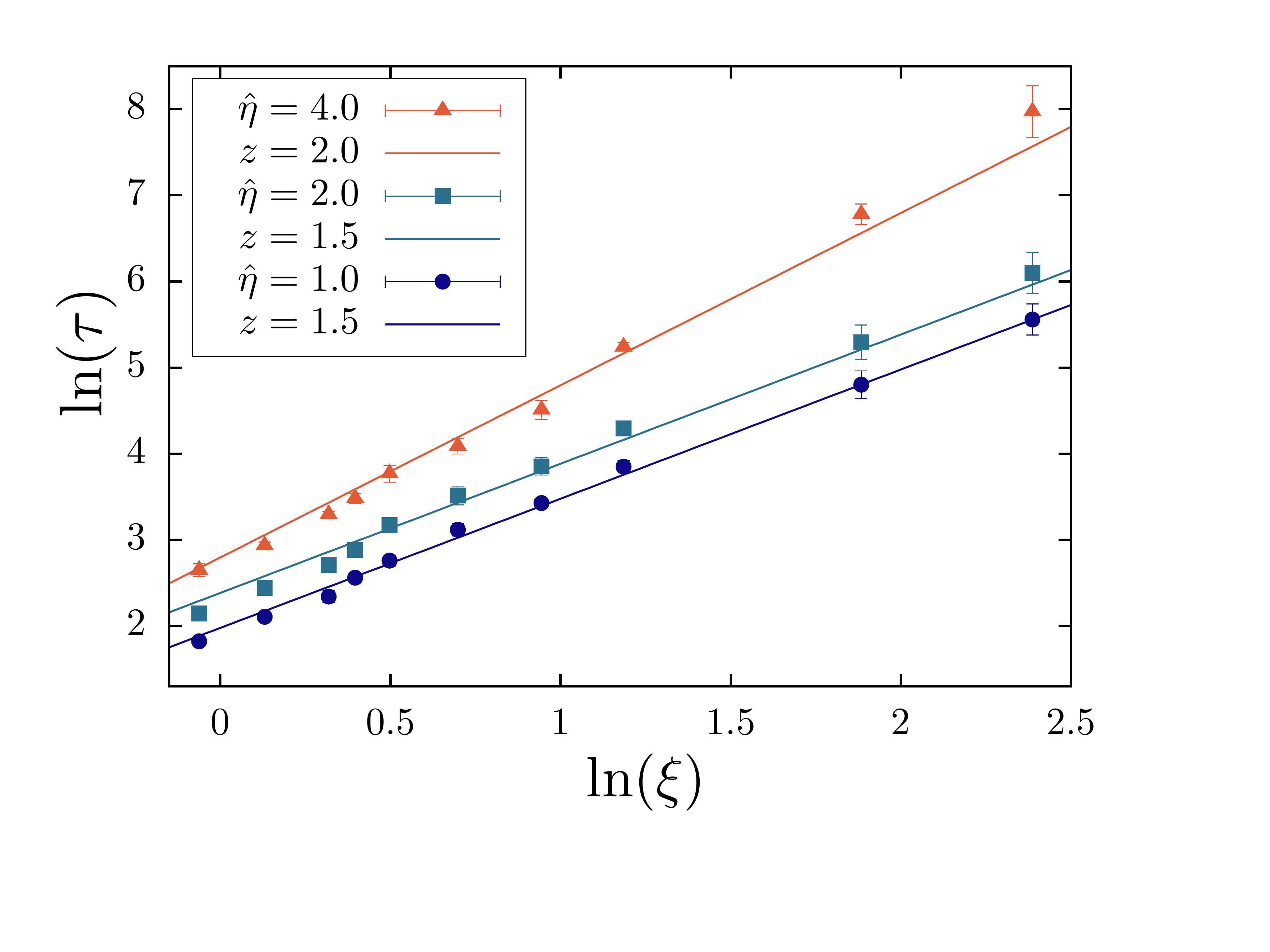}
    \caption{\textbf{Dynamic critical exponents.} Relaxation time vs correlation length in $d=3$, for $L=20$, $N=8000$, and $T\in[1.48:2.00]$,   at various values of the friction coefficient, $\hat \eta = 1.0, 2.0, 4.0$. Each point is an average over $10$ samples, apart from the lowest $T$ (largest $\xi$ and $\tau$) at $\hat \eta=4$, for which we have 4 samples (one such sample takes 7 days to run on a i7-8700-3.20GHz CPU workstation).  Lines are the best fit to $z=1.5$ (low friction - $\hat \eta=1.0, 2.0$) and $z=2$ (large friction - $\hat \eta=4.0$). }
    \label{fig:fitz}
\end{figure}

The numerical findings fully confirm this scenario. In Fig.\ref{fig:fitz} we show results for three different sets of simulations, respectively for $\hat \eta = 1,2,4$. We cannot use larger values for the effective dissipation, because the maximum relaxation time becomes too long to equilibrate the system.
For the smaller values ($\hat \eta = 1,2$), the data are in good agreement with a dynamic critical exponent $z=3/2$, while for $\hat \eta=4$ the characteristic time scales with the correlation length with an exponent $z=2$. We therefore conclude that the ISM exhibits a dynamic crossover in critical behavior, as predicted by the RG approach.

\begin{figure}[t]
    \centering
    \includegraphics[scale = 0.25]{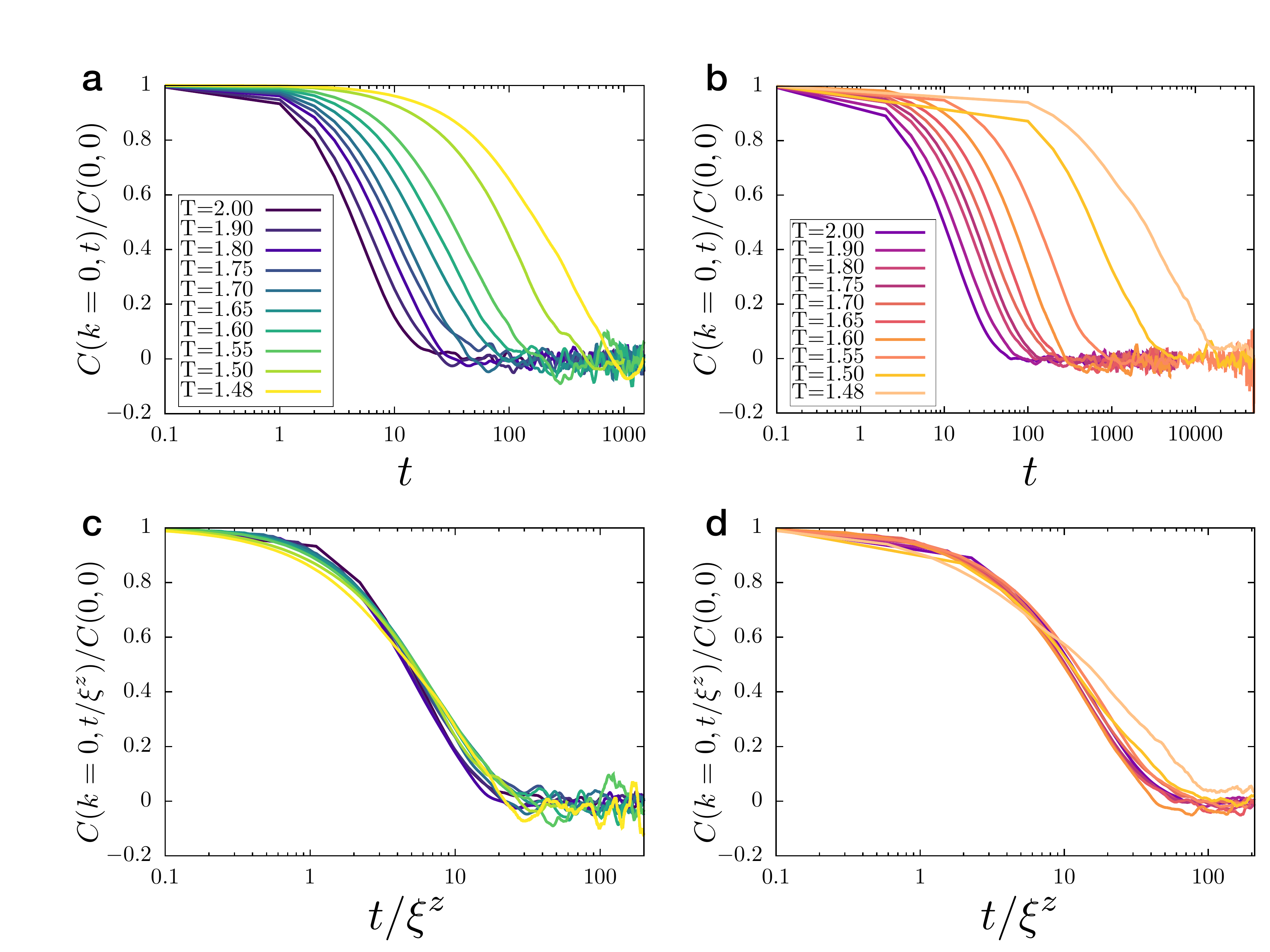}
    \caption{ \textbf{Dynamic scaling for correlations.} Test of the dynamic scaling hypothesis on the dynamic correlation functions at $k=0$. Upper panels: spatio-temporal correlation functions at various values of the temperature for  $\hat \eta = 1$ (panel a) and $\hat \eta =4$ (panel b). Lower panels: same curves plotted  as a function of  $t/\xi^z$ with, respectively, $z=1.5$ (panel c) and $z=2$ (panel d): in both cases the functions verify the dynamic scaling hypothesis.}
    \label{fig:dynscaling}
\end{figure}

To further support the existence of two distinct regimes in dynamical behavior, we tested  the full dynamic scaling hypothesis (\ref{dynscal}) on the dynamic correlation functions. In Fig. \ref{fig:dynscaling}, upper panels,  we display the normalized $C(k=0,t)$ for all the temperatures that we analyzed, and for two  different values of $\hat\eta$. In the lower panels, we report the same curves but plotted as a function of the rescaled variable $t/\xi^z$, where we used the values of $z$ obtained from the previous analysis. The figure shows that dynamic scaling is nicely verified, but with different exponents ($z=3/2$ and $z=2$, respectively) at small and large values of the friction coefficient.

\subsection{Natural swarms and inertial dynamics}
Both theoretical computations and numerical simulations describe a dynamic crossover between two different critical regimes, which is ruled by the interplay of the correlation length $\xi$ vs. the conservation length scale ${\mathcal R}_0$.  
Our analysis has important consequences when considering systems of finite size. What we have shown is that, even in presence of dissipation, the critical behavior of the system can be ruled by a conservative critical dynamics with exponent $z=3/2$ (in $d=3$) in an extensive region of parameters (case ${\cal R}_0> \xi^{3/4}$; red region in Fig.\ref{fig:protocols}). This result is particularly relevant if we think back at the biological motivation of our study: explaining experimental data in natural swarms of insects. As discussed in Section II, swarms exhibit dynamic scaling, but with an exponent smaller than the one  predicted by models of collective motion with a purely dissipative dynamics. This is why we considered the ISM in the first place: to put back inertial  terms in the dynamical equations and to understand whether they can produce a $z<2$ on the collective scale of living groups. The answer to this question is therefore yes. The exponent that we get in the conservative region, $z=1.5$ is not yet the value observed in the data ($z=1$), but it is a big step forward as compared to the prediction of the Vicsek model ($z=2$). This strongly indicates that the ISM captures an important ingredient - inertia - absent in previous models.

\begin{figure}[t]
    \centering
    \includegraphics[width=0.5 \textwidth]{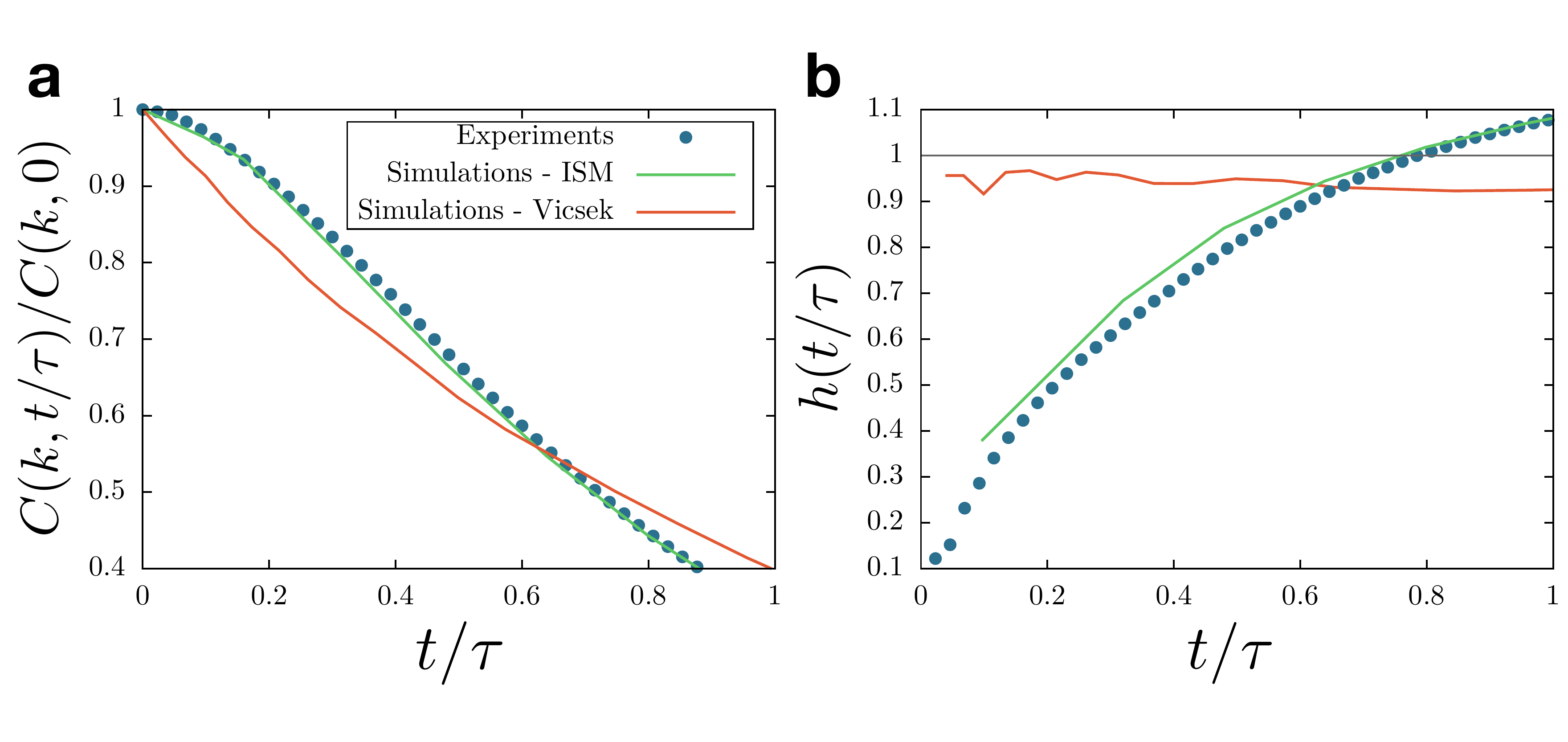}
    \caption{\textbf{Inertial behavior: Experiments vs Models.} a) Normalized dynamical correlation functions $C(k,t)/C(k,0)$ at nonzero values of $k$; in all three cases $k$ has been chosen in such a way to have $k\xi=1$, to reproduce the scaling situation found in experiments on natural swarms \cite{cavagna2017dynamic} (Vicsek swarm $k=0.717$, natural swarm $k=0.798$ and ISM $k=0.673$). b) The relaxation form factor $h (t/\tau) \equiv \dot{C}(t/\tau)/C(t/\tau)$, goes to $1$ for overdamped exponential relaxation, while it goes to $0$ for inertial relaxation \cite{cavagna2017dynamic}. The fixed-network ISM reproduces the correlation form of real swarms in a rather compelling way.}
    \label{fig:inertia}
\end{figure}

Numerical simulations of the ISM also reproduce another feature measured in natural swarms, which is not reproduced by previous models. Experimental correlation functions in natural swarms display a concave shape at short times, incompatible with the exponential relaxation predicted by the Vicsek model \cite{cavagna2017dynamic}. The ISM on the other hand displays the same kind of behavior as in the swarms data. To show this,   
in Fig.\ref{fig:inertia}a we compare the dynamical relaxation of natural swarms with simulations of the ISM and of the Vicsek model in the paramagnetic phase. We can see that ISM reproduces the curvature of the experimental correlation for $t \to 0$, contrary to the Vicsek model.  The consistency between ISM and natural swarms becomes even more striking when we compute the relaxation form factor \cite{cavagna2017dynamic},
\begin{equation}
    h (t/\tau) \equiv \frac{\Dot{C}(t/\tau)}{C(t/\tau)}  \ .
\end{equation}
The limit of this function for $t \to 0$ is equal to $1$ if the dynamics is purely exponential, as in the case of the Vicsek model. On the other hand, an inertial dynamics approaches zero for small times: this is the case of ISM, and of natural swarms. We therefore conclude that the ISM in the paramagnetic phase qualitatively well describes the  inertial dynamics of natural swarms. 


\section{Conclusions}

We have performed a one-loop RG calculation of the critical dynamics of a statistical system with inertial non-dissipative couplings, in presence of a dissipative term which violates the conservation law of the symmetry generator. Our calculation was motivated by recent experiments on the collective dynamics of natural swarms of insects \cite{cavagna2017dynamic}, although the dynamical field equations we studied are relevant also for BEC systems with terms weakly violating the symmetry in the Hamiltonian \cite{bec_magnets, eisenstein2004bose, klaers2010bose}. We find that the RG flow has two fixed points, a conservative yet unstable one, and a dissipative stable fixed point, associated to the dynamical critical exponents $z=d/2$ and $z=2$, respectively. The crossover between the two fixed points is regulated by a conservation length scale, ${\cal R}_0$: for scales much larger than ${\cal R}_0$, the dynamics is ruled by the dissipative fixed point, while for scales smaller than ${\cal R}_0$ critical slowing down is governed by the conservative fixed point. Numerical simulations on the microscopic model confirm our results.

The crossover length scale, ${\cal R}_0$, is determined by the ratio between the transport coefficient, $\lambda_0$, and the effective friction, $\eta_0$, of the spin field. If the coarse-grained parameter $\eta_0$ is certainly connected to its microscopic counterpart, $\hat \eta$, in the original model, the same cannot be said for the transport coefficient, as in the original microscopic model  there is no transport term. The interesting fact, then, is that the conservative transport term, $\lambda_0 \nabla^2 \vm$, is explicitly generated by the renormalization group, through the spin self-energy $\Pi$ at one loop. Therefore, we are in one of those rare cases in which a crucial length scale of the system, i.e. ${\cal R}_0$, cannot be guessed purely on the basis of dimensional analysis of the microscopic equations of motion (possibly with some renormalized anomalous dimensions). Of course, one could have guessed (admittedly rather smartly) that the presence of a symmetry and conservation law, albeit violated by $\eta_0$, should require a conservative transport term. But in case our intuition were not so good, the RG would require by itself the existence of such term, and therefore the emergence of a crossover length scale, thus confirming its power in dictating what is relevant and what is not in strongly correlated systems.

The fact that the crossover length scale ${\cal R}_0$ is larger the smaller the dissipation has important consequences for biological systems. Real biological groups always have finite size, hence in order to study their behaviour we cannot just take for granted the hydrodynamic limit (infinitely large times and distances), but we have to cope with the actual size of the system. In both flocks and swarms, experiments have shown that dissipative terms are rather weak \cite{attanasi+al_14, cavagna2017dynamic}, hence suggesting that the conservation length scale ${\cal R}_0$ is quite large. Under these circumstances one may have a conservation length scale that is {\it larger} than the system's size, ${\cal R}_0> L$. In this case, one expects to find a dynamical critical exponent equal to that of the fully conservative RG fixed point, namely $z=3/2$ in $d=3$, and a  dynamic correlation function with strong signature of non-exponential inertial relaxation. Thanks to this finite-size critical crossover, the fully conservative phenomenology should hold at all practically attainable values of the correlation length, which is always limited by the system's size.

From the point of view of the comparison between theory and experiments in natural swarms, our calculation therefore puts us in a semi-satisfactory situation. Certainly we can say that the form of the dynamical correlation functions of natural swarms, and in particular the non-exponential inertial nature of the short times dynamics, is rendered by the ISM in a much more compelling way than the Vicsek model (Fig.7); actually, our simulations show almost no quantitative difference between theory and experiments in this respect. Concerning the dynamical critical exponent, $z$, the situation is still open, although we would say that the result of the present calculation  - namely $z=3/2$ in finite-size weakly-damped $3d$ swarms - definitely goes in the right direction. Experiments give $z\approx 1$, even though values up to $z=1.2$ would probably be acceptable, given the noise in the data \cite{cavagna2017dynamic}. On the other hand, the Vicsek model, and in fact any model dominated at short times by purely dissipative dynamics, gives $z\approx 2$. This is quite understandable, as all these models belong (at equilibrium) to the same dynamical universality class as classic Heisenberg (Model A of \cite{hohenberg1977theory}), which has $z=2$ at one loop level, with very small corrections at two loops  \cite{hohenberg1977theory}. Moreover, when off-equilibrium (self-propelled) effects are taken into account, numerical simulations performed over time and space scales comparable to real swarms still give $z\approx 2$ \cite{cavagna2017dynamic}, completely incompatible with the data. The present calculation, on the other hand, shows that, once non-dissipative terms are introduced in the dynamics, and provided that dissipation is not too strong, the dynamical critical exponent changes already at one loop, giving $z=3/2$ in three dimensions. This is a value significantly closer to the experimental exponent than that of purely dissipative models. Hence, it seems to us that non-dissipative terms are important to reproduce the correct critical dynamics of real swarms.

Of course, one must now ask how to bridge the gap between the one-loop RG exponent, $z=3/2$, and the experimental value, $z\approx 1$. There are several possibilities. First, one should try to have more statistics in the experiments, possibly with larger swarms, to check whether or not the data are really inconsistent with $z=3/2$; work in this direction requires considerable technical effort on the experimental side (in particular, higher definition and faster acquisition systems). Secondly, one may hope that a two-loop calculation improves things. We are not very optimistic in this respect, though. Normally, two-loop corrections to the exponents are quite small, so it seems hard to bridge the gap between $1.5$ and $1$ in this way; furthermore, in the non-dissipative case the value $z=3/2$ is actually valid at all order of the perturbative series, courtesy of the Ward identities generated by the symmetry \cite{halperin1976renormalization}. Although in our case there is dissipation, we suspect that, as long as the system is in the proximity of the conservative RG fixed point, $z=3/2$ will resist any attempt to be perturbatively changed.

Finally, there is the third and most promising source of corrections to $z$, namely off-equilibrium effects due to the self-propulsion of the individuals. Even though these are not sufficient to change the critical exponent in the Vicsek model, it could be that the compound effect of having non-dissipative inertial couplings {\it and} a self-propelled dynamics, further shifts the exponent in the correct direction. Studying this case from the theoretical point of view (i.e. by using RG), will be quite non-trivial, as one needs to use the approach of Toner and Tu \cite{toner_review}, including in the theory one extra field, the density, coupled to velocity and spin, much as it has been done in \cite{cavagna2015silent} for the low temperature phase. However, at low temperature one could exploit the spin-wave expansion to linearize the equations, while close to $T_c$, which is the case of interest for swarms, one needs to fully take into account the non-linearities through the RG. Performing even a one loop calculation with three fields (which become six once we use the Martin-Siggia-Rose representation) really does not look like a piece of cake. Still, one should try. In the meanwhile, numerical simulations of the full-fledged self-propelled ISM close to criticality should be performed, to see from the data if there is case for hope. 

Actually, we have some reasons to be optimistic. The fact that models with non-dissipative terms have dynamical critical exponent $z$ significantly smaller than the purely dissipative value $2$ may be interpreted as a critical counterpart of the linear spin-wave behaviour at low $T$: in this regime, `second sound' modes propagate linearly, with dispersion relation $\omega = c k$ \cite{cavagna_review}. Naively, this relation would suggest $z=1$ for these systems, but this is not the case, because close to $T_c$ parameters renormalize, so that that the second sound speed, $c$, goes to zero as some function of $k$; this RG-induced $k$ dependence changes the exponent from the trivial $1$, to the final $z=3/2$ in this kind of models \cite{hohenberg1977theory}. Despite this correction, though, the exponent remains significantly lower than the purely dissipative $2$, as a relic of the low-temperature spin-wave dynamics. We may hope that a similar mechanism will be at work when self-propulsion will be taken into account. The first obvious effect of self-propulsion on a system with non-dissipative mode-coupling terms is to produce ballistic (i.e. linear) motion of each individual, even in the disordered collective phase. The dynamic critical exponent does not measure the motion of the individuals, of course, but rather the relaxation law of the velocity fluctuations; however, similarly to what happens with the renormalization of linear spin-waves, one may hope that some relic of the ballistic regime creeps into the critical phase calculation of $z$, thus lowering it below the static equilibrium value, $3/2$, eventually bringing it closer to the experimental value. Further experimental, numerical and theoretical effort will tell whether this educated guess is just wishful thinking or not.


\section*{Acknowledgements}
We warmly thank Enzo Branchini, Erwin Frey and Luca Peliti for important discussions, Lara Benfatto, Claudio Castellani and Jose G. Lorenzana for suggestions about BEC systems, and Stefania Melillo for help with the experimental data of \cite{cavagna2017dynamic}.  
This work was supported by ERC Advanced Grant RG.BIO (contract n. 785932) to AC, and ERANET-CRIB Grant to AC and TSG.  TSG was also supported by grants from CONICET, ANPCyT and UNLP (Argentina). 

%
\appendix
\section{Perturbation expansion}\label{app:msr}

\subsection{Martin-Siggia-Rose formalism}


%
The Martin-Siggia-Rose (MSR) formalism is a method to write stochastic differential equations as a field theory formulated using path integrals. The core idea is that, when computing thermal averages, of all possible field configurations only  those 
satisfying the original equations do contribute.
One can select such configurations using a Dirac delta functional: suppose that we want to select only configurations of the field $\bpsi$  that satisfy the equation $\ \mathcal{F} (\bpsi) -\boldsymbol{\theta}= 0$, where $ \mathcal{F} (\bpsi)$ generically describes the deterministic part of the equation (i.e. time derivatives, differential operators as well as interaction terms, coupling with other fields etc), and $\boldsymbol{\theta}$ is the stochastic noise. If $\det( \delta\mathcal{F}/\delta \bpsi)=1$, which is the case for stochastic Langevin equations in the Ito representation \cite{cardy1996scaling,de2006random}, we can write:
\begin{equation}
    1 = \int \mathcal D \bpsi(\bx,t) \ \delta( \mathcal{F} (\psi(\bx,t))-\boldsymbol{\theta}(\bx,t)) 
\end{equation}
We can introduce the field  $\hat \bpsi(x,t)$ and use the integral representation of the Dirac delta functional:
\begin{equation}
    1 = \int \mathcal D\bpsi  \mathcal D \hat \bpsi
e^{\left \{-i \int d^d x \int dt \ \hat \bpsi(\bx,t) \left [ \mathcal {F} (\bpsi(\bx,t))-\boldsymbol{\theta}(\bx,t)\right ] \right\}}
\end{equation}
The field $\hat \bpsi$ may also be interpreted as a Lagrange multiplier, since it is introduced to select given values, or rather configurations, of the field $\bpsi$. In our case, since (\ref{EqOfMotion1}) and \eqref{EqOfMotion2} are two coupled equations for two distinct fields, we need to implement two delta functions, and therefore introduce two auxiliary fields $\hat \bpsi$ and $\hat\vm$. The above identity can then be averaged over the distribution of the stochastic noises, leaving with an effective functional measure that can be used to compute thermal averages, i.e.
\begin{align}
1 &= \int \mathcal D\bpsi  \mathcal D \hat \bpsi \ \mathcal D\vm  \mathcal D \hat \vm e^{-\mathcal{S}[\bpsi,\hat \bpsi,\vm,\hat \vm]} \\
\langle f \rangle & = \int \mathcal D\bpsi  \mathcal D \hat \bpsi \ \mathcal D\vm  \mathcal D \hat \vm f  e^{-\mathcal{S}[\bpsi,\hat \bpsi,\vm,\hat \vm]} \ ,
\label{msr-measure}
\end{align}
where $f$ is a generic function of the fields. After standard manipulations \cite{cardy1996scaling}, we get,
\begin{widetext}
\begin{equation}
\mathcal{S}[\bpsi,\hat \bpsi,\vm,\hat \vm] = \mathcal{S}_{0,\psi}[\hat \bpsi,\bpsi] + S_{0,s}[\hat \vm,\vm] 
+ \mathcal{S}_I [\bpsi,\hat \bpsi,\vm,\hat \vm]  \quad \quad
\label{actions}
\end{equation}
Here $\mc S_{0,\psi}$ and $\mc S_{s,0}$ are Gaussian free actions respectively for the field $\bpsi$ and $\vm$ and are given by:
\begin{equation}
\begin{split}
\mc S_{0,\psi}  = \int \ddk k  \dw\omega\hat \psi_\alpha(-\bk,-\omega)[-i\omega + \Gamma_0(k^2+r_0^2)] \psi_\alpha(\bk,\omega)
+ \Gamma_0 \hat \psi_\alpha(-\bk,-\omega) \hat \psi_\alpha(\bk,\omega) \qquad \qquad \quad \qquad \qquad
\end{split}
\label{free1}
\end{equation}	
\begin{equation}
\begin{split}
\mc S_{0,s} = \int  \ddk k \dw \omega \hat s_\alpha (-\bk,-\omega)\left[-i\omega +(\eta_0 +\lambda_0 k^2)\right]s_\alpha(\bk,\omega) +  (\eta_0 +\lambda k^2)\hat s_\alpha(-\bk,\omega) \hat s_\alpha(\bk,\omega) \qquad \quad \qquad
\end{split}
\label{free2}
\end{equation}
where greek letters stand for space coordinates and repeated indexes are summed. The interaction term involves both $\vm$, $\hat \vm$ and $\bpsi$, $\hat \bpsi$ and is given by: 	
\begin{equation}
\begin{split}
\mc S_I = -&g_0 \epsilon_{\alpha \beta \gamma} \int \ddk {k_1} \ddk {k_2} \dw {\omega_1}  \dw {\omega_2}(k_2^2-k_1^2) 
\psi_\gamma(\bk_2,\omega_2) \psi_\beta(\bk_1,\omega_1) \hat s_\alpha(-\bk_1-\bk_2,-\omega_1-\omega_2) 
\\ 
-&\frac{g_0}2 \epsilon_{\alpha \beta \gamma} \int \ddk {k_1}  \ddk {k_2} \dw {\omega_1} \dw {\omega_2}\hat \psi_\alpha(\bk_1,\omega_1)  \psi_\beta(\bk_2,\omega_2) s_\gamma(-\bk_1-\bk_2,-\omega_1-\omega_2)
\\
-& 2 \Gamma_0 u_0
 \int \ddk {k_1}\ddk {k_2} \ddk {k_3}
   \dw {\omega_1} \dw {\omega_2}  \dw {\omega_3}
\hat \psi_\alpha(\bk_1,\omega_1)  \psi_\alpha(\bk_2,\omega_2)\psi_\alpha(\bk_3,\omega_3) \psi_\alpha(-\bk_1-\bk_2-\bk_3,-\omega_1-\omega_2-\omega_3)
\end{split}
\end{equation}
From the free part of the action (\ref{free1})(\ref{free2}) we immediately read the expressions  for the bare propagators and correlation functions for the effective field theory, which coincide with Eqs. \eqref{freepropPsi} \eqref{freepropS} and (\ref{freecorrePsi})(\ref{freecorreS}):
\begin{align}
\mean{ \psi_\alpha(-\bk,-\omega) \hat \psi_\beta(\bk,\omega)} &=\delta_{\alpha \beta} G_{0,\psi} (\bk,\omega)=\delta_{\alpha \beta} \left [ -i \omega + \Gamma_0 (k^2 + r_0) \right ]^{-1}\\
\mean{\psi_\alpha(-\bk,-\omega) \psi_\beta(\bk,\omega)} &= \delta_{\alpha \beta} C_{0,\psi}(\bk,\omega) = 2 \delta_{\alpha \beta} \Gamma_0 | G_{0,\psi}|^2 \\
\mean{ s_\alpha(-\bk,-\omega) \hat s_\beta(\bk,\omega)} &=\delta_{\alpha \beta} G_{0,s} (\bk,\omega) = \delta_{\alpha \beta} \left [ -i \omega +( \eta_0 + \lambda_0 k^2)\right ]^{-1} \\
\mean{s_\alpha(-\bk,-\omega) s_\beta(\bk,\omega)} &= \delta_{\alpha \beta} C_{0,s} (\bk,\omega) = 2\delta_{\alpha \beta} (\eta_0 + \lambda_0 k^2) | G_{0,s}|^2
\end{align}
\end{widetext}
 These functions are the building blocks of the perturbative expansion: full correlation functions and propagators can be written in terms of these bare averages. At this point, standard Feynman rules can be applied to carry out the perturbation theory.

 From the form of the interacting part of the action, we can see that there are two kinds of dynamic vertices, namely:
\begin{equation}\feynmandiagram[small,baseline=(d.base),horizontal=d to b] {
	a[particle=\( \hat \psi_\alpha \)] --  b [dot] -- c[particle=\(  \psi_\beta \)] ,b -- [boson] d [particle=\(s_\gamma \)],
};-g_0 \epsilon_{\alpha \beta \gamma}s_\gamma(1) \hat \psi_\alpha(2)\psi_\beta(-1-2) 
\label{vertex1}
\end{equation}

\begin{equation}
\feynmandiagram[small,baseline=(d.base),horizontal=d to b] {
	a[particle=\(  \psi_\gamma \)] --  b [empty dot] -- c[particle=\(  \psi_\beta \)] ,b -- [boson] d [particle=\(\hat s_\alpha \)],
};-\frac{g_0}{2} \epsilon_{\alpha\beta\gamma}(k_2^2-k_1^2) \psi_\beta(1)\psi_\gamma(2)\hat s_\alpha(-2-1) 
\label{vertex2}
\end{equation}
Here we are representing with a solid line the fields $\bpsi$, $\hat \bpsi$, and with  wavy lines the fields $\vm $,$\hat \vm $; for clarity, we are indicating with $\pm n$ the dependence of the
fields on wave-number and frequency: $\psi_\alpha (\pm n) = \psi_\alpha(\pm \bk_n, \pm \omega_n)$. Beside these two dynamic vertices there is also the vertex related to the static interaction coupling $u_0$. Since we are focusing on the contributions of purely dynamic origin to the perturbative expansion we are not concerned with that vertex in our discussion. We just give for granted that the perturbation expansion related to $u_0$ gives back the terms of the equilibrium theory and refer the reader to the standard literature for more details \cite{cardy1996scaling,amit2005field}

At this level, we should notice that the second vertex carries with it an important factor $(k_2^2 -k_1^2)$. This is a consequence of the reversible couplings between the field $\boldsymbol{\psi}$ and $\vm$ present in the equations of motion.
Its origin lies in the symmetries of the system: the spin is the generator of rotations of the order parameter and, consequently, the reversible couplings between the fields occur just via a cross product. The field $\hat \vm$ therefore couples only with $(\nabla \boldsymbol{\psi})^2$; for this reason the vertex is proportional to $(k_2^2-k_1^2)$, and this implies that every diagram with an  $\hat \vm (\bk=0,\omega)$  external line is null.

\subsection{Perturbation expansion at one loop}
To compute average quantities with the measure (\ref{msr-measure}) one proceeds as usual to develop the exponential contribution due to the interaction action, being left with a perturbation expansion where only free propagators and free correlations appear, connected to each other through the various interaction vertices.
When building the full averages in such a way, we have to take into account that both $\mean{\hat \bpsi \bpsi}_0$ and $\mean{\bpsi \bpsi}_0$ are non zero. To graphically distinguish between them, we will represent the propagators with an arrow and the correlation functions with a line, since propagators are time ordered while correlation functions are not.  We will use the same rules also for propagators and correlation functions of $\vm$, but using wavy lines. 
It is more convenient to write down the perturbative expansion of $G$ using the Dyson equation \cite{ryder1996quantum}:
\begin{align}
G_{\psi}^{-1}(\bk,\omega)_{\alpha \beta} &= G_{0,\psi}^{-1}(\bk,\omega)\delta_{\alpha \beta} - \Sigma_{\alpha \beta} (\bk,\omega) \\
G_{s}^{-1}(\bk,\omega) _{\alpha \beta} &=  G_{0,s}^{-1}(\bk,\omega) \delta_{\alpha \beta}- \Pi_{\alpha \beta} (\bk,\omega) 
\end{align}
For which, we use the following diagrammatic notation:

\begin{equation}
G_{0, \psi_{\alpha,\beta}} =  \quad 
\feynmandiagram[layered layout,small,horizontal = a to b]{
	a -- [fermion] b};   \qquad 
C_{0, \psi_{\alpha,\beta}} =  \quad 
\feynmandiagram[layered layout,small,horizontal = a to b]{
	a --  b};
\end{equation}
\begin{equation}
G_{0, s_{\alpha, \beta}} =  \quad 
\feynmandiagram[layered layout,small,horizontal = a to b]{
	a -- [charged boson]b};   \qquad 
C_{0, \psi_{\alpha,\beta}} =  \quad 
\feynmandiagram[layered layout,small,horizontal = a to b]{
	a -- [photon] b};
\end{equation}

\begin{equation}
\Sigma_{\alpha, \beta} \qquad =  \qquad
\feynmandiagram[baseline,layered layout,small,horizontal = a to b]{
	a[particle =\( \quad \ \psi_\alpha\)] -- b[ blob]
	--c[particle =\( \hat \psi_\beta \quad \  \)]}; 	
\end{equation}

\begin{equation}
\Pi_{\alpha, \beta} \qquad =  \qquad
\feynmandiagram[baseline,layered layout,small,horizontal = a to b]{
	a[particle =\( \quad \ s_\alpha\)] --[photon] b[ blob]
	--[photon]c[particle =\( \hat s_\beta \quad \ \)]}; 	
\end{equation}
Here the blob indicates the sum of all 1PI diagrams with an incoming $\bpsi$ (or $\vm$) field and an out coming $\hat \bpsi$ (or $\hat \vm$) field and with amputated external legs: namely, the self-energies $\Sigma_{\alpha \beta}$ and $\Pi_{\alpha \beta}$.

The diagrammatic expressions for the self-energies of $\vm$ and $\boldsymbol{\psi}$ at one loop are:
\begin{align}
\Sigma_{\alpha\beta}=&
\feynmandiagram[baseline,layered layout,small,horizontal= o1 to i1]{
	o1 [particle=\( \qquad \ \psi_\alpha \)]-- i1[empty dot] --[half left] i2[dot] --[anti charged boson,half left] i1 ,
	i2 --o2[dot,particle=\(\hat \psi_\beta \quad \ + \)]
};    
\feynmandiagram[baseline,layered layout,small,horizontal= o1 to i1]{
	o1[particle =\(\qquad  \psi_\alpha \)] -- i1[dot] --[half left,fermion] i2[dot] --[boson,half left] i1 ,
	i2 --o2[particle=\(\hat \psi_\beta \qquad \ \)]
}; \qquad
\label{Diagrams_Sigma} \\
\Pi_{\alpha\beta} = &
\feynmandiagram[baseline,layered layout,small,horizontal= o1 to i1]{
	o1[particle=\(\qquad \ s_\alpha\)] --[photon] i1[ dot] --[half left,fermion] i2[empty dot] --[half left] i1 ,
	i2 -- [photon]o2[particle=\(\hat s_\beta\qquad \ \)]
};
\label{Diagrams_Pi}
\end{align}
where external legs are amputated. It is possible to translate these diagrams into integrals using standard Feynman diagrams rules:
\begin{widetext}
\begin{equation}
\begin{split}
\Sigma_{\alpha\beta}( \bk,\omega) &= 
-2g_0^2\delta_{\alpha\beta} \int \ddk p  \int \dw  {\omega^\p} \biggl[ G_{0,\psi}(\bp,\omega^\p) C_{0,s}(\bk-\bp,\omega-\omega^\p) 
+(k^2-p^2) C_{0,\psi}(\bp,\omega^\p) G_{0,s}(\bk-\bp,\omega-\omega^\p) \biggr] \qquad \label{sup-self1}
\end{split}
\end{equation}
\begin{equation}
\begin{split}
\Pi_{\alpha\beta}(\bk,\omega)& = - 2 g_0^2\delta_{\alpha\beta}   \int \ddk p  \int \dw  {\omega^\p} \biggl[C_{0,\psi}(\bp,\omega^\p) G_{0,\psi}(\bk-\bp,\omega-\omega^\p) ((\bk-\bp)^2-p^2)
 \biggr]
\end{split}
\end{equation}
Performing the frequency integration, we get the following expressions for the self energies:
\begin{equation}
\begin{split}
\Sigma_{\alpha\beta}(\bk,\omega)= &
-2 g_0^2 \delta_{\alpha\beta}\int \ddk p  
\frac {(k^2+r_0)}{(p^2+r_0)(-i\omega + \Gamma_0(p^2+r_0) +\lambda_0(\bk-\bp)^2+\eta_0 ) }
\end{split}
\end{equation}
\begin{equation}
\begin{split}
\Pi_{\alpha\beta}(\bk,\omega) = 
 -g_0^2 \delta_{\alpha\beta}\int \ddk p \frac 1 {(p^2+r_0)((\bk-\bp)^2+r_0)} \frac{ [p^2-(\bk-\bp)^2]^2}{(-i\omega + \Gamma_0(p^2+(\bk-\bp)^2 +2r_0))} \\
\end{split}
\end{equation}
\end{widetext}
We thus find that the self-energies only have, as expected, a diagonal non-zero contribution for $\alpha=\beta$. We shall therefore drop the coordinate index and simply indicate them as $\Sigma$ and $\Pi$, as in Eqs.(\ref{sigma})(\ref{Pi}) in the main text. When the perturbative corrections are calculated integrating over the shell, as in the RG approach we use in Section IV, the integrals are performed between $\Lambda/b$ and $\Lambda$. On the the other hand, in the Callan-Symanzik approach of Section VI all $p$ integrals are performed between $0$ and $\infty$.

\newpage
\section{Vertex corrections}\label{app:g0}
It can be shown that the dynamic coupling constant $g_0$ has no perturbative contributions at all orders of perturbation theory. At one loop the correction $\Delta g^{(1)}_{\alpha \beta \gamma}$, of order $g_0^3$, to vertex (\ref{vertex1}) comes from these two diagrams:
\begin{equation}
 \Delta g^{(1)}_{\alpha \beta \gamma}  = 
 \feynmandiagram[baseline={(current bounding box.center)},small,horizontal= o1 to i1] {
	o1[particle=\(
s_\gamma\)] --[photon,edge label = \(k_3 \)] i1[dot]--[fermion ]i2[dot]--[photon]i3[dot]--[anti fermion]i1,
	i2--[edge label = \(k_1 \)]p2[particle=\(\hat\psi_\alpha\)],
	i3--[edge label = \(k_2 \)]p3[particle=\(\psi_\beta\)],
}; + \quad
\feynmandiagram[baseline={(current bounding box.center)},small,horizontal= o1 to i1] {
	o1[particle=\(s_\gamma\)] --[photon,edge label = \(k_3\)] i1[dot]--[]i2[empty dot]--[charged boson]i3[dot]--[anti fermion]i1,
	i2--[edge label = \(k_1\)]p2[particle=\(\hat\psi_\alpha\)],
	i3--[edge label = \(k_2\)]p3[particle=\(\psi_\beta\)],
}; 
\end{equation}
After integration over the internal lines we get expressions of the kind:
\begin{equation}
\Delta g^{(1)}_{\alpha \beta \gamma} =\epsilon_{\alpha_1 \beta_1 \gamma} \epsilon_{\alpha_2 \beta \gamma_2}\epsilon_{\alpha \beta_3 \gamma_3}\tau_{\alpha_1 \beta_1 \alpha_2 \gamma_2 \beta_3 \gamma_3}(k_1,k_2,k_3) \ ,
\end{equation}
where we are summing over all repeated indices, and the tensor $\tau$ only depends on the internal indices and on the external momenta. At zero incoming momentum and frequency, because of the symmetry under exchange of the two internal lines of the field $\psi$,  $\tau$ becomes a symmetric tensor. In particular $\tau(0,0,0)$  is symmetric under exchange of indices $\alpha_1$ and $\beta_1$. Therefore the contraction between $\epsilon_{\alpha_1 \beta_1 \gamma}$ and $\tau$ is zero.
Other possible one loop corrections may come from both static and dynamic vertices of the kind
\begin{equation}
\Delta g^{(2)}_{\alpha \beta \gamma} = 
\feynmandiagram[baseline={(current bounding box.center)},small,horizontal= o1 to i1] {
	o1[particle=\(s_\gamma\)] --[photon] i1[dot]--[fermion,half left ] i2[dot]--[half left]i1,
	i2--p2[particle=\(\hat\psi_\alpha \)],
	i2--p3[particle=\(\psi_\beta\)],
};
\end{equation}
which is of order $g_0 u_0$. Also in this case, at zero incoming momentum and frequency, the correction $\Delta g^{(2)}_{\alpha \beta \gamma}$ is zero by symmetry. 

It is possible to extend this reasoning to all orders in perturbation theory; the full perturbative expansion $\Delta g_{\alpha \beta \gamma}$ of vertex (\ref{vertex1})   satisfies the following diagrammatic equation
\begin{equation}
\Delta g_{\alpha \beta \gamma}= 
 \feynmandiagram[small,baseline=(d.base),horizontal=d to b] {
	a[particle=\(  \hat \psi_i \)] --  b [blob] -- c[particle=\(  \psi_j \)] ,b -- [boson] d [particle=\(s_\gamma \)],
}; \qquad \qquad \\  
\end{equation}

\begin{equation}
\Delta g_{\alpha \beta \gamma}= 
\feynmandiagram[baseline={(current bounding box.center)},small,horizontal= o1 to i1] {
	o1[particle=\(s_\gamma\)] --[photon] i1[dot]--[fermion,half left ] i2[blob]--[half left]i1,
	i2--p2[particle=\(\hat\psi_\alpha \)],
	i2--p3[particle=\(\psi_\beta\)]
};+ \quad \feynmandiagram[baseline={(current bounding box.center)},small,horizontal= o1 to i1] {
	o1[particle=\(s_\gamma\)] --[photon] i1[dot]--[fermion,half left ] i2[blob]--[anti fermion,half left]i1,
	i2--p2[particle=\(\hat\psi_\alpha \)],
	i2--p3[particle=\(\psi_\beta\)]
}; 
\end{equation}
where we explicitly described the possible ways in which the external $s_\gamma$ line can attach to the correction diagram. The result is zero for the same symmetry reason as above;  the dynamic coupling constant $g_0$ therefore has no perturbative corrections at all orders in perturbation theory.

\section{Ward Identities}\label{app:wi}
The fact that $g_0$ has no corrections at all orders in perturbation theory is related to the presence of Ward Identities relating response functions (or, equivalently vertex functions \cite{ryder1996quantum}) of different order. These identities derive from the fact that the spin is the generator of the rotational symmetry of the order parameter. In absence of dissipation the global spin is conserved. In this case, if the system is prepared in an equilibrium state with global polarization  $\langle \boldsymbol{\Phi}\rangle $ the effect of an homogeneous field $\bH(t)$ coupled to the spin is simply to rotate the polarization, i.e. 
\begin{equation}
\frac{d\langle \boldsymbol{\Phi}\rangle}{dt}=g_0 {\bH} \times \langle \boldsymbol{\Psi}\rangle
\end{equation}
Let us now consider a more complex situation where we apply two fields: the first one, $\bh(\bx,t)$, coupled to the local order parameter, and the second, 
$\bH(t)$, coupled to the spin. The first field will generate a space dependent local polarization $\langle \bpsi(\bx,t)\rangle$, the second field will simply homogeneously rotate  such local polarizations. If there is dissipation, and the global spin is not conserved, the field $\bH(t)$ will also change the value of the global spin, giving a further contribution to the rotation frequency of the $\langle \bpsi(\bx,t)\rangle$. Let us focus on the parts of both fields that are uniquely due to the presence of $\bH(t)$; from Eqs.(\ref{EqOfMotion1})(\ref{EqOfMotion2}) we get
\begin{align}
\frac{d\langle \delta\bpsi(\bx,t)\rangle}{dt}&=g_0 \left ( {\bH}(t) - \delta \vm(t)\right )\times  \langle\bpsi(\bx,t)  \rangle\\
\frac{d\langle \delta\vm(t)\rangle}{dt}&=-\eta_0 \delta \vm(t) + \eta_0 \bH(t)
\end{align}
where $\delta \vm$ is the change of spin per volume. Integrating both equations, we get
\begin{equation}
\begin{split}
\langle \delta\psi_\alpha(\bx,t)\rangle = &g_0 \epsilon_{\alpha \beta \gamma} \int_0^t dt''  \langle \psi_\gamma(\bx,t'') \rangle  \left [ H_\beta(t'')  \right.\\
 &-  \eta_0 \int_0^{t''} dt' e^{-\eta_0(t-t')}  H_\beta(t') \left. \right]  
\end{split}
\end{equation}
Both sides in this expression implicitly also depend on $\bh(\bx,t)$. Let us then derive with respect to this last field and then set it to zero. We get:
\begin{equation}
\begin{split}
&\frac{d\langle \delta\psi_\alpha(\bx,t)\rangle}{dh_\delta(\bx_1,t_1)}\biggr\rvert_{h=0}=g_0 \epsilon_{\alpha \beta \gamma}   \int_0^t dt''  H_\beta(t'')\\
&  \left [ R^h_{\gamma \delta}(\bx,t''; \bx_1,t_1) -  \eta_0  \int_{t''}^{t} dt' e^{-\eta_0(t'-t'')} R^h_{\gamma \delta}(\bx,t';\bx_1,t_1) \right ]  \ 
\label{wi}
\end{split}
\end{equation}
 where we relabelled integration variables in the second integral for future convenience. Here, $R^h_{\gamma \delta}(\bx,t;\bx_1,t_1)=\partial \langle \psi_\gamma(\bx,t)\rangle/\partial h_\delta(\bx_1,t_1)\vert_{h,H=0}$ is the linear response of the order parameter to its conjugate field. Using  response theory, the l.h.s. of (\ref{wi}) can also be written as 
\begin{equation}
\begin{split}
&\frac{d\langle \delta\psi_\alpha(\bx,t)\rangle}{dh_\delta(\bx_1,t_1)}\biggr\rvert_{h=0}=\\
 &\quad\quad\quad \int_0^t dt'' d\bx'' H_\beta(t'') R^{hH}_{\alpha\beta \delta}(\bx,t; \bx_1,t_1; \bx'',t'')  \ ,
\label{wi2}
\end{split}
\end{equation}
where now in the r.h.s  $R^{hH}_{\alpha\beta\delta}(\bx,t; \bx_1,t_1; \bx'',t'')=\partial^2\langle \psi_\alpha(\bx,t)\rangle/(\partial h_\delta(\bx_1,t_1)\partial H_\beta(\bx'',t''))\vert_{h,H=0} $ is the non-linear quadratic response. Equating the r.h.s. of (\ref{wi})(\ref{wi2}) we finally get
\begin{equation}
\begin{split}
\int d\bx'' R^{hH}_{\alpha\beta \delta}(\bx,t; \bx_1,t_1; \bx'',t'') = g_0 \epsilon_{\alpha \beta \gamma} \left[ R^h_{\gamma \delta}(\bx,t''; \bx_1,t_1) \right.\\
\left. -\eta_0    \int_{t''}^t dt'  e^{-\eta_0(t'-t'')}  R^h_{\gamma \delta}(\bx,t';\bx_1,t_1) \right ]  
\end{split}
\end{equation}
with $t_1<t''<t$. For $\eta_0=0$ this relation corresponds to the Ward identity reported for model E in \cite{halperin1976renormalization}.

\section{Shell integration}\label{app:shell}

To perform a RGT, as described in the main text, we need to implement two different steps: integration of short wavelength fluctuations, and rescaling. 
To this end, once fixed the coarse-graining factor $b$, it is convenient to rewrite the fields as the sum of two distinct components, one fluctuating on short wavelengths $\Lambda/b<k<\Lambda$ and and the other on larger ones $0<k<\Lambda/b$, i.e.
\begin{equation}
\bpsi(\bk,\omega)=\bpsi^<(\bk,\omega)+\bpsi^>(\bk,\omega)
\end{equation}
At this point, one integrates out explicitly from Eq.(\ref{msr-measure}) the $\bpsi^>$ fields, to remain with a measure and a new effective action that only depend on the $\bpsi^<$ fields. To perform this integration one proceeds, again, using perturbation theory. The basic ingredients of this perturbation expansion (free propagators and vertices) are the same as the ones discussed in the previous sections, the difference being that they refer to $\bpsi^>$ fields only, while the $\bpsi^<$ are kept fixed as external sources. The perturbation series therefore consists in diagrams with external $\bpsi^<$ legs and internal loops integrated over $>$ propagators. It can be recasted in exponential form, as usual, by only retaining one particle irreducible diagrams. These diagrams, that have external $\bpsi^<$ fields attached, will therefore modify the original terms appearing in the action. For example, for the Gaussian part of the action we get
\begin{widetext}
\begin{equation}
\mc S^{<}_{0,\psi}  = \int^{\Lambda/b} \ddk k  \dw\omega\hat \bpsi^<(-\bk,-\omega)[-i\omega + \Gamma_0(k^2+r_0^2)+\Sigma_b(\bk,\omega)] \bpsi^<(\bk,\omega)
+ \Gamma_0 \hat \psi_\alpha(-\bk,-\omega) \hat \psi_\alpha(\bk,\omega) \ ,
\end{equation}
\end{widetext}
where $\Sigma_b$ has the same expressions as in Eq.(\ref{sup-self1}), but where integrals are performed only in the shell $\Lambda/b<k<\Lambda$. From this expression we immediately see that the behavior of the self-energy $\Sigma_b$ at small $k$ effectively modifies the coefficient of $k^2$. We are then left with a free part of the action similar to the original one, but where integrals run only up to $\Lambda/b$. The second step of the RGT, namely the rescaling of $k$, $\omega$, $\bpsi$ and $\hat\bpsi$, has the purpose of reinstating momentum integrals over their original integration range. At one loop the renormalization of the field is trivial (i.e. related to its physical dimensions) and has therefore not been addressed explicitly in the main text. The result is a new action formally of the same kind as the original one but with a new renormalized kinetic coefficient $\Gamma_b$. A similar procedure can be applied also to the free action of the field $\bsp$, and to the interacting part. All the coefficients and coupling constants will get renormalized by the shell integration and rescaling. If we call ${\mathcal P}$ the set of all parameters entering the action, i.e. ${\mathcal P}\equiv \{ r_0,u_0,\Gamma_0,\eta_0,\lambda_0,g_0\}$, a RGT will therefore imply
\begin{equation}
\begin{split}
\mathcal{P}& \longrightarrow \mathcal{P}_b\\
 \mc S({\mathcal P}) &\longrightarrow \mc S_b =  \mc S(\mathcal{P}_b)
\end{split}
\end{equation}
Multiple iterations of the RGT therefore define a flow in the space of parameters, i.e. in the space of the statistical models defined by the action (\ref{actions}).


\bibliographystyle{apsrev4-1}
\bibliography{general_cobbs_bibliography_file}

\end{document}